\documentclass[12pt,a4paper]{article}
\usepackage{graphicx}
\usepackage{subeqnarray}
\begin{document}
%
%                         _________________ 
%                         |               |   
%                         |               |
%                         |  VERSIONE  B  |
%                         |               |
%                         _________________
%
%                      
%
\def\astrobj#1{#1}
\newenvironment{lefteqnarray}{\arraycolsep=0pt\begin{eqnarray}}
{\end{eqnarray}\protect\aftergroup\ignorespaces}
\newenvironment{lefteqnarray*}{\arraycolsep=0pt\begin{eqnarray*}}
{\end{eqnarray*}\protect\aftergroup\ignorespaces}
\newenvironment{leftsubeqnarray}{\arraycolsep=0pt\begin{subeqnarray}}
{\end{subeqnarray}\protect\aftergroup\ignorespaces}
\newcommand{\diff}{{\rm\,d}}
\newcommand{\appleq}{\stackrel{<}{\sim}}
\newcommand{\appgeq}{\stackrel{>}{\sim}}
\newcommand{\Int}{\mathop{\rm Int}\nolimits}
\newcommand{\Nint}{\mathop{\rm Nint}\nolimits}
\newcommand{\range}{{\rm -}}
%newcommand{\erf}{\mathop{\rm erf}\nolimits}
%\newcommand{\psfc}{\mathop{\rm psfc}\nolimits}
%\newcommand{\Psf}{\mathop{\rm psf}\nolimits}
\newcommand{\displayfrac}[2]{\frac{\displaystyle #1}{\displaystyle #2}}
\def\astrobj#1{#1}
%\begin{titlepage}
%\setcounter{page}{0}
%\headnote{Astron.~Nachr.~000 (2001) 0, 000--000}
%\makeheadline
%
\title{The G-dwarf problem in the Galaxy}
\author{{R.~Caimmi}\footnote{
{\it Astronomy Department, Padua Univ., Vicolo Osservatorio 2,
I-35122 Padova, Italy}
email: roberto.caimmi@unipd.it~~~
fax: 39-049-8278212}
\phantom{agga}}
%
%\medskip
%\small{Dipartimento di Astronomia}}
%
%\date{Received..................................................
%Accepted..................................................}
\maketitle
\begin{quotation}
\section*{}
\begin{Large}
\begin{center}
%\summary

Abstract

\end{center}
\end{Large}
\begin{small}

\noindent\noindent
This paper has two parts: one about observational
constraints, and the other about chemical evolution
models.   In the first part,
the empirical differential metallicity distribution
(EDMD) is deduced from three different samples involving
(i) local thick disk stars derived from Gliese and
scaled in situ samples within the range, $-1.20
\le$[Fe/H]$\le-0.20$ [Wyse, R.F.G., Gilmore, G., 1995,
AJ 110, 2771]; (ii) 46 likely metal-weak thick disk
stars within the range, $-2.20\le$[Fe/H]$\le-1.00$ 
[Chiba, M., Beers, T.C., 2000, AJ 119, 2843]; (iii)
287 chemically selected G dwarfs within 25 pc from
the Sun, with the corrections performed in order to 
take into account the stellar scale height [Rocha-Pinto,
H.J., Maciel, W.J., 1996, MNRAS 279, 447]; in addition
to previous results [Caimmi, R., 2001b, AN 322, 241;
Caimmi, R., 2007, NewA 12, 289] related to (iv) 372 
solar neighbourhood halo subdwarfs [Ryan,S.G., Norris,
J.E., 1991, AJ 101, 1865]; and (v) 268 K-giant bulge 
stars [Sadler, E.M., Rich, R.M., Terndrup, D.M., 1996,
AJ 112, 171].   The metal-poor and metal-rich EDMD
related to the thick disk shows similarities with their
halo and bulge counterparts, respectively.   Then the
thick disk is conceived as made of two distinct regions:
the halo-like and the bulge-like thick disk, and the
related EDMD is deduced.   Under the
assumption that each distribution is typical for
the corresponding subsystem, the EDMD of the thick disk, the
thick + thin disk, and the Galaxy, is determined by weighting 
the mass.   In the second part, models of chemical evolution 
for the halo-like thick disk, the bulge-like thick disk, and
the thin disk, are computed assuming the instantaneous
recycling approximation.
The EDMD data are fitted, to an acceptable extent, by
simple models of chemical evolution implying both
homogeneous and inhomogeneous star formation, provided that
star formation is inhibited during thick disk evolution,
with respect to the thin disk.   The initial
mass function (IMF) is assumed to be a universal
power law, which implies the same value of the
true yield in different subsystems.
The theoretical differential metallicity distribution
(TDMD) is first determined for the halo-like thich disk,
the bulge-like thick disk, and the thin disk separately, 
and then for the Galaxy by weighting the mass.
An indicative comparison is performed between
the EDMD deduced for the disk both in presence
and in absence of [O/Fe] plateau, and its
counterpart computed for (vi) $N=523$ nearby
stars within the range, $-1.5<$[Fe/H]$<0.5$, for
which the oxygen abundance has been determined
both in presence and in absence of the local thermodynamical
equilibrium (LTE) approximation [Ramirez, I.,
Allende Prieto, C., Lambert, D.L., 2007, A\&A 465,
271].   Both distributions are found to exhibit a
similar trend, although systematic differences exist.
In addition, the related empirical age-metallicity 
relation (EAMR) cannot be fitted by the theoretical
age-metallicity relation (TAMR) predicted by the model,
and the reasons for this discrepancy are explained.

\noindent
{\it keywords - 
galaxies: evolution - stars: formation; evolution.}
%END
%\end{titlepage}
\end{small}
\end{quotation}

\section{Introduction} \label{intro}

Abundance trends and the metallicity distribution
within any Galactic subsystem are vital records to
the formation and the evolution of the Milky Way.
In particular, it can be wondered if each component
underwent a separate chemical evolution both in
space and in time, or some fraction of gas was
transferred from an assigned reservoir into another
one, implying violation of mass conservation.   In
the latter alternative, an additional question is
if, and to what extent, simple models of chemical
evolution can be used for each subsystem.   To this
respect, useful indications may be provided taking
into consideration the G-dwarf problem.

In general, the occurrence of a G-dwarf problem is related
to the observation of too few metal deficient G dwarfs (or
of any other selected spectral class) with respect to that 
which could be expected
from the Simple model of chemical evolution (e.g., Searle
and Sargent, 1972; Pagel and Patchett, 1975; Haywood, 2001).
It has been established in different regions of the Galaxy:
the solar neighbourhood (van den Bergh, 1962; Schmidt, 1963)
and, to a lesser extent, the halo (e.g., Hartwick, 1976; 
Prantzos, 2003) and the bulge (e.g., Ferreras et al., 2003).   
In addition, a G-dwarf problem has been detected in both 
bulge-dominated and disk-dominated galaxies (Henry and Worthey, 
1999), which is consistent with the idea that the G-dwarf 
problem is universal (Worthey et al., 1996).

The deficit of metal-poor stars (with respect to the
prediction of the Simple model) may be interpreted
in different ways, such as changes in the initial
mass function (Schmidt, 1963; Adams and Fatuzzo, 1996;
Bromm, 2004; Bromm and Larson, 2004; Larson, 2005),
inflow of unprocessed (Larson, 1974; Lynden-Bell, 1975) 
or processed (Thacker et al., 2002) material from
outside, evolution with inhomogeneous star formation
(Searle, 1972; Malinie et al., 1993; Caimmi, 2000,
2001a, hereafter quoted together as C00%
\footnote{For this reference, two points
may carefully be kept in mind, namely (i) values of
a few parameters must be corrected as explained in
Caimmi (2001b), Sect.\,3, second paragraph,
and (ii) the majority of figures do not correspond
to the caption, as explained in the erratum (Caimmi,
2001a).};
Caimmi, 2001b, hereafter quoted as C01; Oey, 2003; 
Karlsson, 2005; Caimmi, 2007, hereafter quoted as
C07), instantaneous recycling approximation
(Haywood, 2001), 
or overlooking the thick disk population (Haywood,
2006).
For additional alternatives and further details
refer to earlier attempts (Pagel and Patchett, 1975;
Pagel, 1989).

Inhomogeneous (i.e. implying inhomogeneous
star formation) models of chemical evolution succeed in
both providing a solution to the G-dwarf problem
and reproducing substantial scatter exhibited by
the empirical age-metallicity relation (EAMR), with 
regard to the solar neighbourhood (e.g., Malinie et al., 
1993; C00), halo stars (C01), and bulge stars (C07).
The current paper aims to investigate
if inhomogeneous simple models of chemical
evolution are also consistent with the metallicity
distribution in (i) the Galactic disk, conceived
as made of three distinct subsystems: the metal-poor
thick disk, the metal-rich thick disk, and the thin
disk, and (ii) the Galaxy, conceived as made of five
distinct subsystems: the above mentioned three, the
halo, and the bulge.   In any case, the related
metallicity distribution is deduced (by weighting 
the mass) from the data belonging to selected samples
(assumed to be representative of the whole subsystem).
%With this in mind, the paper continues to discover 
%what constraints are related to the formation
%and the evolution of the Galaxy.

Although the Galaxy is dominated (by mass and star
number) by the two major subsystems, the bulge and the 
thin disk, still the contribution of the remaining
ones, the halo and the thick disk, may be relevant
in particular metallicity ranges.   In addition,
evidence for a halo-bulge and thick disk-thin disk collapse
(Wyse and Gilmore, 1992;  Ibata and Gilmore, 1995)
implies that both the proto-halo and the
proto-thick disk were more massive than their present
counterparts.    The empirical distribution of specific
angular momentum on the above mentioned subsystems,
favours the idea of two different kinds of contraction
(Wyse and Gilmore, 1992; Ibata and Gilmore, 1995).
On the other hand, some
evidence seems to exist on a continuous transition
from an extended $(R\appgeq20\,{\rm kpc})$,
pressure-supported halo, to an inner, flattened
$(R\appleq15\,{\rm kpc})$, rotation-supported halo
(Chiba and Beers, 2000).   The chemical abundance
of thick disk stars suggests a similar history to
those of metal-rich ([Fe/H]$\approx-1.3)$ halo stars,
and the thick disk abundance patterns show excellent
agreement with the chemical abundances observed in
metal-poor bulge stars, suggesting the two populations
were formed from the same reservoir at a common epoch
(Prochaska et al., 2000).

The following samples have been taken as representative.
For the thin disk, $N=287$ chemically selected G dwarfs
within 25 pc from the Sun, with the corrections performed
in order to take into account the stellar scale heights
(Rocha-Pinto and Maciel, 1996), hereafter quoted as the
RM96 sample.
For the halo, $N=372$ kinematically selected subdwarfs
(Ryan and Norris, 1991), hereafter quoted as the RN91
sample.
For the bulge, $N=268$ K-giants in Baade's window
(Sadler et al., 1996), hereafter quoted as the SA96
sample.
For the thick disk, a fictitious sample has been
made from the following sources: (i) a normalized
metallicity distribution (total number of stars
not reported therein) related to a volume complete
sample of local thick-disk stars, derived from Gliese
and scaled in situ samples within the range,
$-1.20\le$[Fe/H]$\le-0.20$ (Wyse and Gilmore, 1995),
hereafter quoted as the WG95 sample, and (ii) $N=
46$ likely metal-weak thick disk stars within the
range, $-2.20\le$[Fe/H]$\le-1.00$ (Chiba and Beers,
2000), hereafter quoted as the CB00 sample.
Under the assumption that both WG95 and CB00
samples represent to an acceptable extent the
related thick disk population, and keeping in
mind that they overlap within the range,
$-1.20\le$[Fe/H]$\le-1.00$, a fictitious sample
of $N=592$ stars, hereafter quoted as the
FS07 sample, can be derived for the thick disk.

The oxygen abundance is deduced from iron abundance,
using two alternative relations involving
the presence (PP) or absence (AP) of [O/Fe] plateau for
sufficiently low [Fe/H] values (C01, C07).   Though
more precise dependences have been found in recent
investigations (e.g., Jonsell et al., 2005; Fulbright
et al., 2005; Garcia Perez et al., 2006; Melendez et
al., 2006; Ramirez et al., 2007), related to different
trends for different Galactic subsystems, still they
are lying between the above mentioned alternatives,
which shall be used to preserve comparison with
earlier results (C01, C07).

The value of solar oxygen abundance also affects
the [O/Fe]-[Fe/H] relation.   Recent determinations
point towards lower values (e.g., Allende Prieto
et al., 2001; Sofia and Meyer, 2001; Asplund et al.,
2004; Melendez, 2004), but the question is still
under debate (e.g., Landi et al., 2007; Socas-Navarro
and Norton, 2007).   A solar oxygen mass abundance,
$(Z_O)_\odot=0.0056$, deduced from Allende-Prieto et
al. (2001), shall be used to preserve comparison
with earlier results (C01, C07).

Due to the above mentioned uncertainties, it has
been preferred to deal with less recent EDMD
determinations to preserve comparison with earlier
work (C00; C01; C07).   More recent, statistically
more significant and less well scrutinised
determinations e.g., the Geneva-Copenhagen survey
(Nordstrom et al., 2004) for the local disk and
the ongoing Hambourg survey performed by the Beers
consortium (e.g., Prantzos, 2007) for the metal-poor halo,
are related to iron abundance instead of oxygen
abundance as in a selected sample of nearby stars
(Ramirez et al., 2007), and for this reason shall
not be used in the current attempt.

The empirical, differential metallicity distribution
(EDMD), $\psi=\Delta N/$ $(N\Delta\phi)$, $\log\phi
=$[O/H], has been determined in earlier attempts for
the RN91 sample (C01) and the SA96 sample (C07), using both
the PP and AP dependence of oxygen abundance on iron
abundance.   The same is done in Section \ref{data}
for the RM96 and the FS07 sample.   In addition, a
putative EDMD is derived for the thick + thin
disk and the Galaxy, respectively, by weighting
the mass of the related subsystems, in Section \ref{imed}.  
A comparison with the
predictions of simple models, involving homogeneous and
inhomogeneous star formation, is given in Section
\ref{osmo}  and \ref{inmo}, respectively.   The discussion 
and the conclusion are the subject of Section \ref{disc} 
and \ref{conc}, respectively.
%Throughout this paper the standard notation [A/B]=$\log(A/B)-
%\log(A_\odot /B_\odot)$ will be used, where $A$, $B$, are the
%abundances of two elements in the sample objects; $A_\odot$,
%$B_\odot$, are solar abundances of these elements; and
%$\log$ represents the decimal logarithm.
%Further investigation on a few specific points and
%additional informations on the model are reported
%in the Appendix, where the main symbols used
%in the text are also listed and defined.

\section{The data} \label{data}
\subsection{The empirical differential metallicity
distribution} \label{EDMD}

In dealing with
simple models of chemical evolution, involving the
assumption of instantaneous recycling, the predicted
metal abundance has to be compared with the observed
oxygen abundance (e.g., Pagel, 1989; C00; C01; C07).
Unfortunately, oxygen is more difficult than
iron to detect, and an empirical formula is needed,
to express the former as a function of the latter.

On this matter a clear
dichotomy appears between authors who
support a plateau in [O/Fe] for stars with
[Fe/H]$\appleq-1$ (e.g., Carretta et al.,
2000); and those who do not (e.g., Israelian
et al., 2001a, 2001b). Further insight can be gained
by referring to specific
proceedings (edited by Barbuy et al.,
2001).   

Taking the above alternatives as limiting
situations, the
following formulae (C01; C07) shall be used:
\begin{lefteqnarray}
\label{eq:gra}
&& \left[\frac {\rm O}{\rm H}\right]=
\cases{\left[\displayfrac{\rm Fe}{\rm H}\right]+0.6~;
& $\left[\displayfrac{\rm Fe}{\rm H}\right]\le-1.2$ ; 
$\quad\left[\displayfrac {\rm O}{\rm H}\right]\le-0.6$ ; \cr
& \cr
\displayfrac12\left[\displayfrac{\rm Fe}{\rm H}\right]~;
& $\left[\displayfrac{\rm Fe}{\rm H}\right]\ge-1.2$ ; 
$\quad\left[\displayfrac {\rm O}{\rm H}\right]\ge-0.6$ ;\cr} \\
&& \nonumber \\
\label{eq:isa}
&& \left[\frac {\rm O}{\rm H}\right]=\frac23\left[\frac
{\rm Fe}{\rm H}\right]~;
\end{lefteqnarray}
in presence or in absence of [Fe/H] plateau,
respectively.   The value used for the oxygen 
solar abundance, $(Z_{\rm O})_\odot=0.0056\mp
0.0006$, has been deduced from a number ratio,
$\log(n_O/n_H)=-3.31\mp0.05$, or $A_O=\log(n_O
/n_H)+12=8.69\mp0.05$, corresponding to the 
solar photosphere line at 6300 \AA (Allende-Prieto
et al., 2001), assuming a hydrogen abundance,
$X=0.71$ (Anders and Grevesse, 1989).   For 
further details refer to earlier work (C01).
The assumed
oxygen abundance agrees with recent results
(e.g., Melendez, 2004; Asplund et al., 2004, 2005; 
Shchukina et al., 2005; Ramirez et al., 2007)
in contrast with earlier determinations (e.g.,
Sauval et al., 1984; Grevesse et al., 1984;
Anders and Grevesse, 1989) which yielded
higher values $(A_O\approx8.9)$, but the
question is still under debate (e.g., Landi
et al., 2007; Socas-Navarro and Norton, 2007).

A number of [O/H]-[Fe/H] dependences lying
between those
expressed by Eqs.\,(\ref{eq:gra}) and 
(\ref{eq:isa}), but different for different
Galactic subsystems, have been derived from
recent investigations (e.g., Bensby et al.,
2004; Jonsell et al., 2005; Fulbright et al., 2005;
Garcia Perez et al., 2006; Melendez et al.,
2006; Ramirez et al., 2007).
To allow comparison with previous
results related to halo subdwarfs (C01)
and K-giants in Baade's window (C07),
the solar oxygen abundance assumed therein
(Allende Prieto et al., 2001) shall be
used in the current attempt together with
Eqs.\,(\ref{eq:gra}) and (\ref{eq:isa}),
hereafter quoted as ``in presence''
and ``in absence'' of [O/Fe] plateau,
respectively, with regard to sufficiently 
low metallicities, [Fe/H]$\appleq-1$.

Of course, the validity of Eqs.\,(\ref{eq:gra}) 
and (\ref{eq:isa}) shall be restricted to an
epoch where only pop.\,II stars were present,
[Fe/H]$<-4$ say, when the formation of extremely
metal-deficient ([Fe/H]$<-4)$ but oxygen
overabundant (e.g., Christlieb et al., 2002;
Bromm and Loeb, 2003; Iwamoto et al., 2005; 
Frebel et al., 2006) stars had already been
occurred in presence of pop.\,III stars.

While observations are related to logarithmic
number abundances,
[A/H]$=\log$(A/H) $-\log$(A/H)$_\odot$,
models of chemical evolution
deal with mass abundances, $Z_{\rm A}=
M_{\rm A}/M$,
where A is a generic element heavier than
He, $M_{\rm A}$ is the total mass under the form
of element, A, and $M$ is the total mass
under the form of any element.   The following 
relation (Pagel, 1989;
Malinie et al., 1993; Rocha-Pinto and Maciel,
1996; C00; C01; C07):
\begin{equation}
\label{eq:lgfi}
\log\phi=\log\frac {Z_{\rm O}}{(Z_{\rm O})_\odot}=
\left[\frac{\rm O}{\rm H}\right]~~;
\end{equation}
holds to a good extent%
\footnote{The oxygen mass abundance, $Z_
{\rm O}$,
has been defined as $O$ in C00 and C01, to
simplify the notation therein.}.
For a formal derivation refer to earlier work
(C07, Appendix 1).

Let $\Delta\log\phi=\Delta$[O/H]=[O/H]$^+-$
[O/H]$^-$
be a logarithmic, oxygen abundance bin deduced
from $\Delta$[Fe/H] by use of Eq.\,(\ref{eq:gra})
or (\ref{eq:isa}).   The related, oxygen 
abundance bin is:
\begin{leftsubeqnarray}
\slabel{eq:fiba}
&& \Delta\phi=\Delta^+\phi+\Delta^-\phi~;\quad\Delta^
\mp\phi=\vert\phi-\phi^\mp\vert~~; \\
\slabel{eq:fibb}
&& \phi^\mp=\exp_{10}\left[\frac{\rm O}{\rm H}\right]^\mp~;\quad
\phi=\frac{\phi^++\phi^-}2~~;
\label{seq:fib}
\end{leftsubeqnarray}
where in general, $\exp_\xi$ defines the power
of basis $\xi$ and, in particular, $\exp$
defines the power of basis e, according to the
standard notation.   To take also the RN91 sample
into consideration, bins in [Fe/H]
equal to 0.2 dex shall be used (C01, C07).
%
%(e.g., Ryan and Norris, 1991; Huchra et al., 1991;
%Perrett et al., 2002).

Following earlier attempts (Pagel, 1989; Malinie 
et al., 1993; Rocha-Pinto and Maciel, 1996), the 
comparison
between model predictions and observations shall
be performed using the differential instead of
the cumulative metallicity distribution, as it
is a more sensitive test (Pagel, 1989) and
allows direct comparison between different
samples (Rocha-Pinto and Maciel, 1996).   The
occurrence of a sensitivity error in performing
observations, precludes the use of a proper
differential notation in dealing with the
empirical differential metallicity distribution
(hereafter referred to as EDMD%
\footnote{It is quoted as EGD in C00 and C01.}).
This being the reason why the bin length cannot
be lower than
the sensitivity error, and differential ratios,
i.e. first derivatives, $\diff N/\diff\phi$,
must be replaced by increment ratios, $\Delta N
/\Delta\phi$, where $\Delta\phi$ is the bin
length, $\Delta N$ the number of sample objects
within the selected bin, and $N$ is the total
number of sample objects.

Accordingly, the EDMD in a selected class of
objects is defined as:
\begin{equation}
\label{eq:psi}
\psi(\phi\mp\Delta^\mp\phi)=\log\frac{\Delta N}{N\Delta
\phi} ~;
%\quad\Delta\phi=\Delta^+\phi+\Delta^-\phi~;
\end{equation}
where the increment ratio, $\Delta N/
\Delta\phi$, used in earlier attempts
(Pagel, 1989; Malinie et al., 1993) has
been replaced by its normalized
counterpart, $\Delta N/$ $(N\Delta\phi)$,
used in more recent investigations
(Rocha-Pinto and Maciel, 1996; C00;
C01; C07), to allow comparison between
different samples.
The uncertainty on $\Delta N$
has been evaluated from Poisson errors (e.g., 
Ryan and Norris, 1991), as $\sigma_{\Delta N}=
(\Delta N)^{1/2}$, and the related uncertainty 
in the EDMD is (e.g., C01; C07):
\begin{leftsubeqnarray}
\slabel{eq:psiera}
&& \Delta^\mp\psi=\vert\psi-\psi^\mp\vert=
\log\left[1\mp\frac{(\Delta N)^
{1/2}}{\Delta N}\right]~~; \\
\slabel{eq:psierb}
&& \psi^\mp=\log\frac{\Delta N\mp(\Delta N)^{1/2}}{N\Delta\phi}~~;
\label{seq:psier}
\end{leftsubeqnarray}
where $\psi^-\rightarrow-\infty$ in the limit
$\Delta N\rightarrow1$.   For further details
refer to earlier work (C01).

The [Fe/H]-[O/H] dependence and corresponding
mean fractional oxygen abundance, $\phi$,
and half bin width, $\Delta^\mp\phi$, in
presence of [O/Fe] plateau (PP), according
to Eq.\,(\ref{eq:gra}), and in absence of
[O/Fe] plateau (AP), according to 
Eq.\,(\ref{eq:isa}), respectively,
are shown in Tab.\,\ref{t:OFe} for
the metallicity range of interest.
\begin{table}
\caption[par]{The [Fe/H]-[O/H] dependence
and corresponding
mean fractional oxygen abundance, $\phi$,
and half bin width, $\Delta^\mp\phi$, in
presence (PP) and in absence (AP) of [O/Fe] 
plateau.   
%To save space, $F$ stands for [Fe/H] and $O$ for 3[O/H].
}
\label{t:OFe}
\begin{center}
\begin{tabular}{rrrrrrrrrr}
\multicolumn{2}{c|}{}
& \multicolumn{4}{c|}{PP}
& \multicolumn{4}{c}{AP} \\
\hline\noalign{\smallskip}
\multicolumn{1}{c}{[Fe/H]$^-$} &
\multicolumn{1}{c}{[Fe/H]$^+$} &
\multicolumn{1}{c}{3[O/H]$^-$} &
\multicolumn{1}{c}{3[O/H]$^+$} &
\multicolumn{1}{c}{$\phi$} & \multicolumn{1}{c}{$\Delta^\mp\phi$} &
\multicolumn{1}{c}{3[O/H]$^-$} &
\multicolumn{1}{c}{3[O/H]$^+$} &
\multicolumn{1}{c}{$\phi$} & \multicolumn{1}{c}{$\Delta^\mp\phi$} \\
\noalign{\smallskip}
\hline\noalign{\smallskip}
1.2 & 1.4 & 1.8 & 2.1 & 4.496 & 0.515 & 2.4 & 2.8 & 7.443 & 1.134 \\
1.0 & 1.2 & 1.5 & 1.8 & 3.572 & 0.409 & 2.0 & 2.4 & 5.476 & 0.834 \\
0.8 & 1.0 & 1.2 & 1.5 & 2.837 & 0.325 & 1.6 & 2.0 & 4.028 & 0.613 \\
0.6 & 0.8 & 0.9 & 1.2 & 2.254 & 0.258 & 1.2 & 1.6 & 2.963 & 0.451 \\
0.4 & 0.6 & 0.6 & 0.9 & 1.790 & 0.205 & 0.8 & 1.2 & 2.180 & 0.332 \\
0.2 & 0.4 & 0.3 & 0.6 & 1.422 & 0.163 & 0.4 & 0.8 & 1.604 & 0.244 \\
0.0 & 0.2 & 0.0 & 0.3 & 1.129 & 0.129 & 0.0 & 0.4 & 1.180 & 0.180 \\
$-$0.2 & 0.0 & $-$0.3 & 0.0 & 0.897 & 0.103 & $-$0.4 & 0.0 & 0.868 
& 0.132 \\
$-$0.4 & $-$0.2 & $-$0.6 & $-$0.3 & 0.713 & 0.087 & $-$0.8 & $-$0.4 & 
0.638 & 0.097 \\
$-$0.6 & $-$0.4 & $-$0.9 & $-$0.6 & 0.566 & 0.065 & $-$1.2 & $-$0.8 & 
0.470 & 0.071 \\
$-$0.8 & $-$0.6 & $-$1.2 & $-$0.9 & 0.450 & 0.052 & $-$1.6 & $-$1.2 & 
0.345 & 0.053 \\
$-$1.0 & $-$0.8 & $-$1.5 & $-$1.2 & 0.357 & 0.041 & $-$2.0 & $-$1.6 & 
0.254 & 0.039 \\
$-$1.2 & $-$1.0 & $-$1.8 & $-$1.5 & 0.284 & 0.032 & $-$2.4 & $-$2.0 & 
0.187 & 0.028 \\
$-$1.4 & $-$1.2 & $-$2.4 & $-$1.8 & 0.205 & 0.046 & $-$2.8 & $-$2.4 & 
0.137 & 0.021 \\
$-$1.6 & $-$1.4 & $-$3.0 & $-$2.4 & 0.129 & 0.029 & $-$3.2 & $-$2.8 & 
0.101 & 0.015 \\
$-$1.8 & $-$1.6 & $-$3.6 & $-$3.0 & 0.081 & 0.018 & $-$3.6 & $-$3.2 & 
0.074 & 0.011 \\
$-$2.0 & $-$1.8 & $-$4.2 & $-$3.6 & 0.051 & 0.012 & $-$4.0 & $-$3.6 & 
0.055 & 0.008 \\
$-$2.2 & $-$2.0 & $-$4.8 & $-$4.2 & 0.032 & 0.007 & $-$4.4 & $-$4.0 & 
0.040 & 0.006 \\
$-$2.4 & $-$2.2 & $-$5.4 & $-$4.8 & 0.020 & 0.005 & $-$4.8 & $-$4.4 & 
0.030 & 0.004 \\
$-$2.6 & $-$2.4 & $-$6.0 & $-$5.4 & 0.013 & 0.003 & $-$5.2 & $-$4.8 & 
0.022 & 0.003 \\
$-$2.8 & $-$2.6 & $-$6.6 & $-$6.0 & 0.008 & 0.002 & $-$5.6 & $-$5.2 & 
0.016 & 0.002 \\
$-$3.0 & $-$2.8 & $-$7.2 & $-$6.6 & 0.005 & 0.002 & $-$6.0 & $-$5.6 & 
0.012 & 0.002 \\
$-$3.7 & $-$3.0 & $-$9.3 & $-$7.2 & 0.002 & 0.002 & $-$7.4 & $-$6.0 & 
0.007 & 0.003 \\
\noalign{\smallskip}
\hline
\end{tabular}
\end{center}
\end{table}

As in earlier attempts (Pagel, 1989; Malinie
et al., 1993; Rocha-Pinto and Maciel, 1996;
C00; C01; C07), the bin sizes in normalized oxygen
abundance, $\phi$, correspond to uniform bin
sizes in [Fe/H], according to Eqs.\,(\ref
{eq:lgfi}) and (\ref{seq:fib}) and Tab.\,\ref
{t:OFe}, which implies non uniform bin sizes,
$\Delta\phi=2\Delta^\mp\phi$, in the following
tables and figures.

\subsection{Metallicity distribution in the thick
disk} \label{EDKD}

Both the WG95 and CB00 samples are considered to
infer the metallicity distribution in the
thick disk.   The total number of sample
objects is not specified in the former case,
as the related abundance distribution is
normalized to unity (Wyse and Gilmore, 1995).
A fictitious value, $N=39$, is deduced from
Tab.\,3 therein, and shall be used in performing
calculations.   The metallicity range is $-1.20
\le$[Fe/H]$\le-0.20$ for the WG95 sample and $-2.20
\le$[Fe/H]$\le-1.00$ for the CB00 sample.   The 
related EDMD, using Eqs.\,(\ref{eq:lgfi})-(\ref
{seq:psier}), is listed in 
Table \ref{t:KD} both in presence and in
absence of [O/Fe] plateau.
\begin{table}
\caption[par]{The empirical, differential
metallicity distribution (EDMD) in the thick
disk, deduced from the CB00 sample ($N=46$,
top panel) and the WG95 sample ($N=39$ deduced
from a normalized distribution, bottom panel),
both in presence (PP) and in 
absence (AP) of [O/Fe] plateau.}
\label{t:KD}
\begin{center}
\begin{tabular}{rrrrrrr}
\multicolumn{2}{c|}{PP}
&\multicolumn{2}{c|}{AP} \\
\hline\noalign{\smallskip}
\multicolumn{1}{c}{$\phi$} & \multicolumn{1}{c}{$\phantom{0}\psi$} &
\multicolumn{1}{c}{$\phi$} & \multicolumn{1}{c}{$\phantom{0}\psi$} &
\multicolumn{1}{c}{$\Delta^-\psi$} & \multicolumn{1}{c}{$\Delta^+\psi$} &
\multicolumn{1}{c}{$\Delta N$}  \\
\noalign{\smallskip}
\hline\noalign{\smallskip}
0.284 &    0.810 & 0.187 &    0.868 & 0.113     & 0.090 & \phantom{}19 \\
0.205 &    0.328 & 0.137 &    0.668 & 0.176     & 0.125 & \phantom{1}9 \\
0.129 &    0.051 & 0.101 &    0.337 & 0.374     & 0.198 & \phantom{1}3 \\
0.081 &    0.480 & 0.074 &    0.694 & 0.257     & 0.161 & \phantom{1}5 \\
0.051 &    0.656 & 0.055 &    0.832 & 0.257     & 0.161 & \phantom{1}5 \\
0.032 &    0.890 & 0.040 &    0.957 & 0.257     & 0.161 & \phantom{1}5 \\
      &          &       &	    &           &       &              \\
0.713 &    0.013 & 0.638 & $-$0.034 & 0.206     & 0.139 & \phantom{1}7 \\
0.566 &    0.295 & 0.470 &    0.257 & 0.165     & 0.119 & \phantom{}10 \\
0.450 &    0.392 & 0.345 &    0.384 & 0.165     & 0.119 & \phantom{}10 \\
0.357 &    0.495 & 0.254 &    0.517 & 0.165     & 0.119 & \phantom{}10 \\
0.284 & $-$0.397 & 0.187 & $-$0.339 & $\infty$ & 0.301 & \phantom{0}1 \\
0.205 & $-$0.555 & 0.137 & $-$0.214 & $\infty$ & 0.301 & \phantom{0}1 \\
\noalign{\smallskip}
\hline
\end{tabular}
\end{center}
\end{table}

\begin{figure*}[t]
\begin{center}
%\centerline{\psfig{file=EGDK4A.ps,height=130mm,width=140mm}}
\includegraphics[scale=0.8]{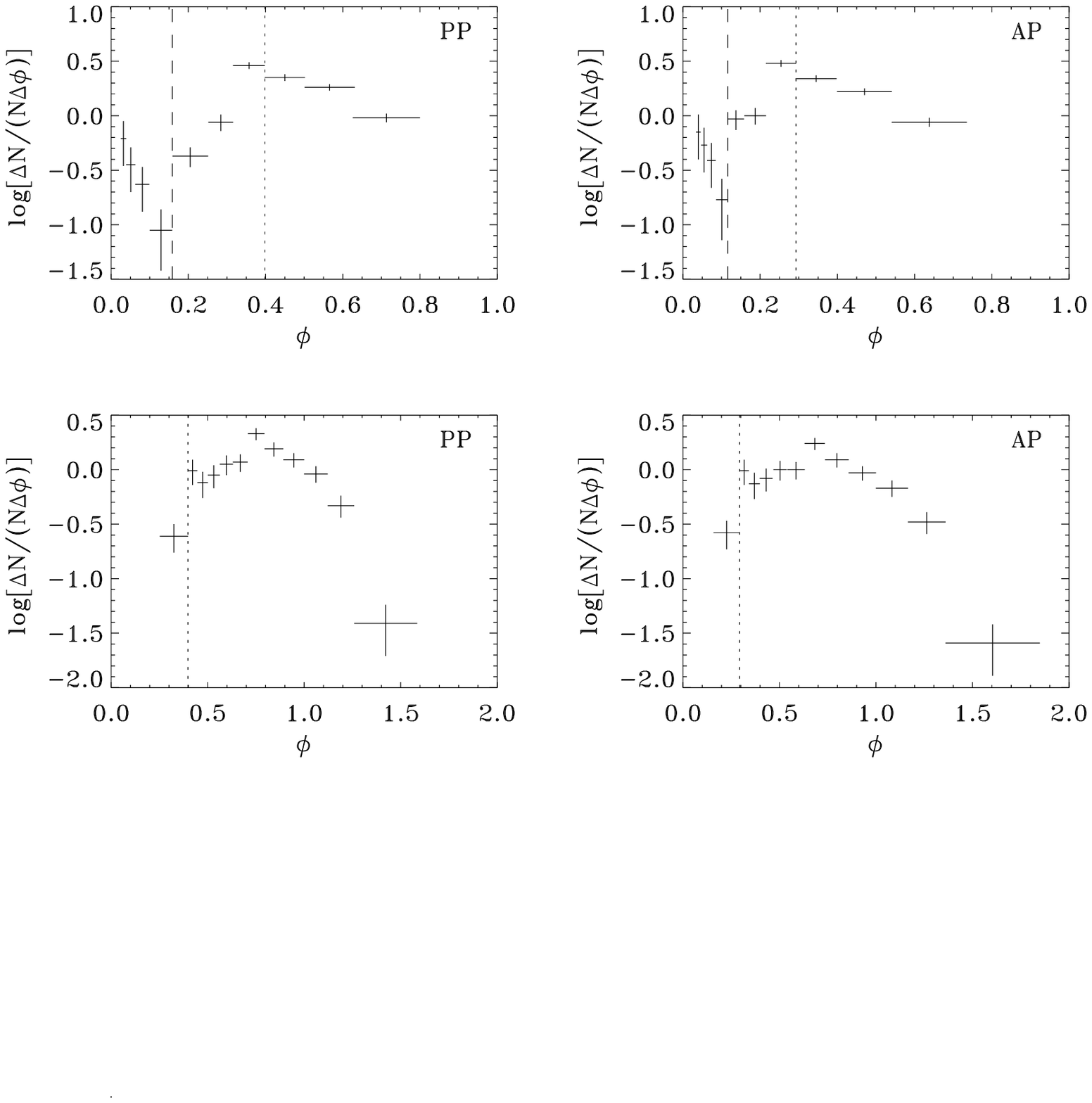}
\caption[KDNA]{The empirical, differential metallicity
distribution (EDMD) in thick (top panels) and thin (bottom
panels) disk, both in presence (PP) and in absence (AP) of
[O/Fe] plateau.   Plots have been deduced from the FS07 sample
for the thick disk, and from the RM96 sample for the thin disk.
The dotted vertical line marks the transition from
halo to bulge/disk globular cluster morphological
type, [Fe/H]$=-$0.8.   The dashed vertical line marks the
transition from halo-like to bulge-like thich disk,
[Fe/H]$=-$1.4.}
\label{f:KDNA}
\end{center}
\end{figure*}

\begin{figure*}[t]
\begin{center}
\includegraphics[scale=0.8]{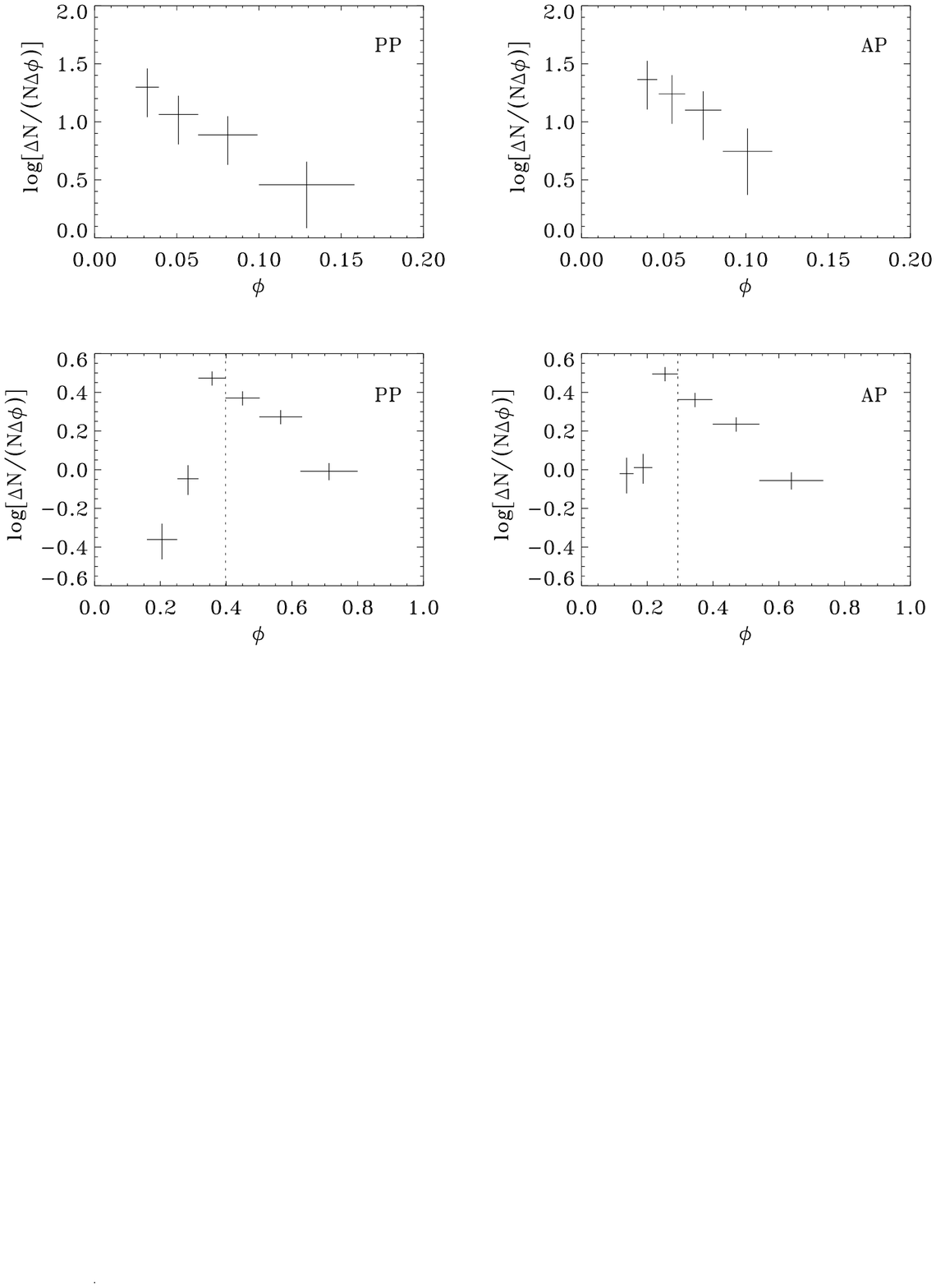}
%\centerline{\psfig{file=EGDK4B.ps,height=130mm,width=140mm}}
\caption[KDNA]{The empirical, differential metallicity
distribution (EDMD) in halo-like (top panels) and bulge-like 
(bottom panels) thick disk, both in presence (PP) and in 
absence (AP) of [O/Fe] plateau.   The dotted vertical line 
marks the transition from
halo to bulge/disk globular cluster morphological
type, [Fe/H]$=-$0.8.}
\label{f:KHKB}
\end{center}
\end{figure*}

The samples under consideration overlap within 
the range, $-1.4\le$[Fe/H]$\le-1.0$, which makes
a renormalization be possible.   The result is a
fictitious sample: $N=592$, and metallicity range,
$-2.2\le$[Fe/H]$\le-0.2$.      
The related EDMD is listed in Table \ref{t:HBKD}
and plotted in Fig.\,\ref{f:KDNA} (top panels) both
in presence and in absence of [O/Fe] plateau.
\begin{table}
\caption[par]{The empirical, differential
metallicity distribution (EDMD) in the thick
disk, deduced from the FS07 sample ($N=592$),
both in presence (PP) and in absence (AP) of 
[O/Fe] plateau.   Bottom and top panels
correspond to halo-like and bulge-like thick
disk, respectively.   The related EDMD, 
calculated for the halo-like ($N=18$) and the bulge-like
($N=574$) thick disk subsample, is denoted as
$\psi_S$.}
\label{t:HBKD}
\begin{center}
\begin{tabular}{rrrrrrrrr}
\multicolumn{3}{c|}{PP}
&\multicolumn{3}{c|}{AP} \\
\hline\noalign{\smallskip}
\multicolumn{1}{c}{$\phi$} & \multicolumn{1}{c}{$\phantom{0}\psi$} &
\multicolumn{1}{c}{$\phantom{0}\psi_S$} &
\multicolumn{1}{c}{$\phi$} & \multicolumn{1}{c}{$\phantom{0}\psi$} &
\multicolumn{1}{c}{$\phantom{0}\psi_S$} &
\multicolumn{1}{c}{$\Delta^-\psi$} & \multicolumn{1}{c}{$\Delta^+\psi$} &
\multicolumn{1}{c}{$\Delta N$}  \\
\noalign{\smallskip}
\hline\noalign{\smallskip}
0.713 &    0.006 & $-$0.008 & 0.638 & $-$0.070 & $-$0.056 & 0.046 & 0.042 & \phantom{1}98 \\
0.566 &    0.261 &    0.273 & 0.470 &    0.218 &    0.235 & 0.038 & 0.035 & \phantom{}140 \\
0.450 &    0.361 &    0.370 & 0.345 &    0.352 &    0.362 & 0.038 & 0.035 & \phantom{}140 \\
0.357 &    0.461 &    0.473 & 0.254 &    0.485 &    0.495 & 0.038 & 0.035 & \phantom{}140 \\
0.284 & $-$0.070 & $-$0.047 & 0.187 & $-$0.009 &    0.011 & 0.083 & 0.070 & \phantom{1}33 \\
0.205 & $-$0.378 & $-$0.361 & 0.137 & $-$0.033 & $-$0.020 & 0.102 & 0.082 & \phantom{1}23 \\
      &          &          &       &          &          &       &       &               \\
0.129 & $-$1.062 &    0.458 & 0.101 & $-$0.784 &    0.745 & 0.374 & 0.198 & \phantom{14}3 \\
0.081 & $-$0.640 &    0.887 & 0.074 & $-$0.429 &    1.101 & 0.257 & 0.161 & \phantom{14}5 \\
0.051 & $-$0.440 &    1.063 & 0.055 & $-$0.296 &    1.240 & 0.257 & 0.161 & \phantom{14}5 \\
0.032 & $-$0.240 &    1.298 & 0.040 & $-$0.162 &    1.364 & 0.257 & 0.161 & \phantom{14}5 \\
\noalign{\smallskip}
\hline
\end{tabular}
\end{center}
\end{table}
The trend exhibited looks like its counterpart related
to the Galactic spheroid (C07).   Having in mind it is
something else a mere coincidence, the fictitious FS07
sample is conceived as made of two fictitious subsamples
related to different subsystems: a halo-like thick disk
$(N=18)$ and a bulge-like thick disk $(N=574)$, in the
metallicity range, $-2.2\le$[Fe/H]$\le-1.4$ and
$-1.4\le$[Fe/H]$\le-0.2$, respectively.   The
corresponding EDMD is listed as $\psi_S$ in Table 
\ref{t:HBKD} and plotted in Fig.\,\ref{f:KHKB},
top and bottom panel, respectively, both in
presence (left panels) and in absence (right panels)
of [O/Fe] plateau.

The results from recent attempts involving
a more refined abundance derivation (e.g.,
Reddy et al., 2006; Ramirez et al., 2007)
are not used in the current paper, in
absence (to the knowledge of the author)
of their counterparts related to the halo,
the bulge, and the metal-weak thick disk.

\subsection{Metallicity distribution in the 
thin disk} \label{EDND}

As in an earlier attempt (C00), the RM96 sample
($N=287$ and $N=325.064$ after corrections due 
to stellar scale heigth; Rocha-Pinto and Maciel, 
1996) is considered in the metallicity range,
$-1.2\le$[Fe/H]$\le0.2$.   The determination of
the EDMD is repeated here owing to a different
value of the assumed solar oxygen and, in addition,
the absence of [O/Fe] plateau is considered.

The results, obtained by use of Eqs.\,(\ref
{eq:lgfi})-(\ref{seq:psier}), are listed in 
Table \ref{t:ND} both in presence and in
absence of [O/Fe] plateau.
\begin{table}
\caption[par]{The empirical, differential
metallicity distribution (EDMD) in the thin
disk, deduced from the RM96 sample ($N=287$ and
$N=325.064$ after corrections due to stellar
scale heigth; Rocha-Pinto and Maciel, 1996)
both in presence (PP) and in 
absence (AP) of [O/Fe] plateau.}
\label{t:ND}
\begin{center}
\begin{tabular}{rrrrrrr}
\multicolumn{2}{c|}{PP}
&\multicolumn{2}{c|}{AP} \\
\hline\noalign{\smallskip}
\multicolumn{1}{c}{$\phi$} & \multicolumn{1}{c}{$\phantom{0}\psi$} &
\multicolumn{1}{c}{$\phi$} & \multicolumn{1}{c}{$\phantom{0}\psi$} &
\multicolumn{1}{c}{$\Delta^-\psi$} & \multicolumn{1}{c}{$\Delta^+\psi$} &
\multicolumn{1}{c}{$\Delta N$}  \\
\noalign{\smallskip}
\hline\noalign{\smallskip}
1.422 & $-$1.412 & 1.604 & $-$1.587 & 0.295 & 0.174 & \phantom{5}4.110 \\
1.129 & $-$0.166 & 1.180 & $-$0.309 & 0.061 & 0.054 & \phantom{}57.410 \\
0.897 &    0.136 & 0.868 &    0.027 & 0.048 & 0.043 & \phantom{}91.490 \\
0.713 &    0.225 & 0.638 &    0.149 & 0.049 & 0.044 & \phantom{}89.127 \\
0.566 &    0.004 & 0.470 & $-$0.038 & 0.072 & 0.062 & \phantom{}42.593 \\
0.450 & $-$0.064 & 0.345 & $-$0.073 & 0.089 & 0.074 & \phantom{}28.951 \\
0.325 & $-$0.614 & 0.226 & $-$0.576 & 0.151 & 0.112 & \phantom{}11.843 \\
\noalign{\smallskip}
\hline
\end{tabular}
\end{center}
\end{table}
The number of objects in each metallicity bin has
been calculated as $\Delta N=\Delta N_0/f+\delta
(\Delta N)_1$, where $\Delta N_0$ is the sample
number, $f$ and $\delta(\Delta N)_1$ are correction
factors for the stellar scale height and observational
errors plus cosmic scatter, respectively (Rocha-Pinto
and Maciel, 1996).   Accordingly, the total number
of objects reads $N=325.064$.   The related plots are 
shown in Fig.\,\ref{f:KDNA} (bottom panels) both in
presence (left panels) and in absence (right panels)
of [O/Fe] plateau.   The dotted vertical
line marks the transition from halo to bulge/disk
globular cluster morphological type (Mackey and van
den Bergh, 2005).   For further details refer to
earlier work (C07).
The distribution could be bimodal with the occurrence
of two maxima, related to low and intermediate oxygen
abundance, respectively.   The former maximum, if
real, takes place near the transition from halo to
bulge/disk globular cluster morphological type, as
for old halo globular clusters and bulge K-giants
in Baade's window (C07).

Metallicity distributions with similar features are
shown by other samples of thin disk stars (e.g., 
Wyse and Gilmore, 1995; J\o\phantom{}rgensen, 2000; Kotoneva
et al., 2002; Reddy et al., 2003).   It has been
questioned that the maximum of the distribution
could be lowered by selection effects (Haywood,
2001).   In fact, using different criteria, the
maximum rises up to about the solar abundance
(e.g., Favata et al., 1997; Haywood, 2001, 2006),
but no general consensus seems still to exist on
the selection of sample objects.   For this reason,
the RM96 sample shall be considered in the current
paper.

\section{Inferred metallicity distribution} 
\label{imed}

Let the procedure used for the Galactic spheroid
(C07) be generalized to a system made of $n_S$
subsystems, where the EDMD related to a representative
sample is available for each subsystem.   The
total mass is:
\begin{equation}
\label{eq:Ms}
M=\sum_{i=1}^{n_S}M_i=\sum_{i=1}^{n_S}\frac{M_i}MM~~;
\end{equation}
where $M_i$ is the mass of $i$-th subsystem.

Let $N$ be the total number of long-lived
(i.e. life time longer than the age of the
system) stars in the system,
and $\Delta N$ the number of long-lived
stars within a selected metallicity bin.
The relative frequency, $\Delta N/N$,
reads:
\begin{equation}
\label{eq:Ns}
\frac{\Delta N}N=\sum_{i=1}^{n_S}\frac{\Delta N_i}N=\sum_{i=1}^{n_S}
\frac{N_i}N\frac{\Delta N_i}{N_i}~~;
\end{equation}
where $\Delta N_i$ and $N_i$ are the
number of long-lived stars within a
selected metallicity bin and the total
number, respectively, with regard to
$i$-th subsystem.
Then the relative frequency, $\Delta N/N$,
related to the system, is expressed as a
mean of the relative frequencies, $\Delta 
N_i/N_i$, related to the subsystems,
weighted by the factors, $N_i/N$.

Under the assumption of a universal
initial mass function (IMF) for star generation, the
number ratios, $N_i/N$, may safely
be related to  the mass ratios,
$M_i/M$, as:
\begin{equation}
\label{eq:NM}
\frac{N_i}N=\frac{M_i}M~~;
\end{equation}
for further details refer to earlier
work (C07).

The combination of Eqs.\,(\ref{eq:Ns})
and (\ref{eq:NM}) yields: 
\begin{equation}
\label{eq:NMs}
\frac{\Delta N}N=\sum_{i=1}^{n_S}\frac{M_i}M\frac
{\Delta N_i}{N_i}~~;
\end{equation}
where the relative frequencies, $\Delta N_i
/N_i$, may be deduced from
representative samples, which allows an
evaluation of the relative frequency, 
$\Delta N/N$.   The related uncertainty may
be obtained
using the standard formula of linear
propagation of errors%
\footnote{Linear propagation instead of
quadratic propagation has been preferred
to maximize the errors.   The symbol,
$\Delta(\Delta N)/N$, would be more
germane to this respect, on the
left-hand side of Eq.\,(\ref{eq:DNMs}),
but $\sigma_{\Delta{\rm N}/{\rm N}}$
is used to avoid confusion.}
together with
evaluation of Poisson errors (e.g.,
Ryan and Norris, 1991), $\sigma_{\Delta
{\rm N}_i}=(\Delta N_i)^
{1/2}$.   The result is:
\begin{equation}
\label{eq:DNMs}
\sigma_{\Delta{\rm N}/{\rm N}}=\sum_{i=1}^{n_S}\frac{M_i}M
\frac{(\Delta N_i)^{1/2}}{N_i}~~;
\end{equation}
which may explicitly be calculated for
assigned samples and mass ratios.

The EDMD related to the system results 
from the combination of Eqs.\,(\ref{eq:psi}), 
(\ref{seq:psier}), and (\ref{eq:NMs}), as:
\begin{leftsubeqnarray}
\slabel{eq:psisa}
&& \psi=\log\sum_{i=1}^{n_S}\frac{M_i}M\frac
{\Delta N_i}{N_i\Delta\phi}~~; \\
\slabel{eq:psisb}
&& \Delta^\mp\psi=\left\vert\log\left[1\mp
\frac{\sigma_{\Delta{\rm N}/{\rm N}}}{\Delta N/N}\right]
\right\vert~~; \\
\slabel{eq:psisc}
&& \psi^\mp=\log\sum_{i=1}^{n_S}\left[\frac{M_i}M\frac
{\Delta N_i\mp(\Delta N_i)^{1/2}}{N_i\Delta\phi}\right]~~;
\label{seq:psis}
\end{leftsubeqnarray}
which may be particularized to the cases
of interest.

\subsection{The thick disk} \label{KD}

The following assumptions have been
made in Subsection \ref{EDMD} for
deriving the halo-like and the bulge-like
thick disk EDMD: (i) the parent samples
WG95 and CB00 represent to a comparable,
acceptable extent the thick disk within
the metallicity ranges, $-1.4\le$[Fe/H]
$\le-0.2$ and $-2.2\le$[Fe/H]$\le-1.0$,
respectively, to allow renormalization
with respect to the common range, $-1.4
\le$[Fe/H]$\le-1.0$; (ii) objects
belonging to the resulting FS07 sample,
whose EDMD cannot be fitted by simple
models of chemical evolution, come from
different subsystems, whose EDMD can be
fitted by simple models of chemical
evolution, namely the halo-like ([Fe/H]
$\le-1.4$) and the bulge-like ([Fe/H]
$>-1.4$) thick disk.
In this view, the EDMD related to the thick
disk depends, via Eqs.\,(\ref{seq:psis}), 
on the halo-like to bulge-like thick disk
mass ratio, $M_{\rm HK}/M_{\rm BK}$, which (to the
knowledge of the author) is not available
at present. 

It has been suggested that the local density
of the metal-weak thick disk population ([Fe/H]
$\le-1$) represents less than one percent of
that of the canonical thick disk (Martin and
Morrison, 1998) and, in any case, it appears
to be a minor constituent of the entire thick 
disk population (Beers et al., 2002).   It
remains unclear whether the metal-weak thick
disk population is properly considered an
actual subsystem of the canonical thick disk
with a distinct EDMD, or whether it corresponds,
in reality, to the low-metal tail of the canonical
thick disk EDMD (Beers et al., 2002).   The
former alternative has been chosen in the current
attempt, according to the above considerations.

\begin{figure*}[t]
\begin{center}
\includegraphics[scale=0.8]{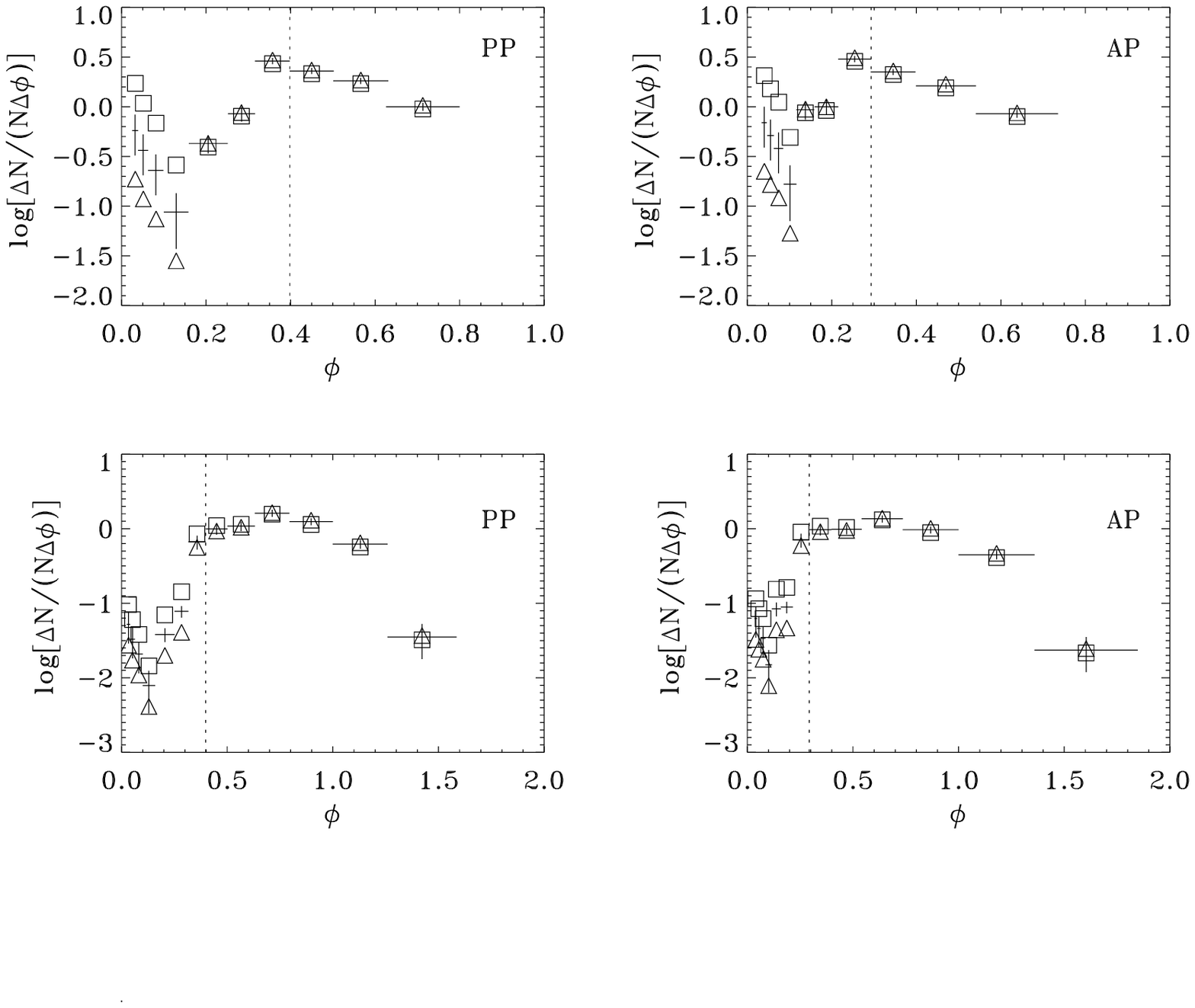}
%\centerline{\psfig{file=EGDK5B.ps,height=130mm,width=140mm}}
\caption[KDNA]{The empirical, differential metallicity
distribution (EDMD) in halo-like + bulge-like thick disk
(top panels) and halo-like + bulge-like thick + thin disk
(bottom panels), both in presence (PP) and in 
absence (AP) of [O/Fe] plateau.   Different symbols
are related to different halo-like to bulge-like thick 
disk mass ratio, $M_{\rm HK}/M_{\rm BK}$, for top panels, and 
to different thick to thin disk mass ratio, $M_{\rm KD}/M_{\rm ND}$,
for bottom panels.   Caption of symbols: top panels,
$M_{\rm HK}/M_{\rm BK}=18/574$ (crosses); 0.01 (triangles);
0.10 (squares); bottom panels, $M_{\rm KD}/M_{\rm ND}=0.10$
(crosses); 0.05 (triangles); 0.20 (squares).   In addition,
$M_{\rm HK}/M_{\rm BK}=18/574\approx0.03136$ for bottom panels.
The dotted vertical line marks the transition from
halo to bulge/disk globular cluster morphological
type, [Fe/H]$=-$0.8.}
\label{f:KDND}
\end{center}
\end{figure*}

Different values of the mass ratio, $M_{\rm HK}/M_{\rm BK}$,
make points related to halo-like and bulge-like
thick disk (on the left and on the right, respectively,
of the vertical dashed line in Fig.\,\ref{f:KDNA}, top
panels) EDMD shift vertically one with respect to the
other: increasing values make the halo-like side of
the distribution rise and the bulge-like side lower,
and vice versa.   The special case, $M_{\rm HK}/M_{\rm BK}=
N_{\rm HK}/N_{\rm BK}=18/574\approx0.03136$, leaves the
thick disk EDMD (Fig.\,\ref{f:KDNA}, top panels)
unchanged, as plotted in Fig.\,\ref{f:KDND} (top
panels, crosses).   The cases, $M_{\rm HK}/M_{\rm BK}=
0.01$ (triangles) and 0.10 (squares) are also
represented for comparison, where the error
bars are suppressed to avoid confusion.

\subsection{The disk} \label{D}

The disk is conceived as made of three
main subsystems, namely (i) the halo-like
thick disk; (ii) the bulge-like thick disk;
and (iii) the thin disk.   Accordingly, the
EDMD related to the disk depends, via
Eqs.\,(\ref{seq:psis}), on the halo-like to
bulge-like thick disk mass ratio, $M_{\rm HK}/
M_{\rm BK}$, and the thick to thin disk mass
ratio, $M_{\rm KD}/M_{\rm ND}$, both unknown at
present.   The effect of the former ratio
on the EDMD has been discussed in Subsection
\ref{KD}.

The cases, $M_{\rm KD}/M_{\rm ND}=0.10$, 0.05, and
0.20, are represented in Fig.\,\ref{f:KDND}
(bottom panels) as crosses, triangles, and
squares, respectively, where the error bars
in the last two alternatives are suppressed
to avoid confusion.   In any case, $M_{\rm HK}/
M_{\rm BK}=18/574\approx0.03136$ has been adopted.

\subsection{The Galaxy}\label{G}

The Galaxy is conceived as made of six main
subsystems, namely: (i) globular clusters
(GC); (ii) field halo stars (FH); (iii)
field bulge stars (FB); (iv) field halo-like
thick disk stars (HK); (v) field bulge-like
thick disk stars (BK); and (vi) field thin
disk stars (ND).   As in C07, the following
mass values are assumed:
\begin{leftsubeqnarray}
\slabel{eq:Ma}
&& M_{\rm GC}=0.001 {\rm M}_{10}~~;\qquad
M_{\rm FH}=0.1 {\rm M}_{10}~~;\qquad
M_{\rm FB}={\rm M}_{10}~~; \\
\slabel{eq:Mb}
&& M_{\rm KD}=M_{\rm HK}+M_{\rm BK}~~;\qquad
M_{\rm FD}=M_{\rm KD}+M_{\rm ND}=5.8{\rm M}_{10}~~; \\
\slabel{eq:Mc}
&& M_{\rm G}=M_{\rm GC}+M_{\rm FH}+M_{\rm FB}+M_{\rm FD}=6.901
{\rm M}_{10}~;\quad{\rm M}_{10}=10^{10}{\rm m}_\odot~;
\label{seq:M}
\end{leftsubeqnarray}
where the halo-like to bulge-like thick disk 
mass ratio, $M_{\rm HK}/M_{\rm BK}$, and the thick to
thin disk mass ratio, $M_{\rm KD}/M_{\rm ND}$, remain
undetermined and must be fixed.

\begin{figure*}[t]
\begin{center}
\includegraphics[scale=0.8]{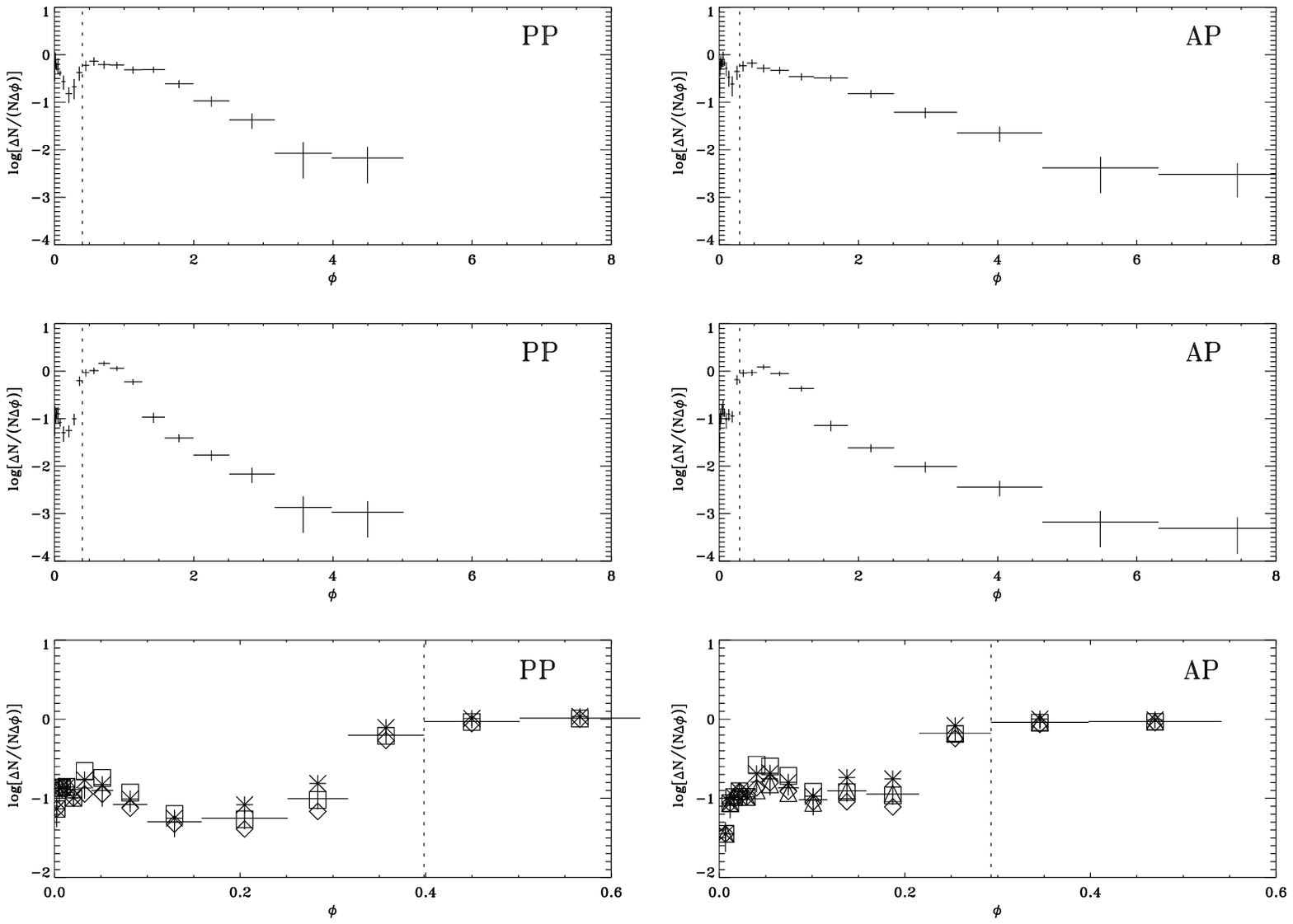}
%\centerline{\psfig{file=EGDK5A.ps,height=130mm,width=140mm}}
\caption[KDNA]{The empirical, differential metallicity
distribution (EDMD) in the Galaxy (middle panels) for
halo-like to bulge-like thick disk mass ratio, 
$M_{\rm HK}/M_{\rm BK}=18/574$, and
thick disk to thin disk mass ratio, $M_{\rm KD}/M_
{\rm ND}=0.1$, both in presence (PP) and in absence
(AP) of [O/Fe] plateau.   The EDMD in the Galactic
spheroid (C07) is shown in top panels for comparison.
Additional cases related to different mass ratios are
also represented (bottom panels) where the error bars
have been suppressed to avoid confusion.   Caption of
symbols: $M_{\rm HK}/M_{\rm BK}=0.01$ (triangles),
0.10 (squares), where $M_{\rm KD}/M_{\rm ND}=0.1$;
$M_{\rm KD}/M_{\rm ND}=0.05$ (diamonds), 0.20 (asterisks),
where $M_{\rm HK}/M_{\rm BK}=18/574$.
Crosses representing error bars are related to the
reference case, $M_{\rm HK}/M_{\rm BK}=18/574\approx0.03136$
and $M_{\rm KD}/M_{\rm ND}=0.10$.
The dotted vertical line marks the transition from
halo to bulge/disk globular cluster morphological
type, [Fe/H]$=-$0.8.}
\label{f:SFGA}
\end{center}
\end{figure*}

The case, $M_{\rm HK}/M_{\rm BK}=18/574\approx0.03136$
and $M_{\rm KD}/M_{\rm ND}
=0.1$, is represented in Fig.\,\ref{f:SFGA}
(middle panels) and compared with its counterpart
related to the Galactic spheroid (top panels,
C07).   Additional cases, $M_{\rm HK}/M_{\rm BK}=0.01$
(triangles), $0.1$ (squares),
with $M_{\rm KD}/M_{\rm ND}=0.1$; and $M_{\rm KD}/M_{\rm ND}
=0.05$ (diamonds), 0.2 (asterisks), with
$M_{\rm HK}/M_{\rm BK}=18/574$; are also presented in
Fig.\,\ref{f:SFGA} (bottom panels).
The related error bars are not shown to avoid 
confusion.   

It is apparent that the effect of changing
the halo-like to bulge-like thick disk mass
ratio, $M_{\rm HK}/M_{\rm BK}$, and the thick to
thin disk mass ratio, $M_{\rm KD}/M_{\rm ND}$,
within reasonable ranges, yields an
uncertainty on the EDMD which is of
the same order as the Poisson error.
Accordingly, from this point on the choice:
\begin{equation}
\label{eq:rM}
M_{\rm HK}/M_{\rm BK}=18/574\approx0.03136~~;\qquad
M_{\rm KD}/M_{\rm ND}=0.1~~;
\end{equation}
shall be assumed as reference case.

The EDMD in the Galaxy looks like its
counterpart related to the Galactic
spheroid (halo + bulge), which is
bimodal.   It is apparent that (closed
or open) simple models of chemical
evolution cannot provide a satisfactory
explanation to a bimodal EDMD, and a
different model is needed (C07).   In
this view, let the main Galactic components
i.e. the halo, the bulge, the thick disk,
and the thin disk, be supposed to undergo
distinct chemical evolutions.   Let the
thick disk be supposed to be substructured
into two subsystems, the halo-like and the
bulge-like thick disk which, in turn, 
undergo distinct chemical evolutions.
The related theoretical differential 
metallicity distribution (hereafter
referred to as TDMD) is calculated in
a selected spectral class of long-lived
stars, and the resulting TDMD is compared
with its empirical counterpart.   To this
aim, simple models implying both homogeneous
and inhomogeneous star formation shall be used,
as in earlier attempts (C01; C07).

\section{Homogeneous simple models}
\label{osmo}

Homogeneous simple models with (inhibiting
star formation) gas (Hartwick, 1976) have
widely been discussed in earlier attempts 
(C01; C07), and the interested reader is
addressed therein for further details.
What is relevant for the current paper,
shall be reported in the following.

Homogeneous simple models of the kind
considered imply four
main assumptions, namely (i) instantaneous
recycling; (ii) homogeneous star formation;
(iii) universal power-law IMF; and (iv)
gas inhibition from forming stars at a
rate proportional to the star formation rate.
Accordingly, the TDMD is represented as a
straight line (e.g., Pagel, 1989; C00; C01; 
C07):
\begin{equation}
\label{eq:psil}
\psi(\phi)=\log\frac{\diff N}{N\diff\phi}=
a\phi+b~~;
\end{equation}
and the explicit expression of the coefficients,
$a$ and $b$, reads:
\begin{lefteqnarray}
\label{eq:a}
&& a=-\frac1{\ln10}\frac{(Z_{\rm O})_\odot}{\hat{p}^{\prime\prime}}~; \\
\label{eq:b}
&& b=\log\left(\frac{\mu_o}{\mu_o-\mu_f}\frac{(Z_{\rm O})_\odot}
{\hat{p}^{\prime\prime}}\right)-a\phi_o~;
\end{lefteqnarray}
where $(Z_{\rm O})_\odot$ is the solar oxygen mass abundance,
$\hat{p}^{\prime\prime}$ the effective (oxygen)
yield, $\mu$ the (allowing star formation) gas mass
fraction, and the indices, $o$ and $f$, denote
the beginning and the end of evolution, respectively.

The combination of Eqs.\,(\ref{eq:a}) and (\ref{eq:b})
yields, after some algebra:
\begin{equation}
\label{eq:mufo}
\frac{\mu_f}{\mu_0}=1+\ln10~a\exp_{10}(-a\phi_0-b)~~;
\end{equation}
and the condition, $\mu_f/\mu_0\ge0$, translates
into:
\begin{equation}
\label{eq:mum0}
b\ge-a\phi_0+\log(-\ln10~a)~~;
\end{equation}
which makes a restriction on (otherwise acceptable)
linear fits to the EDMD.

The oxygen mass abundance, $Z_{\rm O}$, to a large extent, 
may be related to the gas mass fraction, $\mu$, as
(e.g., Hartwick, 1976; Pagel, 1989; C00; C01; C07):
\begin{lefteqnarray}
\label{eq:Oyz}
&& Z_{\rm O}-(Z_{\rm O})_o=-\hat{p}^{\prime\prime}\ln
\frac{\mu}{\mu_o}~~; \\
\label{eq:ypy}
&& \hat{p}^{\prime\prime}=\frac{\hat{p}}{1+\kappa}~~;
\end{lefteqnarray}
where $\hat{p}$, is the true (oxygen) yield and
$\kappa$ is the ratio of (inhibiting star formation)
gas mass fraction to long-lived star and
stellar remnant mass fraction.

It can be seen that a simple model
with (inhibiting star formation) gas, lower mass
limit of long-lived stars, $m_{mf}$, and inhibition
parameter, $\kappa$, is equivalent to a simple
model with the same values of independent parameters
except the above mentioned two, $(m_{mf})_1\le m_
{mf}$ and $\kappa_1=0\le\kappa\,$; and vice versa.
Accordingly, the related (true) yield, $\hat{p}_1$, has
the same value as the effective yield, $\hat{p}^
{\prime\prime}$, expressed by Eq.\,(\ref{eq:ypy}).
For further details refer to earlier papers (C01; C07).

For the models under consideration, the (inhibited
from forming stars) gas mass fraction, $D$, is
related to the long-lived star and stellar remnant
mass fraction, $s$, as (e.g., Hartwick, 1976; C01;
C07):
\begin{equation}
\label{eq:D}
D(t)-D_o=\kappa[s(t)-s_o]~~;
\end{equation}
and mass conservation reads:
\begin{equation}
\label{eq:MC}
\mu(t)+s(t)+D(t)=1~~;
\end{equation}
where $\mu$ is the (allowing star formation) gas
mass fraction.   The combination of Eqs.\,(\ref
{eq:D}) and (\ref{eq:MC}) yields:
\begin{lefteqnarray}
\label{eq:sm}
&& s-s_o=\frac{\mu_0-\mu}{1+\kappa}~~; \\
\label{eq:Dm}
&& D-D_o=\frac{\kappa(\mu_0-\mu)}{1+\kappa}~~;
\end{lefteqnarray}
finally, the related EDMD can be determined
following a standard procedure (e.g., Pagel
and Patchett, 1975; Caimmi, 1981), as performed
in C01.

With the aim of applying the above results
to the chemical evolution of the Galaxy,
the value of the true normalized yield, $\hat{p}/
(Z_{\rm O})_\odot$, the long-lived star and stellar remnant mass
fraction in a star generation, $\alpha$, and the
lower mass limit of long-lived stars, $m_{mf}$, shall be
determined from a linear fit to the EDMD in the disk solar
neighbourhood (C00) with regard to a solar
oxygen mass abundance, $(Z_{\rm O})_\odot=0.0056$ (C01),
and inhibition parameter, $\kappa=0$.   Positive
or negative values of $\kappa$, inferred from
linear fits to the EDMD in a specified Galactic
subsystem, would imply inhibited or enhanced star
formation, respectively, in comparison to the disk solar
neighbourhood.

Two extreme values of the power-law IMF exponent,
$p$, shall be considered, namely: (i) $p=2.9$,
which is a fit to the IMF determined by
Scalo (1986) or Miller and Scalo (1979),
for $m\appgeq {\rm m}_\odot$ (Wang and Silk, 1993),
and provides a good
approximation also in terms of oxygen production
(Wang and Silk, 1993),
and (ii) $p=2.35$, which coincides with the IMF
determined by Salpeter (1955).   More recent
fits to the Scalo (1986) IMF yield a gentler
slope, $p=2.7$ (e.g., Weidner and Kroupa, 2005).

For fixed values of the normalized yield, $\hat
{p}/Z_{\rm O}$, and fractional oxygen abundance
at the beginning and at the end of evolution,
$\phi_o$ and $\phi_f$, a change in the IMF slope
leaves all the output parameters unaltered
except the lower mass limit of long-lived stars,
$m_{mf}$, the mass fraction of a star generation
locked up in long-lived stars and stellar remnants,
$\alpha$, and the efficiency of star formation
rate, $C$, where the product, $\alpha C$, is also
left unchanged (C01, Appendix C, Theorem 2).
Accordingly, values of the above mentioned
parameters, related to IMF exponents within the
range, $2.35<p<2.9$, lie between their
counterparts related to $p=2.35$ and $p=2.9$.

The characteristic stellar masses shall be
taken as in earlier attempts (C00; C01; C07),
namely: upper star mass limit, $\widetilde{m}
_{Mf}=60$; neutron star progenitor lower mass
limit, $\widetilde{m}_{tr}=9$; long-lived star
upper mass limit,
$\widetilde{m}_{mr}=1$; oxygen synthesising
star lower mass limit, $\widetilde{m}_{mrO}=
10.5$; typical neutron star mass, $\widetilde
{m}_{ns}=1.5$; typical white dwarf mass,
$\widetilde{m}_{wd}=0.6$; where $\widetilde
{m}=m/{\rm m}_\odot$.   For references see
e.g.: $\widetilde{m}_{Mf}$, $\widetilde{m}_
{wd}$ (Wang and Silk, 1993); $\widetilde{m}_
{ns}$ (Prantzos, 1994); $\widetilde{m}_{tr}$
(Prantzos and Silk, 1998); $\widetilde{m}_
{mrO}$ (Pilyugin and Edmunds, 1996).   On
the other hand, the lower mass limit of
long-lived stars, $\widetilde{m}_{mf}$, results
from the model.

Some parameters which remain fixed are listed in
Tab.\,\ref{t:infi}, where the indices, 2.9 and 2.35,
denote the value of the power-law IMF exponent
used in computing the corresponding quantities.
\begin{table}
\caption[pahd]{Values of parameters
which remain fixed for different simple
homogeneous models where star formation is
inhibited.   Star masses are
normalized to the solar value, $\widetilde{m}=
m/{{\rm m}_\odot}$.   The indices, 2.9 and 2.35,
denote the value of the power-law IMF exponent
used in computing the corresponding quantities.}
\label{t:infi}
\begin{center}
\begin{tabular}{ll}
%\multicolumn{1}{c|}{meaning}
%&\multicolumn{1}{c|}{parameter}
%&\multicolumn{2}{|c}{value} \\
%\hline\noalign{\smallskip}
%& & halo & disk \\
%\noalign{\smallskip}
\hline\noalign{\smallskip}
$\hat{p}/(Z_{\rm O})_\odot\cdot10$ & 7.3722 \\
$(\widetilde{m}_{mf})_{2.9}\cdot10$ & 3.4235 \\
$(\widetilde{m}_{mf})_{2.35}\cdot10^3$ & 6.9136 \\
$\alpha_{2.9}\cdot10$ & 7.3666 \\
$\alpha_{2.35}\cdot10$ & 8.9104 \\
$(Z_{\rm O})_\odot\cdot10^3$ & 5.6 \\
$\mu_o$ & 1 \\
$s_o$ & 0 \\
$D_o$ & 0 \\
%$\phi_{o{\rm H}}\cdot10^3$ & 1 \\
%$\phi_{o{\rm B}}$ & 0.20 \\
%$\phi_{f{\rm H}}$ & 1 \\
%$\phi_{f{\rm B}}$ & 5.5 \\
\noalign{\smallskip}
\hline
\end{tabular}
\end{center}
\end{table}
The initial and final normalized oxygen abundance 
assumed for the halo-like thick disk, $\phi_{o{\rm H}}$ 
and $\phi_{f{\rm H}}$, and the bulge-like thick disk, 
$\phi_{o{\rm B}}$ and $\phi_{f{\rm B}}$, are $\phi_
{o{\rm H}}=0.03$; $\phi_{f{\rm H}}=\phi_{o{\rm B}}=
0.16$; $\phi_{f{\rm B}}=0.80$.   The thin disk shall
not be considered, as a large fraction of the related
EDMD cannot be fitted by homogeneous models (see
Fig.\,\ref{f:KDNA}, bottom panels).

For an assigned subsystem, different models 
can be obtained by changing a
single remaining input parameter, chosen to be
either the normalized effective yield, $\hat
{p}^{\prime\prime}/(Z_{\rm O})_\odot$, or the slope of
the TDMD, $a$, which conforms to Eqs.\,(\ref{eq:psil})
and (\ref{eq:a}).
A few cases are listed in Tab.\,\ref{t:homo},
related to the halo-like thick disk (HK) and
the bulge-like thick disk (BK), respectively,
where $\overline{\phi}$
represents the mean oxygen abundance
(normalized to the solar value) of stars
at the end of evolution.   Model HK2 is very
close to the limiting case, $\mu_f=0$, defined
by Eqs.\,(\ref{eq:mufo}) and (\ref{eq:mum0}).
\begin{table}
\caption[pahd]{Values of parameters related
to homogeneous simple models where star formation
is inhibited, corresponding
to linear fits to the empirical differential
metallicity distribution (EDMD) in halo-like (HK) and
bulge-like (BK) thick disk, plotted in Fig.\,\ref
{f:KHKB}.   The mean oxygen abundance
(normalized to the solar value) of stars
at the end of evolution is denoted as $\overline
{\phi}$.   Model HK2 is very
close to the limiting case, $\mu_f=0$, defined
by Eqs.\,(\ref{eq:mufo}) and (\ref{eq:mum0}).
Values of parameters related to inhomogeneous
simple models which fit the linear part of the
EDMD in thin disk (DN), plotted in Fig.\,\ref
{f:KDNA} (bottom panels), are listed for
comparison.  The definition, $Y=\hat{p}^{\prime
\prime}/(Z_{\rm O})_\odot$, is to save space.}
\label{t:homo}
\begin{center}
\begin{tabular}{rrrrrrr}
\multicolumn{1}{c}{} & \multicolumn{1}{c}{HK1} &
\multicolumn{1}{c}{HK2} & \multicolumn{1}{c}{BK1} &
\multicolumn{1}{c}{BK2} & \multicolumn{1}{c}{DN1} &
\multicolumn{1}{c}{DN2} \\
\hline\noalign{\smallskip}
%
%$\hat{p}^{\prime\prime}/(Z_{\rm O})_\odot$
%
$Y$ & 8.3518~E$-$2 & 6.8285~E$-$2 &
3.6191~E$-$1 & 2.0508~E$-$1 & 4.4233~E$-$1 & 4.4233~E$-$1 \\
$-a$ & 5.2000~E$-$0 & 6.3600~E$-$0 & 1.2000~E$-$0 & 2.1176~E$-$0 &
9.8183~E$-$1 & 9.8183~E$-$1 \\
$b$ & 1.3170~E$-$0 & 1.3885~E$-$0 & 8.1340~E$-$1 & 1.3311~E$-$0 &
1.0210~E$-$0 & 9.7643~E$-$1 \\
$\kappa$ & 7.8271~E$-$0 & 9.7962~E$-$0 & 1.0370~E$-$0 & 2.5947~E$-$0 & 
0.0000~E$-$0 & 0.0000~E$-$0 \\
$\mu_f$ & 1.7355~E$-$1 & 7.1092~E$-$2 & 2.7260~E$-$2 & 1.7656~E$-$2 & 
2.9047~E$-$1 & 1.2799~E$-$1 \\
$s_f$ & 9.3627~E$-$2 & 8.6040~E$-$2 & 4.7753~E$-$1 & 2.7327~E$-1$ &
7.0953~E$-$1 & 8.7201~E$-$1 \\
$D_f$ & 7.3283~E$-$1 & 8.4287~E$-$1 & 4.9521~E$-$1 & 7.0907~E$-$1 &
0.0000~E$-$0 & 0.0000~E$-$0 \\
$\overline{\phi}$ & 8.2804~E$-$2 & 8.4469~E$-$2 & 6.2538~E$-$1 &
4.9020~E$-$1 & 5.9389~E$-$1 & 5.0003~E$-$1 \\
\noalign{\smallskip}
\hline
\end{tabular}
\end{center}
\end{table}
Values of parameters related to inhomogeneous
simple models which fit the linear part of the
EDMD in thin disk (DN), plotted in Fig.\,\ref
{f:KDNA} (bottom panels), are listed for
comparison.

It is apparent that halo-like and bulge-like
thick disk models demand (to a different extent)
inhibited star formation to maintain (i)
universal power-law IMF, and (ii) true normalized
yield, $\hat{p}/(Z_{\rm O})_\odot$, 
unchanged with respect
to a value deduced from an acceptable fit to the
linear region of the EDMD in the thin disk solar 
neighbourhood (C00).
More precisely, about 73-80\% and 49-71\% of the 
initial halo-like and bulge-like thick disk gas,
respectively, has to be inhibited from forming 
stars, for providing an acceptable fit to the
related EDMD.   It results in a normalized 
effective yield, $\hat{p}^{\prime\prime}/(Z_
{\rm O})_\odot$, about three-four times
lower in the halo-like thick disk with respect to 
the bulge-like thick disk.   It is worth remembering
that the above difference raises to about one
order of magnitude for the halo and the bulge,
and enhanched (instead of inhibited) star
formation is demanded for the latter.   For
further details refer to earlier work (C07).

\begin{figure*}[t]
\begin{center}
\includegraphics[scale=0.8]{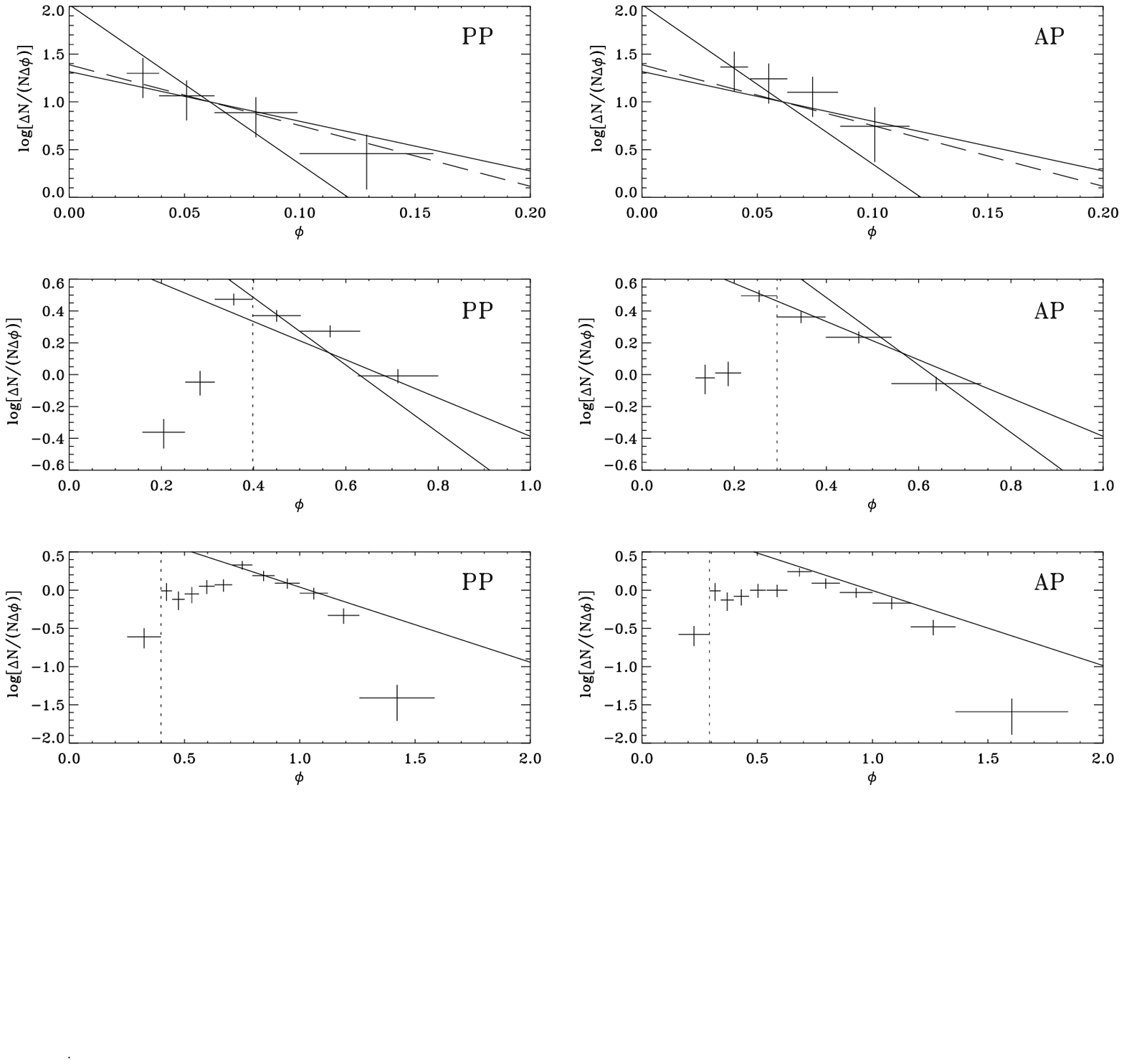}
%\centerline{\psfig{file=EGDK6.ps,height=130mm,width=140mm}}
\caption[EDMD]{Comparison between theoretical (TDMD)
and empirical differential metallicity
distribution (EDMD) in halo-like thick disk (top panels),
bulge-like thick disk (middle panels) and 
thin disk (bottom panels) stars, 
both in presence (left panels) and in absence (right 
panels) of [O/Fe] plateau, respectively.
The straight lines correspond to homogeneous
models, HK1-HK2 (top panels), BK1-BK2 (middle
panels), and to inhomogeneous models, DN1-DN2
(bottom panels), defined in Table \ref{t:homo}.
%related to homogeneous models where $a=-0.98183$;
%$\hat{p}^{\prime\prime}/(Z_{\rm O})_\odot=0.44233$;
%and $b=1.0211$ (PP), $b=0.97643$ (AP).
The dashed
line in top panels (model HK2) is very close to the
limiting case, $\mu_f=0$, defined by 
Eqs.\,(\ref{eq:mufo}) and (\ref{eq:mum0}) and
model HK1 is
represented by the less inclined (in absolute value)
full line.
Accordingly, the linear fits lying between the
dashed line and the more inclined (in absolute
value) full line, must be excluded.   
Crosses represent the data and related
uncertainties, as in Figs.\,\ref{f:KDNA}-\ref{f:KHKB}.
The dotted vertical line marks the transition from
halo to bulge/disk globular cluster morphological
type, [Fe/H]=$-$0.8.}
\label{f:HKND}
\end{center}
\end{figure*}

The TDMD deduced from models HK1-HK2 and 
BK1-BK2
is represented in Fig.\,\ref{f:HKND} and
compared to the corresponding EDMD in
connection with halo-like thick disk
(top panels),
bulge-like thick disk (middle panels),
and thin disk (bottom panels) stars, both in
presence (left panels) and in absence
(right panels) of [O/Fe] plateau.

Though the linear fit to the thin disk EDMD
is derived using inhomogeneous models (C00),
still it can be related to homogeneous models
where $a=-0.98183$;
$\hat{p}^{\prime\prime}/(Z_{\rm O})_\odot=0.44233$;
and $b=1.0211$ (PP), $b=0.97643$ (AP).   The dashed
line in top panels corresponds to model HK2 in Table
\ref{t:homo}, which is very close to the
limiting case, $\mu_f=0$, defined by 
Eqs.\,(\ref{eq:mufo}) and (\ref{eq:mum0}),
and model HK1 is represented by the less inclined (in
absolute value) full line.
Accordingly, the linear fits lying between the
dashed line and the more inclined (in absolute
value) full line, must be excluded.   

For the halo-like thick disk, homogeneous simple
models in the presence of inhibited star formation 
provide an acceptable fit (though in a restricted
range) with values of
input parameters as listed in Tab.\,\ref
{t:infi} and $\phi_{o{\rm H}}=0.03$; $\phi_{f{\rm H}}=
0.16$; $0.068<\hat{p}^{\prime\prime}/
(Z_{\rm O})_\odot<0.084$, both in presence and in
absence of [O/Fe] plateau.

For the bulge-like thick disk, homogeneous simple
models in the presence of inhibited star formation
provide an acceptable fit with values of
input parameters as listed in Tab.\,\ref
{t:infi} and $\phi_{o{\rm B}}=0.16$; $\phi_{f{\rm B}}=
0.80$; $0.205<\hat{p}^{\prime\prime}/
(Z_{\rm O})_\odot<0.362$, both in presence and in
absence of [O/Fe] plateau.

For the thin disk (represented for comparison), 
homogeneous simple models provide no acceptable 
fit with values of input parameters as listed in 
Tab.\,\ref{t:infi} and $\phi_{o{\rm N}}=0.325$,
$\phi_{f{\rm N}}=1.58$, (PP); $\phi_{o{\rm N}}=0.226$,
$\phi_{f{\rm N}}=1.85$, (AP); $\hat{p}^{\prime\prime}/
(Z_{\rm O})_\odot=\hat{p}/(Z_{\rm O})_\odot=0.44233$.

Further inspection of Fig.\,\ref{f:HKND}
shows a marked deficiency in the number of stars
observed below a threshold, [Fe/H]$\approx-1$
for the bulge-like thick disk and [Fe/H]$\approx-0.8$
for the thin disk, which is close to the transition
from halo to bulge/disk globular cluster morphological
type.    In other words, a G-dwarf
problem seems to exist for both the
bulge-like thick disk and the thin disk.

Let $\diff N_i$ be the predicted total 
number of long-lived stars within the
$i$-th subsystem $(1\le i\le n_S)$
belonging to a system, related to a 
fractional oxygen metal abundance,
$\phi\mp\diff\phi/2$, with regard
to a simple homogeneous model of
chemical evolution.   Let $N_i$ be
the predicted total number of 
long-lived stars within the above
mentioned subsystem.
The predicted counterparts for the whole
system, related to the selected models,
are $\diff N=\sum_i\diff N_i$ and
$N=\sum_iN_i$, respectively.

The relative frequency, $\diff N/N$, reads:
\begin{equation}
\label{eq:dNN}
\frac{\diff N}N=\sum_{i=1}^{n_S}\frac{\diff N_i}N=
\sum_{i=1}^{n_S}\frac{N_i}N\frac{\diff N_i}{N_i}~~;
\end{equation}
or, using Eq.\,(\ref{eq:NM}):
\begin{equation}
\label{eq:dNM}
\frac{\diff N}N=
\sum_{i=1}^{n_S}\frac{M_i}M\frac{\diff N_i}{N_i}~~;
\end{equation}
where $M_i$ and $M$ are the mass of the
subsystem and the whole system, respectively.

The TDMD related to the whole system, keeping
in mind Eq.\,(\ref{eq:psil}), is:
\begin{equation}
\label{eq:dpsia}
\psi=\log\frac{\diff N}{N\diff\phi}=\log\sum_{i=1}^{n_S}\frac
{\diff N_i}{N\diff\phi}
%\nonumber \\&& \phantom{\psi}
=\log\sum_{i=1}^{n_S}\frac{N_i}N\frac
{\diff N_i}{N_i\diff\phi}~~;
\end{equation}
and the combination of Eqs.\,(\ref{eq:NM}),
(\ref{eq:psil}), and (\ref{eq:dpsia}) yields:
\begin{equation}
\label{eq:psti}
\psi=\log\sum_{i=1}^{n_S}\frac{M_i}M\exp_{10}(a_i\phi+b_i)~~;
\end{equation}
where the particularization to the thick disk,
the thick + thin disk, and the Galaxy, is defined
by Eqs.\,(\ref{seq:M}) and (\ref{eq:rM}).

Two alternatives shall be considered, K$_1$ and
K$_2$ (halo-like + bulge-like thick disk),
D$_1$ and D$_2$ (thick + thin disk), G$_1$ and
G$_2$ (Galaxy), according if the coefficients, 
$a_i$ and $b_i$,
are taken from cases HK1, BK1, or HK2, BK2, of 
Table \ref{t:homo}; DN1 or DN2 of Table \ref
{t:homo}; H1, B1, or H2, B2, of Table 8 (C07),
respectively.   The last models are related to 
the halo (H) and bulge (B), which are needed in
dealing with the Galaxy.

\begin{figure*}[t]
\begin{center}
\includegraphics[scale=0.8]{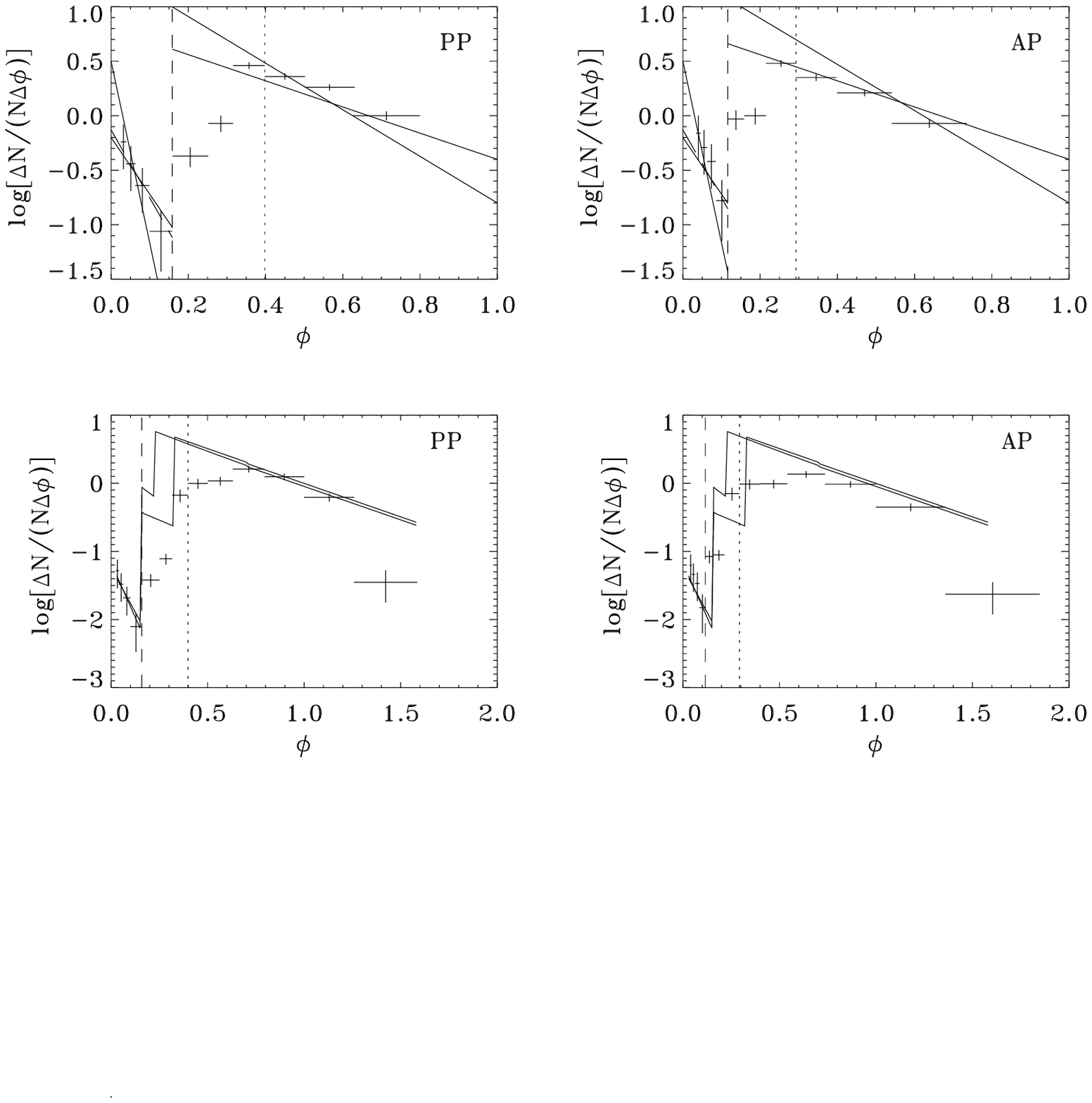}
%\centerline{\psfig{file=EGDK6A.ps,height=130mm,width=140mm}}
\caption[EDMD]{Comparison between theoretical (TDMD)
and empirical differential metallicity
distribution (EDMD) in halo-like + bulge-like
thick disk (top panels) and thick + thin disk 
(bottom panels), 
both in presence (left panels) and in absence (right 
panels) of [O/Fe] plateau, respectively.
The straight lines correspond to homogeneous
models, HK1-HK2 (low oxygen abundances), BK1-BK2
(intermediate oxygen abundances), and DN1-DN2
(high oxygen abundances), defined in Table
\ref{t:homo}.
Crosses represent renormalized data and
related uncertainties,
as in Fig.\,\ref{f:KDND}.
The dotted vertical line marks the transition from
halo to bulge/disk globular cluster morphological
type, [Fe/H]=$-$0.8.   The dashed vertical line marks
the (assumed) transition from halo-like to bulge-like
thick disk.}
\label{f:HKDD}
\end{center}
\end{figure*}

The TDMD related to the halo-like + bulge-like
thick disk and to the thick + thin disk, is 
plotted in Fig.\,\ref
{f:HKDD} (top and bottom panels, respectively)
and compared to its empirical counterpart,
represented in Fig.\,\ref{f:KDND} (crosses),
both in presence (left panels) and in absence
(right panels) of [O/Fe] plateau.
A discontinuity is exhibited by the TDMD,
where bulge-like thick disk or thin disk 
formation is assumed
to start.   

\begin{figure*}[t]
\begin{center}
\includegraphics[scale=0.8]{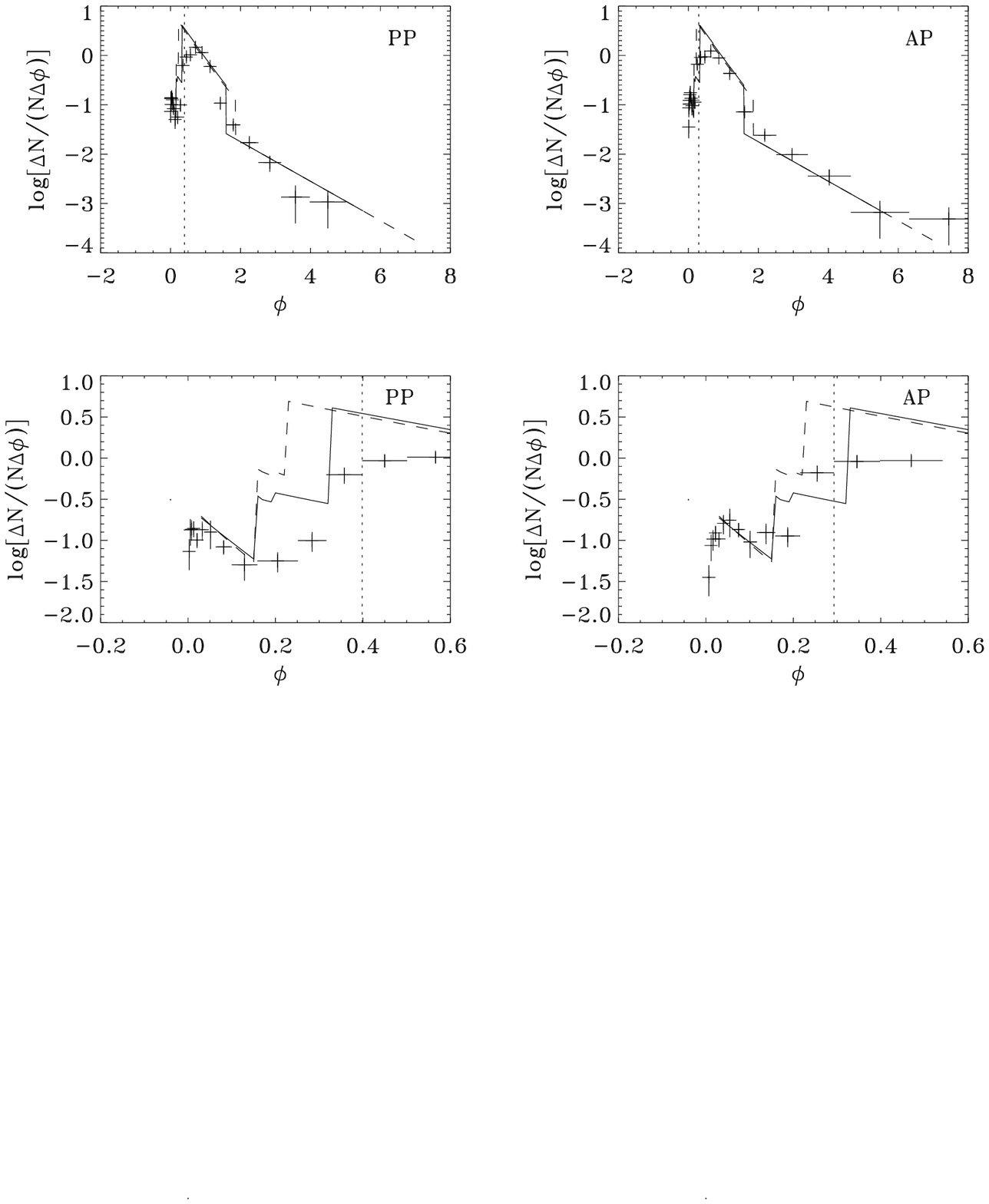}
%\centerline{\psfig{file=EGDK8.ps,height=130mm,width=140mm}}
\caption[EDMD]{Comparison between theoretical (TDMD)
and empirical differential metallicity
distribution (EDMD) in the Galaxy (top panels) and 
zoomed for low oxygen abundance (bottom panels), 
both in presence (left panels) and in absence (right 
panels) of [O/Fe] plateau, respectively.
The curves correspond to models H1, B1, HK1, BK1,
DN1 (full); H2, B2, HK2, BK2, DN2 (dashed) defined 
in: C07, Tables 7 and 8 therein (H, B); Tables
\ref{t:infi} and \ref{t:homo} (HK, BK, DN); all
combined via Eq.\,(\ref{eq:psti}).
The dotted vertical line marks the transition from
halo to bulge/disk globular cluster morphological
type, [Fe/H]=$-$0.8.}
\label{f:HG}
\end{center}
\end{figure*}

The TDMD related to the Galaxy is plotted
in Fig.\,\ref{f:HG} and compared to its
empirical counterpart represented in
Fig.\,\ref{f:SFGA} (middle panels),
both in presence (left panels) and in
absence (right panels) of [O/Fe] plateau.
Discontinuities are exhibited by the TDMD
where bulge, halo-like thich disk, bulge-like 
thick disk, and thin disk formation are assumed
to start.

Although a G-dwarf problem seems to
exist for both the halo, the bulge, the halo-like
thick disk, the bulge-like thick disk, and the
thin disk, the current
model provides a viable interpretation to the
occurrence of three extremum points, two
maxima and one minimum, in the EDMD.

\section{Inhomogeneous, simple models}
\label{inmo}

Inhomogeneous simple models with (inhibiting star
formation) gas have widely been discussed in earlier
attempts (C01; C07), following the basic formulation
of the model (Malinie et al., 1993; C00), and the
interested reader is addressed therein for further
details.   What is relevant for the current paper, 
shall be reported in the following.

In the light of the model, each Galactic subsystem
is conceived
as being structured into a number of discrete, entirely
gaseous, identical regions, and a background of
long-lived stars, stellar remnants, and gas inhibited from
star formation,
which have been generated earlier. The evolution
occurs via a sequence of identical time steps. At the
beginning of each step, star formation stochastically takes
place in a subclass of ``active'' regions, as described
by simple homogeneous models, while the others remain
``quiescent''. At the end of each step, high-mass stars
have died whereas low-mass stars have survived up until today,
according to instantaneous recycling approximation.
In addition, the enriched gas which remains from active regions
is instantaneously mixed with the unenriched gas
within quiescent regions, to form a new set of identical
regions for the next step.

To allow comparison between simple homogeneous and
inhomogeneous models, the current investigation
shall be restricted to the special case of expected
evolution, where the fraction of active regions is
time-independent.   With respect to the general case,
both the theoretical age-metallicity relation (TAMR) 
and the TDMD exhibit minor changes for
the disk solar neighbourhood (C00) and negligible variations
for the halo solar neighbourhood (C01).

In any case,
active regions evolve as in the Hartwick's (1976)
model, yielding the usual relationship between
the newly synthesised oxygen gas mass fraction within
an active region, $\Delta^\ast(Z_{\rm O}^\prime)_{\rm
R}$, and the gas mass fraction within an
active region, $\mu_{\rm R}^\prime$, at the end of each
step in the following way:
\begin{equation}
\label{eq:DZOR}
\Delta^\ast(Z_{\rm O}^\prime)_{\rm R}=-\hat{p}^{\prime
\prime}\ln(\mu_{\rm R}^\prime)~~;
\end{equation}
provided that the IMF is universal and the effects of
initial composition on star evolution are neglected.

The newly synthesised oxygen gas mass fraction, due
to an active region, within the system, $\Delta^\ast
(Z_{\rm O})_{\rm R}$, at the end of each step, is (C00):
\begin{equation}
\label{eq:DZOR1}
\Delta^\ast(Z_{\rm O})_{\rm R}=\frac{\mu_{\rm R}^\prime}
{n[1-\chi(1-\mu_{\rm R}^\prime)]}\Delta^\ast(Z_{\rm O}^
\prime)_{\rm R}~~;
\end{equation}
where $n$ is the number of active and quiescent
regions, $\chi$ the probability of a region being
active, and both remain constant during the
expected evolution.

The newly synthesised oxygen gas mass fraction, due
to all active regions, within the system, $\Delta^\ast
Z_{\rm O}$, at the end of each step, is (C00):
\begin{equation}
\label{eq:DZO1}
\Delta^\ast Z_{\rm O}=k\Delta^\ast(Z_{\rm O})_{\rm R}=
n\chi\Delta^\ast(Z_{\rm O})_{\rm R}~~;
\end{equation}
where $k=n\chi$ is the number of active regions.

The changes in (allowing star formation) gas and
oxygen abundance, related to the $(\ell+1)$-th step,
read (C00; C01):
\begin{lefteqnarray}
\label{eq:mulo}
&& \mu_{\ell+1}=\mu_\ell q=\mu_oq^{\ell+1}~~; \\
\label{eq:ZOlo}
&& (Z_{\rm O})_{\ell+1}=(Z_{\rm O})_\ell+\Delta^
\ast Z_{\rm O}=(Z_{\rm O})_o+(\ell+1)\Delta^
\ast Z_{\rm O}~~; \\
\label{eq:DZO}
&& \Delta^\ast Z_{\rm O}=-\frac{\hat{p}^{\prime\prime}
(1-q)\mu_{\rm R}^\prime\ln(\mu_{\rm R}^\prime)}{(1-\mu_
{\rm R}^\prime)q}~~; \\
\label{eq:q}
&& q=1-\chi(1-\mu_{\rm R}^\prime)~;\quad0\le q\le1~;
\end{lefteqnarray}
where the parameter, $q$, may be considered as an
effective gas
mass fraction within a region at the end of each step,
i.e. the mean gas mass fraction averaged on both
active and quiescent regions.

The combination of Eqs.\,(\ref{eq:DZOR})-(\ref
{eq:q}) yields:
\begin{lefteqnarray}
\label{eq:ZOlp}
&& (Z_{\rm O})_{\ell+1}=(Z_{\rm O})_o-\hat{p}^
\prime[\ln(\mu_{\ell+1})-\ln(\mu_o)]~~; \\
\label{eq:DZOp}
&& \Delta^\ast Z_{\rm O}=-\hat{p}^\prime\ln q~~; \\
\label{eq:ypyq}
&& \hat{p}^\prime=\hat{p}^{\prime\prime}\frac
{(1-q)\mu_{\rm R}^\prime\ln\mu_{\rm R}^\prime}
{(1-\mu_{\rm R}^\prime)q\ln q}~~;\qquad\hat{p}^
\prime\le\hat{p}^{\prime\prime}~~;
\end{lefteqnarray}
where $\hat{p}^\prime$ is an effective yield
due to both inhomogeneous star formation and inhibited star
formation.   In the limit of instantaneous
recycling i.e. chemically homogeneous
interstellar medium (e.g., the Simple model, Hartwick's
model), $\chi\to1$, $q\to\mu_{\rm R}^\prime$,
and $\hat{p}^\prime\to\hat{p}^{\prime\prime}$.

According to Eq.\,(\ref{eq:ZOlp}), 
the dependence of oxygen abundance
on the gas mass fraction in inhomogeneous
models remains unchanged with respect to
homogeneous models provided that: (i) the expected
evolution is considered; (ii) the probability
of a region being active, $\chi_\ell$, the
gas mass fraction within a region, $\mu_{{\rm R}
\ell}^\prime$, and the (true) yield, $\hat{p}_\ell$, during
the $(\ell+1)$-th step, do not depend on the
evolution; and (iii) the effective yield,
$\hat{p}^{\prime\prime}_\ell$, is replaced
by the effective yield, $\hat{p}^{\prime}_\ell$.
In absence of inhibited star formation, it can
be shown that the effective yield, $\hat{p}^
\prime$, takes the same expression as in the
inhomogeneous model of Malinie et al. (1993).
For further details refer to earlier papers (C00, C01).

Although the TDMD cannot be analytically expressed
in the framework of inhomogeneous simple models, it
still can in the special case of
expected evolution, with regard to the starting
point, $\psi(\phi_o)=\psi_o$, and the ending
point of the initial step, $\psi(\phi_o+\Delta^\ast
\phi_o)=\psi_1$, where $\Delta^\ast\phi_o$ is the net
oxygen abundance (normalized to the solar value)
increase in gas component at the end of the first
step.   The related explicit expressions can be
found in previous work [C01; C07, Eqs.\,(39) and
(40) therein].
An approximate expression of the TDMD related
to the first step, $\psi_1$, with the terms up
to the second order retained, can be found in
previous work
[C01; C07, Eq.\,(41) therein].   The last expression
is valid also in the general case, provided
$\Delta^\ast\phi$ is replaced by $\Delta^\ast\phi_o$ and the
probability, $\chi$, by the relative frequency,
$\nu_o=k_o/n_o$, being $k_o$ and $n_o$ the number
of active and all regions, respectively, with
regard to the first step.

The values of some parameters related to the expected
evolution, concerning cases HK1, HK2; BK1, BK2; DN1,
DN2; are listed in Table \ref{t:inho}.
\begin{table}
\caption[pahd]{Values of parameters related
to the expected evolution of inhomogeneous
simple models, in connection
with two different cases for the halo-like
thick disk, HK1 and HK2, for the bulge-like
thick disk, BK1 and BK2, and for the thin
disk, DN1 and DN2, respectively.   The 
indices, 2.9 and 2.35,
denote values related to the power-law IMF
exponent, $p$, in computing the corresponding
quantities.   For the parameter,
$\psi_1$, upper and lower values are
calculated as in C07 by use of Eqs.\,(40)
and (41) therein, respectively.   The
effective yield, $\hat{p}^\prime$, is
related to inhomogenities in oxygen
abundance due to the presence of active
and quiescent regions, whereas oxygen is
uniformly distributed within active regions.
The lower part of the table (last 6 rows)
is related to models with inhibited
star formation.   Parameters not reported
therein have the same value as in the upper part,
with the exception of $\hat{p}$, $\alpha$,
and $m_{mf}$, which are listed in Tab.\,\ref
{t:infi} together with other parameters not
appearing here.   The effective yield, $\hat
{p}^{\prime\prime}$, due to the presence of
(inhibiting star formation) gas within active
regions, is listed as $\hat{p}$ in the upper 
part of the table.
The effective yield, $\hat{p}^\prime$, due to
the presence of both (inhibiting
star formation) gas within active
regions, and (precluding star formation) gas
within quiescent regions, is listed with the
same notation in the upper part of the table.
The index, R, denotes a generic active
region.   The mean oxygen abundance
(normalized to the solar value) of stars
at the end of evolution is denoted as $\overline
{\phi}$.}
\label{t:inho}
\begin{center}
\begin{tabular}{crrrrrr}
\multicolumn{1}{c}{} & \multicolumn{1}{c}{HK1} &
\multicolumn{1}{c}{HK2} & \multicolumn{1}{c}{BK1} &
\multicolumn{1}{c}{BK2} & \multicolumn{1}{c}{DN1} &
\multicolumn{1}{c}{DN2} \\
\noalign{\smallskip}
\hline\noalign{\smallskip}
$\mu_{\rm R}^\prime$               & 2.3356~E$-$1 & 2.0512~E$-$1 & 2.1579~E$-$1 & 1.2933~E$-$1 &    3.7148~E$-$1 &    3.6106~E$-$1 \\
$q$                                & 9.2861~E$-$1 & 7.9327~E$-$1 & 8.4346~E$-$1 & 4.5877~E$-$1 &    9.5175~E$-$1 &    9.2106~E$-$1 \\
$\chi$                             & 9.3147~E$-$2 & 2.6007~E$-$1 & 1.9961~E$-$1 & 6.2162~E$-$1 &    7.6765~E$-$2 &    1.2355~E$-$1 \\
$\hat{p}/(Z_{\rm O})_\odot$        & 8.3518~E$-$2 & 6.8285~E$-$2 & 3.6191~E$-$1 & 2.0508~E$-$1 &    7.3722~E$-$1 &    7.3722~E$-$1 \\
$\hat{p}^\prime/(Z_{\rm O})_\odot$ & 3.8418~E$-$2 & 3.1411~E$-$2 & 1.6648~E$-$1 & 9.4338~E$-$2 &    4.4233~E$-$1 &    4.4233~E$-$1 \\
$\mu_f$                            & 7.4358~E$-$1 & 3.9599~E$-$1 & 5.0613~E$-$1 & 4.4298~E$-$2 &    2.9047~E$-$1 &    1.2799~E$-$1 \\
$\alpha_{2.9}$                     & 9.6108~E$-$1 & 9.6795~E$-$1 & 8.5071~E$-$1 & 9.0955~E$-$1 &    7.3666~E$-$1 &    7.3666~E$-$1 \\
$\alpha_{2.35}$                    & 9.8634~E$-$1 & 9.8880~E$-$1 & 9.4337~E$-$1 & 9.6710~E$-$1 &    8.9104~E$-$1 &    8.9104~E$-$1 \\
$(\widetilde{m}_{mf})_{2.9}$       & 4.1289~E$-$2 & 3.3283~E$-$2 & 1.8306~E$-$1 & 1.0516~E$-$1 &    3.4235~E$-$1 &    3.4235~E$-$1 \\
$(\widetilde{m}_{mf})_{2.35}$      & 2.0404~E$-$5 & 1.1592~E$-$5 & 1.1294~E$-$3 & 2.4576~E$-$4 &    6.9136~E$-$3 &    6.9136~E$-$3 \\
$\Delta^\ast\phi$                  & 2.8456~E$-$3 & 7.2745~E$-$3 & 2.8341~E$-$2 & 7.3509~E$-$2 &    2.1874~E$-$2 &    3.6375~E$-$2 \\
$\Delta^\ast\phi_{\rm R}^\prime$   & 1.2146~E$-$1 & 1.0818~E$-$1 & 5.5498~E$-$1 & 4.1947~E$-$1 &    7.3003~E$-$1 &    7.5101~E$-$1 \\
$\psi_o$                           & 6.3844~E$-$1 & 7.9973~E$-$1 & 4.7970~E$-$2 & 5.0128~E$-$1 & $-$8.3341~E$-$1 & $-$7.1627~E$-$1 \\
$\psi_1(40)$                       & 6.3107~E$-$1 & 7.7680~E$-$1 & 3.1076~E$-$2 & 4.2576~E$-$1 & $-$8.3984~E$-$1 & $-$7.2694~E$-$1 \\
$\psi_1(41)$                       & 6.3104~E$-$1 & 7.7659~E$-$1 & 3.0965~E$-$2 & 4.2344~E$-$1 & $-$8.3985~E$-$1 & $-$7.2698~E$-$1 \\
                                   &              &              &              &              &                 &                 \\
$\kappa$                           & 7.8271~E$-$0 & 9.7962~E$-$0 & 1.0370~E$-$0 & 2.5947~E$-$0 &    0.0000~E$-$0 &    0.0000~E$-$0 \\
$s_{\rm R}^\prime$                 & 8.6829~E$-$2 & 7.3626~E$-$2 & 3.8498~E$-$1 & 2.4221~E$-$1 &    6.2852~E$-$1 &    6.3894~E$-$1 \\
$D_{\rm R}^\prime$                 & 6.7962~E$-$1 & 7.2126~E$-$1 & 3.9923~E$-$1 & 6.2846~E$-$1 &    0.0000~E$-$0 &    0.0000~E$-$0 \\
$\overline{\phi}$                  & 7.6505~E$-$2 & 7.0371~E$-$2 & 3.6920~E$-$1 & 3.0277~E$-$1 &    5.9389~E$-$1 &    5.0003~E$-$1 \\
$s_f$                              & 2.9049~E$-$2 & 5.5946~E$-$2 & 2.4245~E$-$1 & 2.6586~E$-$1 &    7.0953~E$-$1 &    8.7201~E$-$1 \\
$D_f$                              & 2.2737~E$-$1 & 5.4806~E$-$1 & 2.5142~E$-$1 & 6.8984~E$-$1 &    0.0000~E$-$0 &    0.0000~E$-$0 \\
\noalign{\smallskip}
\hline
\end{tabular}
\end{center}
\end{table}

The parameters appearing therein, which equal
their counterparts corresponding to homogeneous
models (Tables \ref{t:infi} and \ref{t:homo},
cases HK and BK),
must be connected with active regions, and
for this reason some corresponding output
parameters show different values.   For
further details refer to earlier work (C07,
Appendix B).
It is worth recalling that cases DN1 and 
DN2 listed in Table \ref{t:homo} for
comparison, are related to inhomogeneous
models listed in Table \ref{t:inho}.

New input parameters are: the effective yield,
$\hat{p}^\prime$, and the normalized oxygen
abundance increase at the end of a step
with regard to the whole system, $\Delta^\ast
\phi$, and to an active region, $\Delta^\ast
\phi_{\rm R}^\prime$, respectively.   In
presence of inhibited (from forming stars)
gas (cases HK and BK), the star
and inhibited gas mass fraction related to
the whole system at the end of evolution,
$s_f$ and $D_f$, are expressed in terms of
their counterparts related to an active region
at the end of a step, $s_{\rm R}^\prime$ and
$D_{\rm R}^\prime$, as (C07):
\begin{lefteqnarray}
\label{eq:ssR}
&& s_f=\frac{1-\mu_f}{1-\mu_{\rm R}^\prime}s_{\rm R}^\prime~~; \\
\label{eq:DDR}
&& D_f=\frac{1-\mu_f}{1-\mu_{\rm R}^\prime}D_{\rm R}^\prime~~;
\end{lefteqnarray}
which are listed in the lower part of
Tab.\,\ref{t:inho}.   For further
details refer to earlier work (C07).

\begin{figure*}[t]
\begin{center}
\includegraphics[scale=0.8]{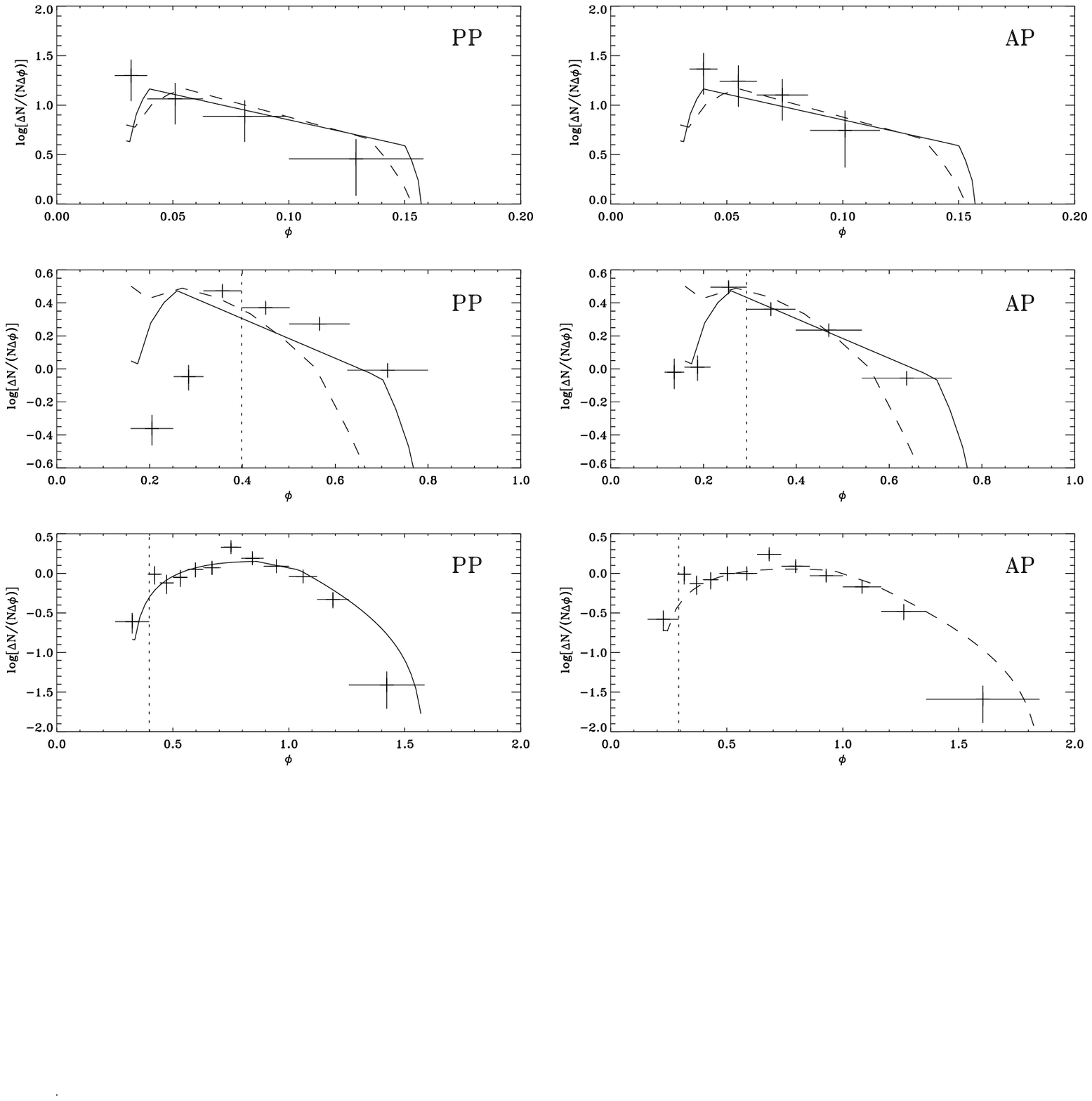}
%\centerline{\psfig{file=EGDK9.ps,height=130mm,width=140mm}}
\caption[EDMD]{Comparison between theoretical (TDMD)
and empirical differential metallicity
distribution (EDMD) in halo-like thick disk (top panels),
bulge-like thick disk (middle panels), and
thin disk (bottom panels) stars,
both in presence (left panels) and in absence
(right panels) of [O/Fe] plateau, respectively.
Full curves correspond to inhomogeneous
models HK1 (top panels), BK1 (middle panels),
and DN1 (left bottom panel), and dashed curves 
to HK2 (top panels), BK2 (middle panels), and
DN2 (right bottom panel),
defined in Table \ref{t:inho}.
Crosses represent the data and related
uncertainties, as in Figs.\,\ref{f:KDNA}-\ref{f:KHKB}.
The dotted vertical line marks the transition from
halo to bulge/disk morphological
type in globular clusters, [Fe/H]=$-$0.8.}
\label{f:IKND}
\end{center}
\end{figure*}

The TDMD related to cases HK1, BK1, DN1,
(full lines) and HK2, BK2, DN2, (dashed 
lines) is
compared in Fig.\,\ref{f:IKND} to the
corresponding EDMD with regard to
halo-like thick disk (top panels), bulge-like
thick disk (middle panels), and thin
disk (bottom panels) stars, both 
in presence (left panels) and in 
absence (right panels) of [O/Fe] plateau.

Inhomogeneous simple models with inhibited star
formation provide an acceptable fit using values
of input parameters listed in Table \ref{t:infi}
and $0.031<\hat{p}^\prime/(Z_{\rm O})_\odot<0.039$
(halo-like thick disk); $0.094<\hat{p}^\prime/
(Z_{\rm O})_\odot<0.167$ (bulge-like thick disk); in
and $\hat{p}^\prime/(Z_{\rm O})_\odot\approx0.442$
(thin disk).

In dealing with active regions, the normalized
oxygen abundance at the end of evolution,
$\phi_f$, has to be replaced with the
normalized oxygen abundance at the end of
a step, $\phi_i+\Delta^\ast\phi_{\rm R}^\prime$.

Further inspection of Fig.\,\ref{f:IKND}
shows that, with respect to homogeneous
simple models plotted in Fig.\,\ref
{f:HKND}, the fit is more or less
unchanged for the halo-like thick disk, 
slightly worsened for the bulge-like thick
disk in presence of [O/Fe] plateau, but
substantially improved for the bulge-like
thick disk in absence of [O/Fe] plateau
and the thin disk.   In addition,
a G-dwarf problem still remains (but
alleviated) only for the bulge-like thick
disk in presence of [O/Fe] plateau.

\begin{figure*}[t]
\begin{center}
\includegraphics[scale=0.8]{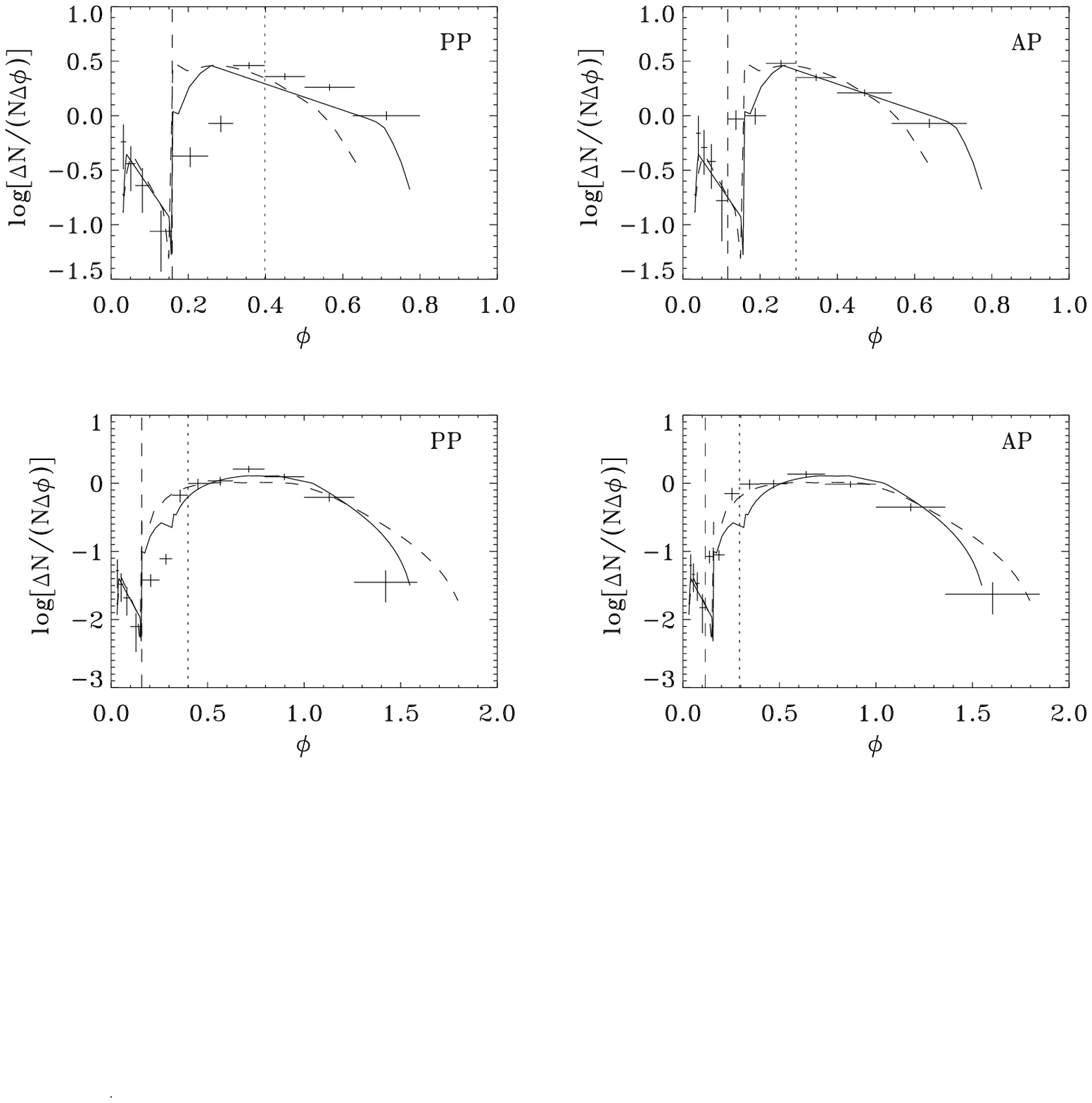}
%\centerline{\psfig{file=EGDK9A.ps,height=130mm,width=140mm}}
\caption[EDMD]{Comparison between theoretical (TDMD)
and empirical differential metallicity
distribution (EDMD) in halo-like + bulge-like
thick disk (top panels) and thick + thin disk
(bottom panels),
both in presence (left panels) and in absence
(right panels) of [O/Fe] plateau, respectively.
Full and dashed curves correspond to models K1,
K2, (top panels), and D1, D2, (bottom panels),
obtained using models HK1, BK1; HK2, BK2; (top
panels); and HK1, BK1, DN1; HK2, BK2, DN2;
(bottom panels); respectively, defined in
Tables \ref{t:infi} and \ref{t:inho}, and combined
via  Eqs.\,(\ref{eq:NM}) and (\ref{eq:dpsia}).
Crosses represent the renormalized data and related
uncertainties, as in Fig.\,\ref{f:HKDD}.
The dotted vertical line marks the transition from
halo to bulge/disk globular cluster morphological
type, [Fe/H]=$-$0.8.   The dashed vertical line
marks the (assumed) transition from halo-like to
bulge-like thick disk.}
\label{f:IKDD}
\end{center}
\end{figure*}

The TDMD related to a system made of
different subsystems, has to be
numerically computed using Eqs.\,(\ref
{seq:M}) and (\ref{eq:dpsia}) for
assigned values of the thick disk
to thin disk mass ratio, $M_{\rm KD}/
M_{\rm ND}$, and halo-like to bulge-like
thick disk mass ratio, $M_{\rm HK}/
M_{\rm BK}$.   The following models
have been considered with regard to
the reference case: K1 and K2 for
the thick disk, using models HK1,
BK1, and HK2, BK2, respectively,
defined in Table \ref{t:inho};
D1 and D2 for the thick + thin disk, using
models HK1, BK1, DN1, and HK2, BK2,
DN2, respectively,
defined in Table \ref{t:inho}.
A comparison with the related EDMD
(Fig.\,\ref{f:KDND}, top and bottom
panels, respectively, crosses) is
made in Fig.\,\ref{f:IKDD} (top
and bottom panels, respectively)
both in presence
(left panels) and in absence (right
panels) of [O/Fe] plateau.   The
trend looks like its counterpart
exhibited by homogeneous simple
models (Fig.\,\ref{f:HKDD}) but
the fit is improved, and similar
considerations can be made.

\begin{figure*}[t]
\begin{center}
\includegraphics[scale=0.8]{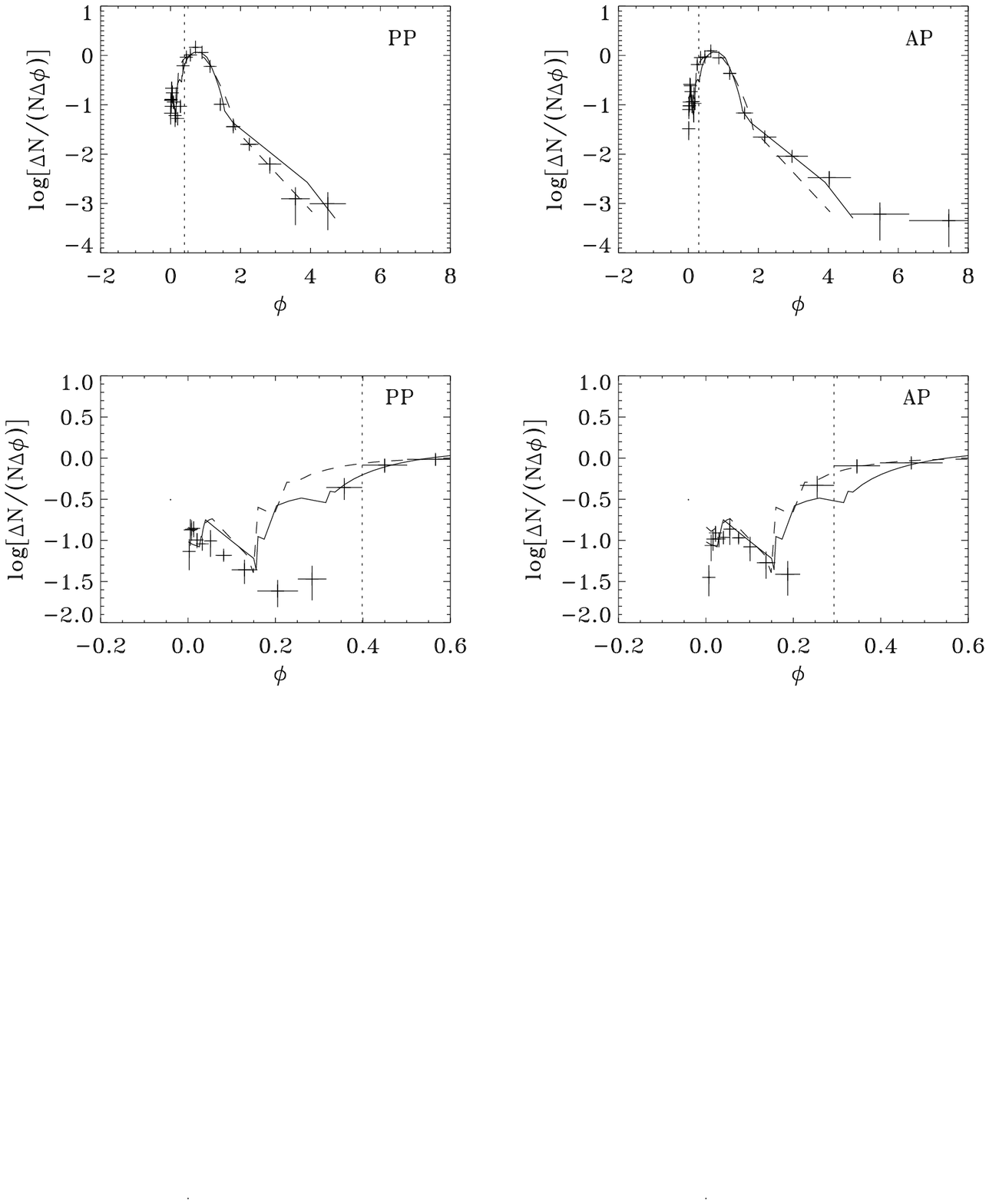}
%\centerline{\psfig{file=EGDK11.ps,height=130mm,width=140mm}}
\caption[EDMD]{Comparison between theoretical (TDMD)
and empirical differential metallicity
distribution (EDMD) in the Galaxy (top panels)
and zoomed for low normalized
oxygen abundances (bottom panels),
both in presence (left panels) and in absence
(right panels) of [O/Fe] plateau, respectively.
Curves correspond to models H1, B1, HK1, BK1,DN1,
(full); H2, B2, HK2, BK2, DN2, (dashed); defined
in C07, Tables 7 and 9 therein (H, B) and Tables
\ref{t:infi} and \ref{t:inho} (HK, BK, DN), all
combined via Eqs.\,(\ref{eq:NM}) and (\ref{eq:dpsia}).
The dotted vertical line marks the transition from
halo to bulge/disk globular cluster morphological
type, [Fe/H]=$-$0.8.}
\label{f:IG}
\end{center}
\end{figure*}

The following models have also been considered:
G1 and G2 for the Galaxy, using models H1, B1,
HK1, BK1, DN1, and H2, B2, HK2, BK2, DN2,
respectively, where models H and B are related
to the halo and the bulge, respectively, and
are defined in C07 (Table 9).   A comparison
with the related EDMD (Fig.\,\ref{f:SFGA}, middle
and bottom panels, crosses) is made 
in Fig.\,\ref{f:IG} both in
presence (left panels) and in absence (right panels)
of [O/Fe] plateau.   The trend looks like its
counterpart exhibited by homogeneous simple models
(Fig.\,\ref{f:HG}) but the fit is improved,
and similar considerations can be made.

In the framework of inhomogeneous models, both
active and quiescent regins must take origin
from (allowing star formation) gas.   An inspection
of Table \ref{t:inho} shows that, for models HK, only
a few percent (2.9-5.6\%) of the total mass has
been converted into (long-lived) stars, leaving
about 74-39\% in (allowing star formation) gas;
and about 22-55\% in (inhibiting star formation)
gas; for models BK, about one quarter (24-27\%)
of the total mass has
been converted into (long-lived) stars, leaving
about 51-4\% in (allowing star formation) gas;
and about 29-69\% in (inhibiting star formation)
gas; for models DN, more than two thirds (71-87\%)
of the total mass has
been converted into (long-lived) stars, leaving
about 29-13\% in (allowing star formation) gas;
and no (inhibiting star formation) gas.

%The above results are consistent with the idea,
%that the total amount of gas left at the end of
%the halo-like thick disk evolution, gave birth
%to the bulge-like thick disk.   In fact, the
%resulting mass ratio, $M_{\rm HK}/M_{\rm BK}$,
%would be equal to a few percents.    The same
%idea, related to the thick and the thin disk,
%would imply a mass ratio, $M_{\rm KD}/M_{\rm ND}$,
%equal to about one quarter, which is much larger
%than a value of about one tenth, estimated from
%observations.   Accordingly, the thick and the
%thin disk underwent distinct chemical evolutions,
% unless a substantial fraction of the proto-thick
%disk contributed to bulge formation.

The mean normalized oxygen abundance within
long-lived stars is $\overline{\phi}=$
0.077-0.070, 0.37-0.30, and 0.60-0.50, in
the halo-like thick disk, the bulge-like
thick disk, and the thin disk, respectively.

It is worth noting that the Galaxy would be
mostly luminous for a IMF exponent, $p=2.9$,
and lower stellar mass limit, $\widetilde{m}_
{\rm mf}\approx0.34$, and mostly dark for $p=2.35$
and $\widetilde{m}_{\rm mf}\approx0.007$.   On the other
hand, the assumption of an universal IMF
implies the same value of the true yield
and lower mass limit of long-lived stars,
for Galactic subsystems.

\section{Discussion} \label{disc}

Although the Galaxy appears to be dominated
(by mass and star number) by the two major
subsystems, the bulge and the thin disk,
still the contrary holds with respect to 
selected metallicity bins.   For instance,
low-metallicity bins are dominated by halo
stars.   Even if the spheroid and the disk
evolved separately (albeit with similar
present metal abundance), the distribution
of specific angular momentum is known to
be consistent with both a halo-bulge
and a thick disk-thin disk (Wyse and Gilmore,
1992; Ibata and Gilmore, 1995)
collapse.   Then in early times a reversed
situation occurred, where the main Galactic
components were the halo and the thick disk,
and gas exchange between the two reservoirs
cannot be excluded.   Some signature of the
formation and the early evolution of the
Galaxy could be found in the EDMD related
to the whole system.    In this view, the
thick disk has to be conceived as pre-existing
with respect to the thin disk.   For a
discussion of different scenarios on the
origin of the thick disk see e.g., Gilmore
et al. (1995).

In brief, the method used in the current
and earlier work (C07) is the following.
An assigned system (e.g., the Galactic
spheroid, the Galactic thick disk, the
Galactic disk, the Galaxy) is conceived
as made of a number of subsystems where
the chemical evolution occurs separately.
Then the resulting EDMD and TDMD are
calculated using Eqs.\,(\ref{eq:NM}),
(\ref{eq:psil}), and (\ref{eq:dpsia}),
i.e. weighting by mass.   The larger
uncertainty concerns the thick disk,
where (i) two different samples (WG95
and CB00) have been renormalized under
the assumption that they are equally
representative of objects belonging
to the metallicity range in common,
and (ii) the resulting sample (FS07),
has been conceived as made of two
different kind of objects, belonging
to the halo-like or the bulge-like
thick disk, according if the related low-metallicity
and the high-metallicity EDMD resembles
its halo and bulge counterpart, respectively.
In this view, the results are merely indicative
and need to be confirmed using more complete
samples.   In any case, the G-dwarf problem
in the bulge-like thick disk, the thick + thin
disk, and the Galaxy, is alleviated for
homogeneous models and much strongly for
inhomogeneous models, with respect to models
where chemical evolution occurs at the same
extent in different subsystems.

The absence of a G-dwarf problem in the
(thick + thin) Galactic disk has been
established in a recent attempt (Haywood,
2006), using the Simple model.   It is not
in contradiction with the results of the
current paper, for the following reasons.
First, scalelength corrections in thin
disk samples have been used in Haywood
(2006), while different samples belonging
to the thick and the thin disk have been
used in the current paper.   Second, the
chemical evolution has been assumed to
occur at the same rate in the whole disk
in Haywood (2006), while a different extent
has been considered in the current paper,
according if the halo-like thick disk, the
bulge-like thick disk, and the thin disk,
are dealt with.

A lower mass limit of long-lived stars,
exceeding the theoretical
Jeans stellar mass $(0.007\le\widetilde{m}_J\le0.01)$,
occurs for a power-law exponent $p=2.9$, which
is a fit to the Scalo (1986) IMF for $m\appgeq
{\rm m}_\odot$, concerning both mass distribution and
oxygen production (Wang and Silk, 1993). A less
steep Salpeter (1955) IMF, implying $p=2.35$, 
is marginally consistent with the theoretical
Jeans stellar mass (Tab.\,\ref{t:infi}).   On
the other hand, the occurrence of stellar
wind would reduce oxygen nucleosynthesis by a factor of 
about 2.5 (e.g., Wang and Silk, 1993) which, in turn, would
raise the lower mass limit of long-lived stars, provided
the (true) yield remains unchanged (e.g., C07, Appendix C).

Within the current model, the lower mass
limit of long-lived stars is related to
a power-law IMF where $2.35\le p\le2.9$
provides a poor fit to low $(\widetilde{m}<1)$
masses (e.g., Tinney, 1993; Burrows et al., 
1993; Binney, 1999; Weidner and Kroupa, 2005).
In particular,
the number of low-mass stars is overestimated
by a power-law IMF in comparison with
current observations.   
The real and the power-law IMF, $\phi
(\widetilde{m})=\widetilde{m}^{-p}$,
can be constrained to yield the same
mass in a star generation, as:
\begin{leftsubeqnarray}
\slabel{eq:lomaa}
&& \int_{(\widetilde{m}_{mf})_{\rm real}}^{\widetilde{m}_{Mf}}
\widetilde{m}\Phi_{\rm real}(\widetilde{m})\diff\widetilde{m}=
\frac{\widetilde{m}_{Mf}^{2-p}-\widetilde{m}_{mf}^{2-p}}{2-p}~; \\
\slabel{eq:lomab}
&& (\widetilde{m}_{mf})_{\rm real}\le\widetilde{m}_{mf}~;
\label{seq:loma}
\end{leftsubeqnarray}
where $\widetilde{m}_{Mf}$ is the upper
mass limit of formed stars.
Accordingly, the parameter, $m_{mf}$,
related to a power-law IMF, has to be
conceived as an effective lower mass
limit of long-lived stars.

The more relevant parameters of inhomogeneous simple 
models related to halo and bulge (C07), halo-like
thick disk, bulge-like thick disk, and thin disk,
are listed in Table \ref{t:phbd}.
\begin{table}
\caption[phbd]{Comparison between parameters
of inhomogeneous simple models, related to
the Galactic halo and bulge (C07, Table 9,
cases H1 and B1), halo-like thick disk, bulge-like
thick disk, and thin disk (Table \ref{t:inho},
cases HK1, BK1,
and DN1).   An universal power-law initial
mass function (IMF) is assumed in all cases, which
leaves other parameters i.e. $\hat{p}$,
$\alpha$, and $m_{mf}$, unchanged.   Gas
and star mass fractions, $\mu_f$, $D_f$,
and $s_f$, are related to the initial mass
with regard to the bulge, where total mass
conservation is violated by gas inflow (C07).
Caption of parameters: $\chi$ - probability
of a region being active; $\mu_R^\prime$ -
(allowing star formation) gas mass fraction
within an active region at the end of a step;
$q$ - effective gas mass fraction within a
region at the end of a step; $\kappa$ - ratio
of gas mass fraction which inhibits star
formation to long-lived star and stellar
remnant mass fraction (inhibition parameter);
$\mu_f$ - (allowing star formation) gas mass
fraction at the end of evolution; $D_f$ -
(inhibiting star formation) gas mass fraction
at the end of evolution; $s_f$ - long-lived
star mass fraction at the end of evolution;
$\hat{p}^{\prime
\prime}/(Z_{\rm O})_\odot$ - effective
oxygen yield related to inhibited or enhanched
star formation, normalized to the solar oxygen
abundance; $\hat{p}^{\prime}/(Z_{\rm O})_\odot$
- effective oxygen yield related to both
inhomogeneous star formation and inhibited or
enhanched star formation, normalized to the
solar oxygen abundance.}
\label{t:phbd}                 
\begin{center}
\begin{tabular}{llllll}
%\multicolumn{1}{c|}{meaning} &
\multicolumn{1}{c|}{parameter}
&\multicolumn{2}{c}{value} \\
\hline\noalign{\smallskip}
%&
& halo & bulge & halo-like  & bulge-like & thin disk \\
%&
&      &       & thick disk & thick disk &           \\
\noalign{\smallskip}
\hline\noalign{\smallskip}
%probability          &
$\chi$                                     & $9.5579~10^{-1}$ & $\phantom{-}8.7545~10^{-1}$ & $9.3147~10^{-2}$ & $1.9961~10^{-1}$ & $7.6765~10^{-2}$ \\
%gas mass fraction    &
$\mu_{\rm R}^\prime$                       & $8.2201~10^{-4}$ & $\phantom{-}1.4139~10^{-1}$ & $2.3356~10^{-1}$ & $2.1579~10^{-1}$ & $3.7148~10^{-1}$ \\
%gas mass fraction    &
$q$                                        & $4.4991~10^{-2}$ & $\phantom{-}2.4833~10^{-1}$ & $9.2861~10^{-1}$ & $8.4346~10^{-1}$ & $9.5175~10^{-1}$ \\
%inhibition parameter &
$\kappa$                                   &  4.5169          & $         - 3.2099~10^{-1}$ &  7.8271          &  1.0370          &  0.0000          \\
%gas mass fraction    &
$\mu_f$                                    & $4.0973~10^{-6}$ & $\phantom{-}3.8030~10^{-3}$ & $7.4358~10^{-1}$ & $5.0613~10^{-1}$ & $2.9047~10^{-1}$ \\
%gas mass fraction    &
$D_f$                                      & $8.1874~10^{-1}$ & $         - 4.7094~10^{-1}$ & $2.2737~10^{-1}$ & $2.5142~10^{-1}$ &  0.0000          \\
%star mass fraction   &
$s_f$                                      & $1.8126~10^{-1}$ & $\phantom{-}1.4671        $ & $2.9049~10^{-2}$ & $2.4245~10^{-1}$ & $7.0953~10^{-1}$ \\
%normalized yield     &
$\hat{p}^{\prime\prime}/(Z_{\rm O})_\odot$ & $1.3363~10^{-1}$ & $\phantom{-}1.0857        $ & $8.3518~10^{-2}$ & $3.6191~10^{-1}$ & $7.3722~10^{-1}$ \\
%normalized yield     &
$\hat{p}^\prime/(Z_{\rm O})_\odot$         & $5.3452~10^{-3}$ & $\phantom{-}7.6001~10^{-1}$ & $3.8418~10^{-2}$ & $1.6648~10^{-1}$ & $4.4233~10^{-1}$ \\
\noalign{\smallskip}
\hline
\end{tabular}
\end{center}
\end{table}
The related values are taken from C07 (Table
9, models H1 and B1), and Table \ref{t:inho}
(models HK1, BK1, and DN1), respectively.
An universal power-law IMF is
assumed,
%$\phi(\widetilde{m})\propto\widetilde{m}^{-p}$,
which makes no change in value of the physical
parameters, $\hat{p}$, $\alpha$, and $m_{mf}$.
Negative values of inhibition parameter,
$\kappa$, and final (inhibiting star
formation) gas mass fraction, $D_f$, related
to the bulge, correspond to enhanced instead
of inhibited star formation.   For further
details refer to earlier work (C07).

For active regions, the probability of
star formation is $\chi\appleq1$ for the halo,
$\chi<1$ for the bulge, and $\chi\ll1$ for both
the thick and the thin disk. Accordingly, the gas
mass fraction remaining at the end of each step,
$\mu_{\rm R}^\prime$, is close to
zero for the halo, about one seventh for the
bulge, one fifth-one fourth for the thick disk,
and one third for the thin
disk.  The effective yield, $\hat{p}^{\prime
\prime}$, is about twice larger in the thin
disk (where it coincides with the true yield)
than in the bulge-like thick disk; ten times
than in the halo-like thick disk; five times
than in the halo; (where gas is partially
inhibited from forming stars); and about two
thirds lower than in the bulge (where star
formation is enhanced by inflowing gas with
same composition as the preexisting gas; for
further details refer to C07).

For the whole system, a similar trend occurs.
The ratio of gas mass fraction at the end and at the
beginning of a step is $q\ll1$ for the halo, $q<1$ for
the bulge, and $q \appleq1$ for both the thick and the
thin disk. Accordingly, the (allowing star formation)
gas mass fraction remaining at the end of the evolution,
$\mu_f$, is close to zero for the halo, about
four thousandths for the bulge, three quarters
for the halo-like thick disk, one half for the
bulge-like thick disk, and one third for the thin
disk.  The effective yield,
$\hat{p}^\prime$, is about three times larger in
the thin disk (where it is due to inhomogeneous
star formation) than in the bulge-like thick disk;
a dozen times than in the halo-like thick disk;
eighty times than in the halo; (where it is due
to both inhomogeneous star formation and inhibition
from forming stars); and about three fifths lower
than in the bulge (where it is due to both inhomogeneous
star formation and enhanced star formation; for further
details refer to C07).   At the end of evolution,
it remains about 18\% of long-lived
stars (including stellar remnants) and 82\% of gas
inhibited from forming stars in the halo;
about 72\% of long-lived stars from primeval
gas and 28\% from inflowed gas in the bulge
(with respect to the final mass, while mass
fractions listed in Table \ref{t:phbd} are
related to the initial mass);
about 0.3\% of long-lived stars, 74\% of
(allowing star formation) gas, and 23\% of
(inhibiting star formation) gas in the
halo-like thick disk;
about 24\% of long-lived stars, 51\% of
(allowing star formation) gas, and 25\% of
(inhibiting star formation) gas in the
bulge-like thick disk;
about 71\% of long-lived stars and 29\% of
gas allowing star formation in the thin disk.

The TDMD related to both homogeneous and
inhomogeneous simple models provides an
acceptable fit to the EDMD related to the
thick disk, the thick + thin disk,
(Figs.\,\ref{f:HKDD} and \ref{f:IKDD};
upper and lower panels, respectively)
and the Galaxy (Figs.\,\ref{f:HG}
and \ref{f:IG}): in particular,
a non monotonic trend is reproduced.
While models assume that star formation
at the beginning of the evolution of
each subsystem starts abruptly with constant
efficiency, the data seem to indicate
a somewhat gradual rate, with increasing
efficiency.   In other words, the
occurrence of some physical process
(not necessarily gas infall) seems
to inhibit star formation in early times
of evolution.   Accordingly, the
history of each Galactic subsystem
(halo, bulge, halo-like thick disk,
bulge-like thick disk, thin disk) could
be conceived
as made of two distinct phases, namely
(i) assembling, where (long-lived) star
formation efficiency is gradually increasing,
and (ii) stabilization, where (long-lived)
star formation efficiency maintains constant.

The assumption that the EDMD of the local
disk is representative of the global disk,
even if in contrast with an inside-out
disk formation, can be considered as a
useful zero-th order approximation.   On
the other hand, nearby stars older than
about 0.2 Gyr come from birth sites which 
span a large range in Galactocentric
distances (e.g., Rocha-Pinto et al., 2000).
The orbital diffusion coefficient deduced
from the observed increase of velocity
dispersion with age implies that presently
local stars have suffered a rms azimuthal
drift from about 2 kpc (for an age of 0.2
Gyr) to many Galactic orbits (for an age
of 10 Gyr); for further details refer to
earlier work 
(Wielen, 1977).   Considerable, but smaller,
drift should occur also on the radial
direction.   In this sense, the star formation
rate inferred for nearby stars is a measure
of the global Milky Way star formation rate,
at least at the sun Galactocentric radius
(Rocha-Pinto et al., 2000), according to
the estimates of the diffusion coefficient
(e.g., Meusinger et al., 1991).

As outlined in an earlier attempt (Wang and
Silk, 1993), radial flows of the gas
(possibly due to the observed spiral density
waves) or different disk star formation
histories (with time scales comparable to
that of chemical evolution) between the
inner end the outer parts of the Galaxy,
may change the local abundances, but the
overall abundance in the disk should not
be affected as long as there is no gain or
loss of material in the disk.   It can also
be noticed (Wang and Silk, 1993) that the
average oxygen abundance in the disk is
roughly the solar value, and can be plausibly
explained by the standard Scalo IMF with
lower and upper star mass limit, $\widetilde
{m}_{mf}=0.1$ and $\widetilde{m}_{Mf}=60$,
respectively, or by a power-law IMF with
exponent, $p$, within the range, $2.35\le
p\le2.9$, and the same $\widetilde{m}_{mf}$ 
and $\widetilde{m}_{Mf}$.   For further
details refer to the parent paper (Wang
and Silk, 1993).

At the end of halo-like and bulge-like
thick disk evolution, the fractional
gas and star mass predicted by the
model are:
\begin{lefteqnarray}
\label{eq:MUKg}
&& \frac{(M_{\rm UK})_{\rm gas}}{(M_{\rm UK})_o}=
(\mu_{\rm UK})_f+(D_{\rm UK})_f~~;\qquad{\rm U}={\rm H},{\rm B}~~; \\
\label{eq:MUKs}
&& \frac{(M_{\rm UK})_{\rm stars}}{(M_{\rm UK})_o}=
(s_{\rm UK})_f~~;\qquad{\rm U}={\rm H},{\rm B}~~;
\end{lefteqnarray}
where $(M_{\rm HK})_o$ and $(M_{\rm BK})_o$
are the initial halo-like and bulge-like
thick disk mass, respectively.   The
combination of Eqs.\,(\ref{eq:MUKg}) and
(\ref{eq:MUKs}) yields:
\begin{equation}
\label{eq:MKgs}
(M_{\rm UK})_{\rm gas}=\frac{(\mu_{\rm UK})_f+(D_{\rm UK})_f}
{(s_{\rm UK})_f}(M_{\rm UK})_{\rm stars}\qquad{\rm U}={\rm H},{\rm B}~~;
\end{equation}
where $(M_{\rm HK})_{\rm stars}$ and
$(M_{\rm BK})_{\rm stars}$ must be
equal to the current halo-like and
bulge-like thick disk mass, respectively.

For models HK1-2, BK1-2, the values
of fractional masses, $(\mu_{\rm UK})_f$,
$(D_{\rm UK})_f$, and $(s_{\rm UK})_f$,
are listed in Table \ref{t:inho}, and
taking a current halo-like thick disk
mass, $M_{\rm HK}=(M_{\rm HK})_{\rm
stars}=(9/296)(29/55){\rm M}_{10}
\approx0.016{\rm M}_{10}$, and a
current bulge-like thick disk mass,
$M_{\rm BK}=(M_{\rm BK})_{\rm
stars}=(287/296)(29/55){\rm M}_{10}
\approx0.511{\rm M}_{10}$, according
to Eqs.\,(\ref{eq:Mb}) and
(\ref{eq:rM}), it is found the following:
\begin{lefteqnarray}
\label{eq:MHKg}
&& (M_{\rm HK})_{\rm gas}=0.5359-0.2705{\rm M}_{10}~~; \\
\label{eq:MBKg}
&& (M_{\rm BK})_{\rm gas}=1.5974-1.3351{\rm M}_{10}~~; \\
\label{eq:MDKg}
&& (M_{\rm DK})_{\rm gas}=2.1333-1.6056{\rm M}_{10}~~;
\end{lefteqnarray}
where $(M_{\rm DK})_{\rm gas}=
(M_{\rm HK})_{\rm gas}+(M_{\rm BK})_
{\rm gas}$ is the thick disk gas mass
budget available for the thin disk
formation, which amounts to about
36\%-28\% of the assumed thin disk
mass, Eq.\,(\ref{eq:Mb}).

The above discrepancy could be
avoided, taking into consideration
mass loss during the thick disk
evolution, in the sense that gas
flows away but chemical evolution
therein occurs as in the remaining
gas.   Accordingly, Eqs.\,(\ref
{eq:MUKg})-(\ref{eq:MKgs}) still
hold but the star mass at the end
of halo-like and bulge-like thick
disk evolution is related to current
halo-like and bulge-like thick disk
mass, as:
\begin{equation}
\label{eq:fU}
(M_{\rm UK})_{\rm stars}=\frac{M_{\rm UK}}{1-f_{\rm U}}~~;\qquad
{\rm U}={\rm H},{\rm B}~~;
\end{equation}
where $(M_{\rm UK})_{\rm stars}$
includes stars both inside and outside
the thick disk, and $f_{\rm H}$, $f_
{\rm B}$, are
the star mass fraction outside the
halo-like and bulge-like thick disk,
respectively.

The boundary conditions to the problem
under discussion are:
\begin{lefteqnarray}
\label{eq:B1}
&& (M_{\rm HK})_{\rm gas}=M_{\rm BK}+M_{\rm ND}~~; \\
\label{eq:B2}
&& (M_{\rm BK})_{\rm gas}=M_{\rm ND}~~;
\end{lefteqnarray}
the gas mass budget at the end of the
halo-like and bulge-like thick disk
must necessarily equal the bulge-like
thick disk + thin disk and the thin
disk mass, respectively, in absence
of mass loss outside the disk.

The combination of Eqs.\,(\ref{eq:MKgs}),
(\ref{eq:fU}), (\ref{eq:B1}), and (\ref
{eq:B2}) yields:
\begin{equation}
\label{eq:fgs}
\frac{\delta_{\rm HU}M_{\rm BK}+M_{\rm ND}}{M_{\rm UK}}(1-f_{\rm U})=
\frac{(\mu_{\rm UK})_f+(D_{\rm UK})_f}{(s_{\rm UK})_f}~~;\qquad
{\rm U}={\rm H},{\rm B}~~;
\end{equation}
where $\delta_{ij}$ is the Kronecker symbol.
Then the halo-like and bulge-like thick
disk mass fraction to be lost during the
evolution is:
\begin{equation}
\label{eq:fUn}
f_{\rm U}=1-\frac{(\mu_{\rm UK})_f+(D_{\rm UK})_f}{(s_{\rm UK})_f}
\frac{M_{\rm UK}}{\delta_{\rm HU}M_{\rm BK}+M_{\rm ND}}~~;\qquad
{\rm U}={\rm H},{\rm B}~~;
\end{equation}
which, for models HK1-2, BK1-2, and mass values
prescribed by Eqs.\,(\ref{eq:Mb}) and (\ref
{eq:rM}), reads:
\begin{lefteqnarray}
\label{eq:fH}
&& f_{\rm H}=0.9074-0.9532~~; \\
\label{eq:fB}
&& f_{\rm B}=0.6970-0.7468~~;
\end{lefteqnarray}
where it can be seen that more
than 90\% of
the halo-like thick disk initial
gas budget and more than 69\% of
the bulge-like thick disk initial
gas budget were removed from their
reservoirs during the evolution.
In other words, the bulge-like
thick disk and the thin disk did
not begin to start abruptly, but
were gradually assembled during
the evolution of their precursors,
as inferred from the features of
the EDMD mentioned above.

The results found in the current paper
rely on the assumption of universal IMF,
but the method used here can be generalized
to the case where different IMFs in space
and/or time are considered.   To this aim,
it is sufficient to realize that the relation,
$M_i=\overline{m}_iN_i$, with regard to
$i$-th subsystem, holds for different mean
stellar masses, $\overline{m}_i$, related
to different IMFs.   Accordingly, Eq.\,(\ref
{eq:NM}) has to be generalized as $N_i/N=
(\overline{m}/\overline{m}_i)(M_i/M)$, where
$\overline{m}$ is the mean stellar mass within
the system.    Different IMFs during the
assembling and stabilization phase have
been considered since a long time (e.g.,
Schmidt, 1963; Caimmi, 1978a,b, 1981, 1982;
C00).

A different IMF in the bulge, with respect
to the disk, has recently been advocated
for explaining the giant star metal distribution
in both the Milky Way and Andromeda bulge
(Ballero et al., 2007a), contrary
to what has been found for the Milky Way bulge
(C07).   This discrepancy could be owing to
two orders of reasons, namely: (i) instantaneous
mixing has been assumed for chemical evolution
in the former case and instantaneous recycling
in the latter and, (ii) the cumulative metallicity
distribution has been used in the former case
and the differential metallicity distribution
in the latter.   In addition, different imput
parameters have been used for the related models.
A comparison between observations and model
predictions with regard to [$\alpha$/Fe] ratios
(e.g., Ballero et al., 2007b), cannot be performed
in the framework of the model used here, as only
oxygen chemical evolution is considered.
Neverthless, it is worth mentioning that Fe
production mainly depends on type Ia supernovae
which, in turn, have binary star progenitors.
Then different IMFs in different subsystems
could be mimiced by different amounts of
binary star fractions.

The results found in the current paper also
rely on the assumed [O/H]-[Fe/H] relation,
Eqs.\,(\ref{eq:gra})-(\ref{eq:isa}).
Data from recent investigations, different
for different Galactic subsystems (e.g.,
Bensby et al., 2004; Jonsell et al., 2005;
Fulbright et al., 2005; Garcia Perez et al.,
2006; Melendez et al., 2006; Ramirez et al.,
2007) yield [O/H]-[Fe/H] relations lying
between their counterparts expressed by
Eqs.\,(\ref{eq:gra}) and (\ref{eq:isa})
which, in turn, allow comparison with earlier results
related to the halo (C01), the bulge (C07),
and the Galactic spheroid (C07).   On the
other hand, the method used in the current
paper holds regardless from the assumed (or
not) [O/H]-[Fe/H] relation.

Although direct [O/H] values are available
from recent observations, still they cannot
be used for the global EDMD in absence of
their counterparts related to each Galactic
subsystem.   For instance, a thick disk
sample investigated by Reddy et al. (2006)
is biased towards low-metallicity objects
similarly to the WG95 sample; a sample
investigated by Ramirez et al. (2007),
hereafter quoted as the RA07 sample, is
made of $N=523$ nearby stars belonging
to the halo, the thick disk, and the thin
disk, within a metallicity range, $-1.5\le$
[Fe/H]$\le0.5$.   Distinct and homogeneous
samples representative of the halo, the
bulge, the thick disk, and the thin disk,
where oxygen abundance is directly inferred,
would considerably improve the results.

Neverthless, an indicative comparison
between the EDMD deduced for the thick
+ thin disk can be made, with regard
to the FS07 sample (Fig.\,\ref{f:KDND},
bottom panels, crosses), and the RA07
sample where oxygen abundance was
directly inferred both in presence
and in absence of the LTE approximation
(Ramirez et al., 2007, Table 6, available
at the CDS).   The local thermodynamic
equilibrium (LTE) approximation well
holds deep in the stellar photosphere,
and a continuous transition occurs up
to complete non-equilibrium (NLTE)
high in the atmosphere.   For further
details refer to specialistic textbooks
(e.g., Gray, 2005,
Chap.\,6).   The related EDMD, obtained
by use of Eqs.\,(\ref{eq:lgfi})-(\ref
{seq:psier}), is listed in Table \ref
{t:RA07} both in presence (LTE) and
in absence (NLTE) of the LTE approximation.
\begin{table}
\caption[par]{The empirical, differential
metallicity distribution (EDMD), 
deduced from the RA07 sample ($N=523$),
both in presence (LTE) and in 
absence (NLTE) of the LTE approximation.}
\label{t:RA07}
\begin{center}
\begin{tabular}{rrrrrrrrrrr}
\multicolumn{3}{c|}{} &
\multicolumn{4}{c|}{LTE} &
\multicolumn{4}{c}{NLTE} \\
\hline\noalign{\smallskip}
\multicolumn{1}{c}{[O/H]} &
\multicolumn{1}{c}{$\phi$} &
\multicolumn{1}{c}{$\Delta^\mp\phi$} &
\multicolumn{1}{c}{$\phantom{0}\psi$} &
\multicolumn{1}{c}{$\Delta^-\psi$} &
\multicolumn{1}{c}{$\Delta^+\psi$} &
\multicolumn{1}{c}{$\Delta N$} &
\multicolumn{1}{c}{$\phantom{0}\psi$} &
\multicolumn{1}{c}{$\Delta^-\psi$} &
\multicolumn{1}{c}{$\Delta^+\psi$} &
\multicolumn{1}{c}{$\Delta N$}  \\
\noalign{\smallskip}
\hline\noalign{\smallskip}
$-$1.2 & 0.064 & 0.007 &          &           &        & \phantom{10}0 & $-$0.882 & $\infty$ & 0.301 & \phantom{10}1 \\
$-$1.1 & 0.080 & 0.009 &          &           &        & \phantom{10}0 &          &           &       & \phantom{10}0 \\
$-$1.0 & 0.101 & 0.012 & $-$1.082 & $\infty$ & 0.301  & \phantom{10}1 &          &           &       & \phantom{10}0 \\
$-$0.9 & 0.127 & 0.015 &          &           &        & \phantom{10}0 & $-$1.182 & $\infty$ & 0.301 & \phantom{10}1 \\
$-$0.8 & 0.160 & 0.018 &          &           &        & \phantom{10}0 & $-$1.282 & $\infty$ & 0.301 & \phantom{10}1 \\
$-$0.7 & 0.201 & 0.023 & $-$1.382 & $\infty$ & 0.301  & \phantom{10}1 & $-$0.683 & 0.257     & 0.161 & \phantom{10}5 \\
$-$0.6 & 0.253 & 0.029 & $-$1.005 & 0.374     & 0.198  & \phantom{10}3 & $-$1.181 & 0.533     & 0.232 & \phantom{10}2 \\
$-$0.5 & 0.318 & 0.036 & $-$1.105 & 0.374     & 0.198  & \phantom{10}3 & $-$0.627 & 0.176     & 0.125 & \phantom{10}9 \\
$-$0.4 & 0.401 & 0.046 & $-$0.837 & 0.206     & 0.139  & \phantom{10}7 & $-$0.403 & 0.113     & 0.090 & \phantom{1}19 \\
$-$0.3 & 0.505 & 0.058 & $-$0.636 & 0.135     & 0.103  & \phantom{1}14 & $-$0.250 & 0.082     & 0.069 & \phantom{1}34 \\
$-$0.2 & 0.635 & 0.073 & $-$0.539 & 0.104     & 0.084  & \phantom{1}22 & $-$0.001 & 0.053     & 0.047 & \phantom{1}76 \\
$-$0.1 & 0.800 & 0.092 & $-$0.176 & 0.058     & 0.051  & \phantom{1}64 & $-$0.027 & 0.048     & 0.044 & \phantom{1}90 \\
$-$0.0 & 1.007 & 0.115 & $-$0.044 & 0.044     & 0.040  & \phantom{}109 & $-$0.104 & 0.047     & 0.042 & \phantom{1}95 \\
   0.1 & 1.267 & 0.145 & $-$0.129 & 0.043     & 0.039  & \phantom{}113 & $-$0.213 & 0.048     & 0.043 & \phantom{1}93 \\
   0.2 & 1.595 & 0.183 & $-$0.269 & 0.045     & 0.041  & \phantom{}103 & $-$0.489 & 0.059     & 0.052 & \phantom{1}62 \\
   0.3 & 2.009 & 0.230 & $-$0.626 & 0.062     & 0.054  & \phantom{1}57 & $-$0.950 & 0.093     & 0.076 & \phantom{1}27 \\
   0.4 & 2.529 & 0.290 & $-$1.139 & 0.104     & 0.084  & \phantom{1}22 & $-$1.637 & 0.206     & 0.139 & \phantom{10}7 \\
   0.5 & 3.183 & 0.365 & $-$2.281 & 0.533     & 0.232  & \phantom{10}2 & $-$2.582 & $\infty$ & 0.301 & \phantom{10}1 \\
   0.6 & 4.008 & 0.459 & $-$2.381 & 0.533     & 0.232  & \phantom{10}2 &          &           &       & \phantom{10}0 \\
\noalign{\smallskip}
\hline
\end{tabular}
\end{center}
\end{table}
Halo contamination may safely be neglected except
at low metallicities, where the global contribution
is also negligible in that (i) the halo star
subsample in the RA07 sample is by far incomplete, and (ii)
the metal-weak thick disk ([Fe/H]$\le-1$) has
not been considered therein.   For further details
refer to the parent paper (Ramirez et al., 2007).

\begin{figure*}[t]
\begin{center}
\includegraphics[scale=0.8]{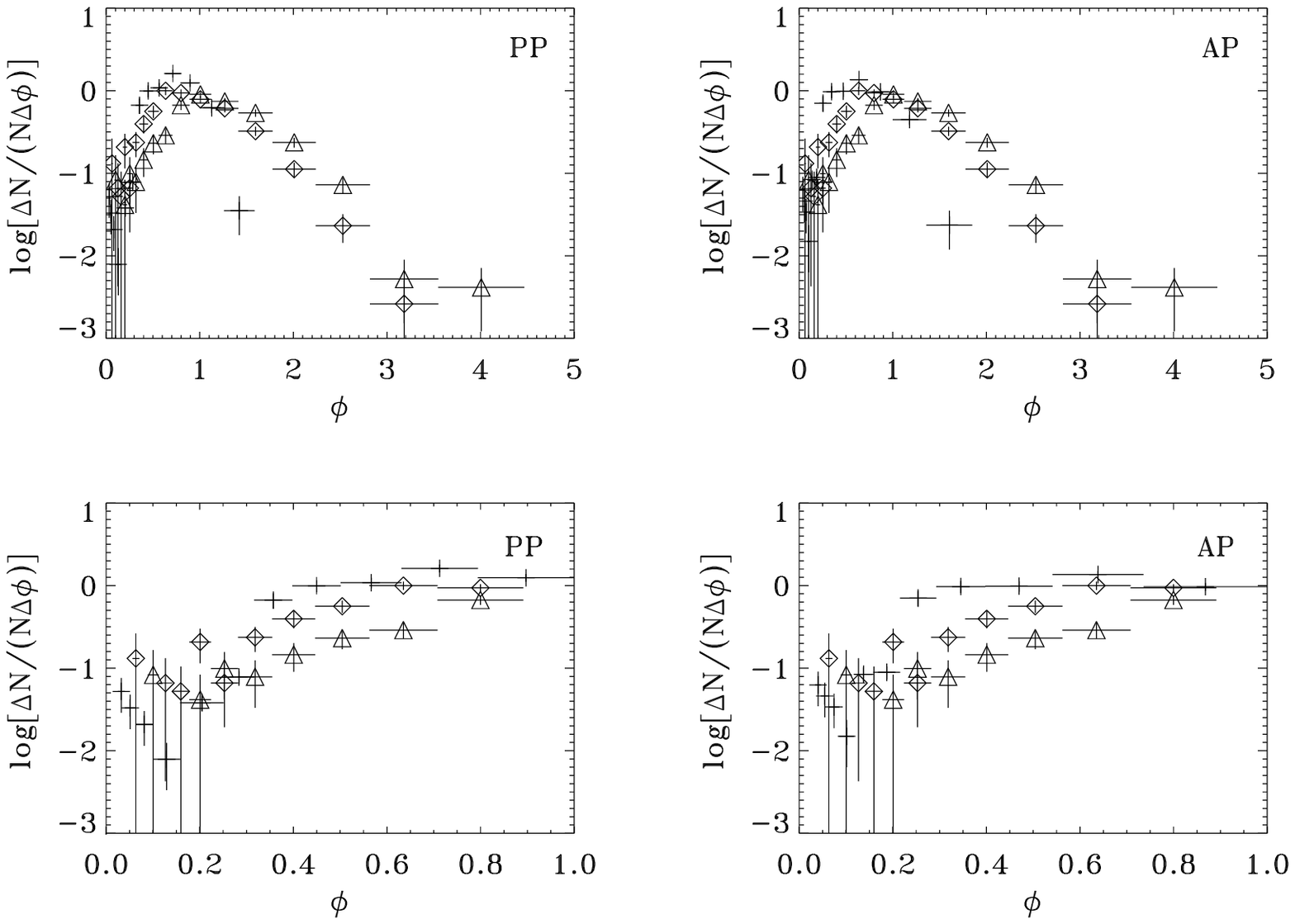}
%\centerline{\psfig{file=EGOH4B.ps,height=130mm,width=140mm}}
\caption[KDNA]{The empirical, differential metallicity
distribution (EDMD) deduced from the RA07 sample $(N=523)$
including nearby halo (by far incomplete), thick disk, and
thin disk stars, both in presence (triangles) and in absence 
(diamonds) of the LTE approximation.   The EDMD deduced for
the thick + thin disk (Fig.\,\ref{f:KDND}, crosses) is also
represented for comparison both in presence (left panels)
and in absence (right panels) of [O/Fe] plateau.   The
low-metallicity tail is zoomed in bottom panels.}
\label{f:RA07}
\end{center}
\end{figure*}

The EDMD deduced from the RA07 sample and listed in
Table \ref{t:RA07}, is compared in Fig.\,\ref
{f:RA07} with its counterpart related to the
thick + thin disk, plotted in Fig.\,\ref{f:KDND}
(bottom panels, crosses), both in presence and
in absence of the LTE approximation in the
former case, and [O/Fe] plateau in the latter.
The low-metallicity tail is zoomed in bottom
panels.   It is apparent that using the LTE
approximation understimates the EDMD for
subsolar oxygen abundances and overstimates
for supersolar oxygen abundances.

Although the EDMD related to the thick + thin
disk (Fig.\,\ref{f:KDND}, bottom panels, crosses)
is systematically lower for low oxygen abundances
and higher for intermediate values (ending at
$\phi\approx1.56$) with respect to the EDMD
deduced from the RA07 sample both in presence
(triangles) and in absence (diamonds) of the
LTE approximation, still a similar trend is
exhibited.

A closer agreement could occur in the
low-metallicity range after removing
halo contamination in the RA07 sample
or, alternatively, what in the current
paper has been defined as the ``halo-like
thick disk'' could be due to halo
contamination in the CB00 sample.
In any case, the RA07 sample cannot
be considered as representative of
the metal-weak ([Fe/H]$\appleq-1$)
thick disk, as mentioned above.
The discrepancy in the intermediate-metallicity
range is reduced in absence of the
LTE approximation, and further
(slightly) reduced in presence
of [Fe/H] plateau (left panels),
with the exception of the more
metal-rich ($\phi\approx1.56$)
bin in the disk EDMD, which is
related to the more metal-rich
bin in the RM96 sample.   Accordingly,
the RM96 sample could be biased
towards high metallicities, and
the above mentioned bin have little
statistical meaning (e.g., Haywood,
2001).   In conclusion,
the disk EDMD plotted in Fig.\,\ref
{f:KDND} (bottom panels) may be
considered as acceptable to a first
extent.

\begin{figure*}[t]
\begin{center}
\includegraphics[scale=0.8]{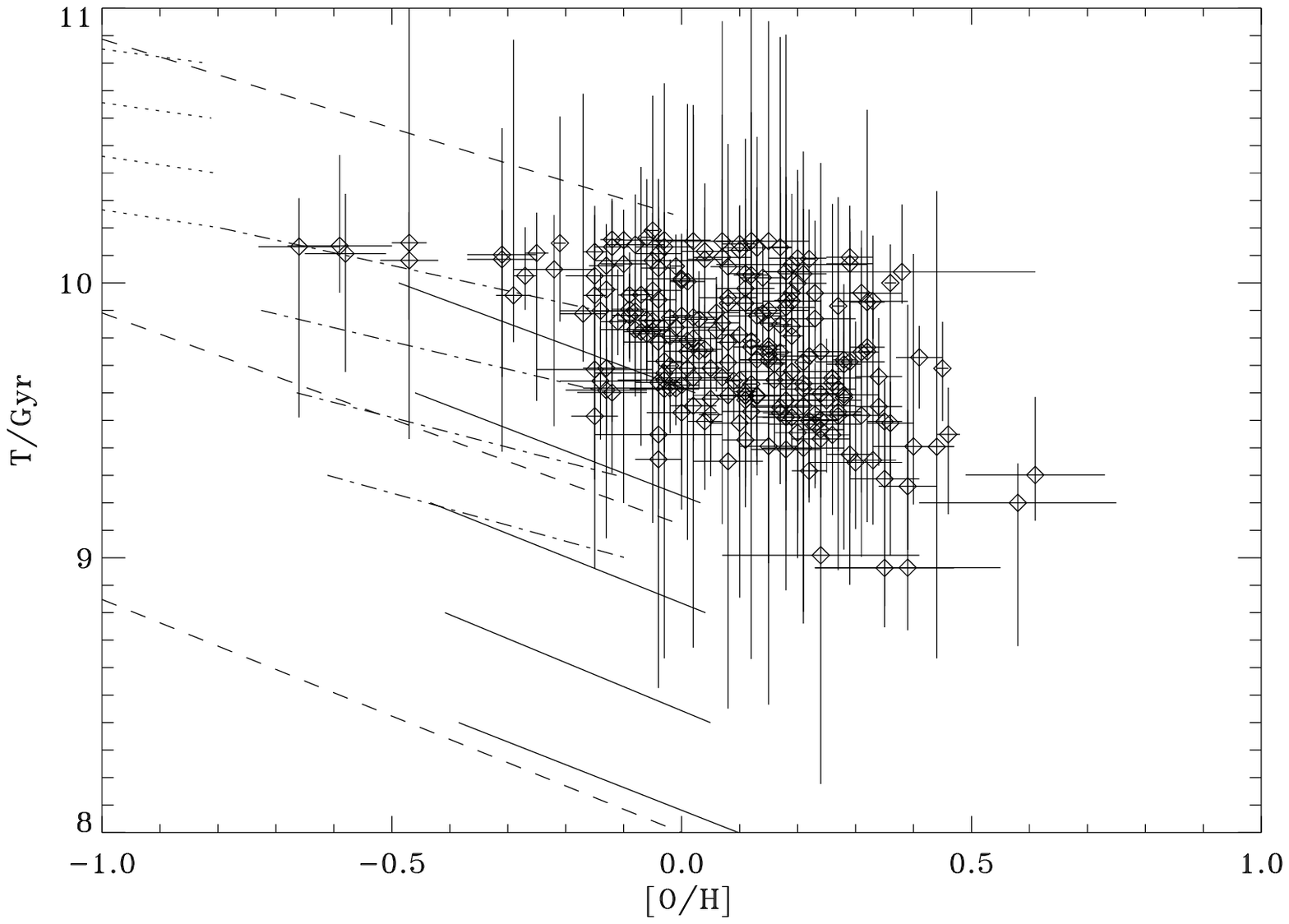}
%\centerline{\psfig{file=EGDK12l.ps,height=130mm,width=140mm}}
\caption[mamr]{Comparison between the empirical (EAMR)
and theoretical age-metallicity relation
(TAMR) in presence of the LTE approximation.   The data are
from a subsample $(N=223)$ of the RA07 sample, for which
ages have been determined (Ramirez et al., 2007).
Dashed, dotted, dash-dotted, and full lines are related
to models
H1 (C07), HK1, BK1, and DN1, respectively, in the special
case of constant star formation rate within active regions.
The halo star formation begins at
([O/H], $T$/Gyr)$=(-$3, 12.5) and ends at (0, 8.0),
within four time steps, $\Delta T$/Gyr$=$1.125 (C07).
The starting point of each step is out of
scale on the left; the first step is out of scale on
the top; and cannot be shown.
The halo-like thick disk star formation begins at ($-$1.52, 11.0) and
ends at ($-$0.80, 10.2), 
within four time steps, $\Delta T$/Gyr$=$0.2.
The starting point of each step is out of
scale on the left, and cannot be shown.
%The last three steps are out of scale on the right,
%and cannot be shown.
The bulge-like thick disk star formation begins at ($-$0.80, 10.2) and
ends at ($-$0.10, 9.0), 
within four time steps, $\Delta T$/Gyr$=$0.3.
The thin disk star formation begins at ($-$0.49, 10.0) and
is still continuing (in the sense that star formation occurs) up
today at (0.20, 0.0) within twenty-five time steps, $\Delta T$/
Gyr$=$0.4.   The last twenty time steps are out scale on the
bottom.}
\label{f:LTE}
\end{center}
\end{figure*}

\begin{figure*}[t]
\begin{center}
\includegraphics[scale=0.8]{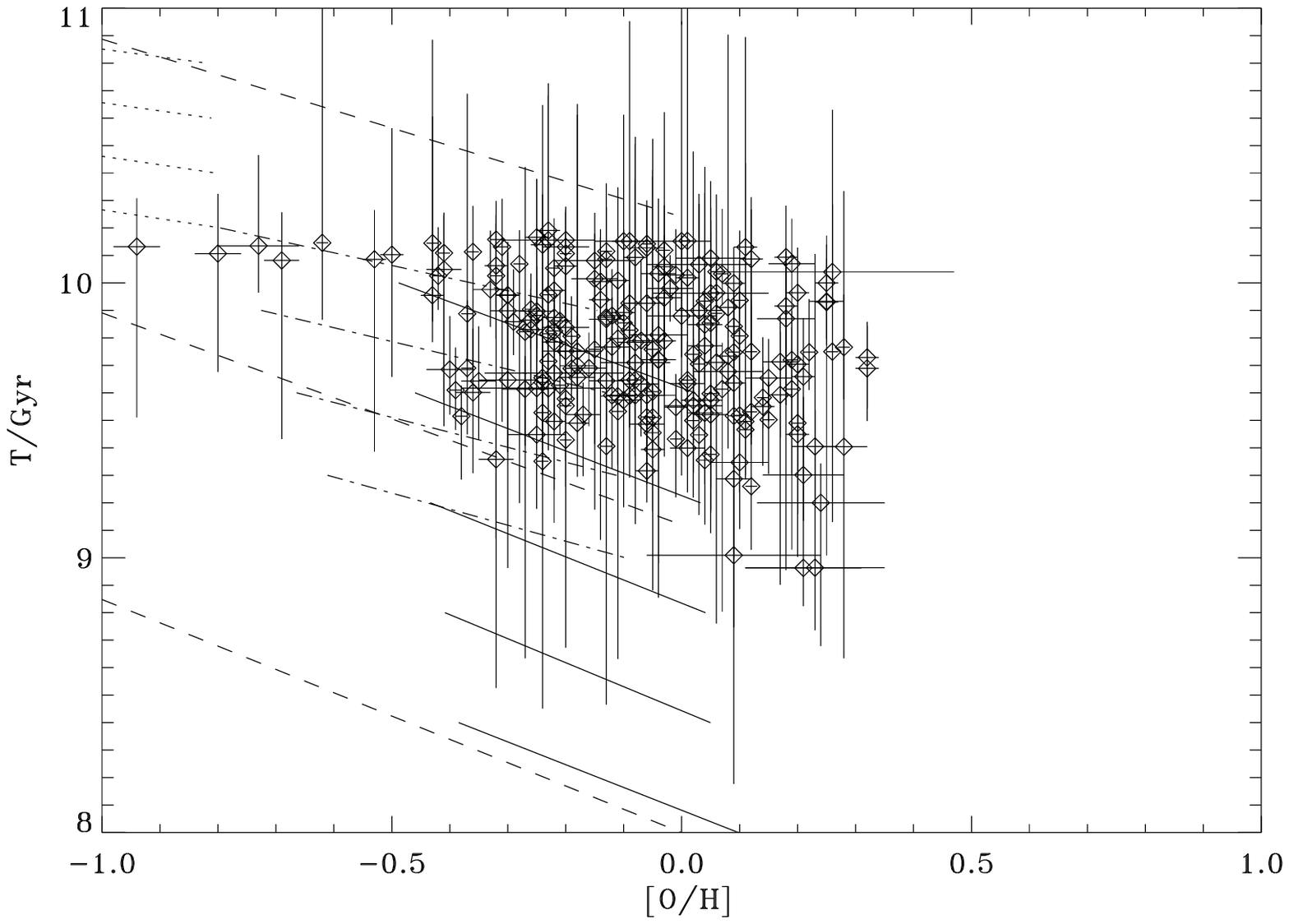}
%\centerline{\psfig{file=EGDK12N.ps,height=130mm,width=140mm}}
\caption[mamr]{Comparison between the empirical (EAMR)
and theoretical age-metallicity relation
(TAMR) in absence of the LTE approximation.   The data are
from a subsample $(N=223)$ of the RA07 sample, for which
ages have been determined (Ramirez et al., 2007).
Dashed, dotted, dash-dotted, and full lines are related
to models
H1 (C07), HK1, BK1, and DN1, respectively, in the special
case of constant star formation rate within active regions.
The halo star formation begins at
([O/H], $T$/Gyr)$=(-$3, 12.5) and ends at (0, 8.0),
within four time steps, $\Delta T$/Gyr$=$1.125 (C07).
The starting point of each step is out of
scale on the left; the first step is out of scale on
the top; and cannot be shown.
The halo-like thick disk star formation begins at ($-$1.52, 11.0) and
ends at ($-$0.80, 10.2), 
within four time steps, $\Delta T$/Gyr$=$0.2.
The starting point of each step is out of
scale on the left, and cannot be shown.
%The last three steps are out of scale on the right,
%and cannot be shown.
The bulge-like thick disk star formation begins at ($-$0.80, 10.2) and
ends at ($-$0.10, 9.0), 
within four time steps, $\Delta T$/Gyr$=$0.3.
The thin disk star formation begins at ($-$0.49, 10.0) and
is still continuing (in the sense that star formation occurs) up
today at (0.20, 0.0) within twenty-five time steps, $\Delta T$/
Gyr$=$0.4.   The last twenty time steps are out scale on the
bottom.}
\label{f:NLTE}
\end{center}
\end{figure*}

The empirical age-metallicity relation
(EAMR), taken from a subsample $(N=223)$
of the RA07 sample, is plotted in
Figs.\,\ref{f:LTE} and \ref{f:NLTE}
in presence and in absence of the
LTE approximation, respectively.
Also represented therein is the temporal
behaviour of (allowing star formation) gas
oxygen abundance, normalized to the solar
value, related to models H1 (C07, Table 9;
dashed lines), HK1 (Table \ref{t:inho}; dotted
lines), BK1 (dash-dotted lines), and DN1
(full lines).   The curves correspond to
the special case of constant star formation
rate within active regions but, in any case,
the extreme points related to the beginning
and the end of a step, are left unchanged.
Then the curves may be considered as representative
of a time-dependent star formation rate.

It is worth recalling that simple models
of chemical evolution used in the current
paper are time-independent, with regard to
the TDMD.   Accordingly, the initial and the
final time, together with the time step in
the case of inhomogeneous models, can be
selected for best fitting the EAMR.

For the halo, according to C07, initial and final
values, ([O/H], $T$/Gyr)$=(-$3, 12.5) and
(0, 8.0), respectively, have been chosen,
together with a time step, $\Delta T$/Gyr$=$
1.125, which implies four steps during the
evolution.   The initial and final values of
fractional oxygen abundance, $(\phi_\ell,
\Delta^*\phi_{\rm R}^\prime)$, at any step, are
$(0.001, 0.9502)$, $(0.0176, 0.9668)$,
$(0.0342, 0.9834)$, $(0.0508, 1)$, corresponding
to constant increments, $\Delta^*\phi=
\phi_\ell-\phi_{\ell-1}=0.01658$ and
$\Delta^*\phi_{\rm R}^\prime=0.9492$.
The related TAMR
is represented by four (the first out of scale)
dashed lines in
Figs.\,\ref{f:LTE} and \ref{f:NLTE}.
Owing to the logarithmic scale on the
horisontal axis, $\log\phi=[{\rm O}/{\rm H}]$,
the ending point of each step is very close to
$[{\rm O}/{\rm H}]=0$, but only the last step
actually ends therein.   The starting point
of each step is out of scale on the left.

For the halo-like thick disk, initial and final
values, ([O/H], $T$/Gyr)$=(-$1.52, 11.0) and
($-$0.80, 10.2), respectively, have been chosen,
together with a time step, $\Delta T$/Gyr$=$
0.2, which implies four steps during the
evolution.   The initial and final values of
fractional oxygen abundance, $(\phi_\ell,
\Delta^*\phi_{\rm R}^\prime)$, at any step, are
$(0.03, 0.1515)$, $(0.03285, 0.1543)$,
$(0.03569, 0.1572)$, $(0.03854, 0.16)$, corresponding
to constant increments, $\Delta^*\phi=
\phi_\ell-\phi_{\ell-1}=0.002846$ and
$\Delta^*\phi_{\rm R}^\prime=0.1215$.
The related TAMR
is represented by four dotted lines in
Figs.\,\ref{f:LTE} and \ref{f:NLTE}.
The starting point
of each step is out of scale on the left.

For the bulge-like thick disk, initial and final
values, ([O/H], $T$/Gyr)$=(-$0.80, 10.2) and
($-$0.61, 9.0), respectively, have been chosen,
together with a time step, $\Delta T$/Gyr$=$
0.3, which implies four steps during the
evolution.   The initial and final values of
fractional oxygen abundance, $(\phi_\ell,
\Delta^*\phi_{\rm R}^\prime)$, at any step, are
$(0.16, 0.7150)$, $(0.1883, 0.7433)$,
$(0.2167, 0.7717)$, $(0.2450, 0.8)$, corresponding
to constant increments, $\Delta^*\phi=
\phi_\ell-\phi_{\ell-1}=0.002834$ and
$\Delta^*\phi_{\rm R}^\prime=0.5550$.
The related TAMR
is represented by four dot-dashed lines in
Figs.\,\ref{f:LTE} and \ref{f:NLTE}.
The starting point of the first step coincides
with the ending point of the last step related
to the halo-like thick disk.

For the thin disk, initial and final
values, ([O/H], $T$/Gyr)$=(-$0.49, 10.0) and
(0.20, 0.0), respectively, have been chosen,
together with a time step, $\Delta T$/Gyr$=$
0.4, which implies twenty-five steps during the
evolution.   The initial and final values of
fractional oxygen abundance, $(\phi_\ell,
\Delta^*\phi_{\rm R}^\prime)$, at the first
five steps, are
$(0.325, 1.0550)$, $(0.3469, 1.0769)$,
$(0.3687, 1.0988)$, $(0.3906, 1.1207)$,
$(0.4125, 1.1425)$, corresponding
to constant increments, $\Delta^*\phi=
\phi_\ell-\phi_{\ell-1}=0.02187$ and
$\Delta^*\phi_{\rm R}^\prime=0.7300$.
The related TAMR
is represented by five full lines in
Figs.\,\ref{f:LTE} and \ref{f:NLTE}.
The remaining twenty steps are 
out of scale on the bottom.

It is apparent that the TAMR related to the
disk cannot provide an acceptable fit to
the EAMR deduced from the RA07 sample both
in presence and in absence of the LTE
approximation.   This is why models of
chemical evolution used in the current
paper have been selected to fit the FS07
sample for the thick disk and the RM96
sample for the thin disk, where a small
amount of oxygen-rich (supersolar)
objects were found using Eqs.\,(\ref{eq:gra})
and (\ref{eq:isa}).   On the contrary, a
consistent number of oxygen-rich objects
is present in the RA07 sample which, on
the other hand, is biased towards
oxygen-poor objects (Ramirez et al.,
2007).

In the future, additional effort should
be devoted in collecting representative
samples of both the halo, the bulge, the
halo-like and bulge-like
thick disk, and the thin disk, where
the oxygen abundance is directly
derived (e.g., Ramirez et al., 2007)
instead of being empirically deduced
from iron abundance, as in Eqs.\,(\ref
{eq:gra}) and (\ref{eq:isa}).   Additional
support to the
distinction between halo-like and
bulge-like thick disk could be provided
from the related distributions of specific
angular momentum.   In particular, the
former and the latter are expected to
exhibit differences as in the Galactic
spheroid with
respect to the Galactic disk (Wyse and
Gilmore, 1992; Ibata and Gilmore, 1995).
It would also be
important to establish if the metal-weak
thick disk (e.g., Martin and Morrison, 1998;
Beers et al., 2002) is the actual precursor
of the metal-enriched thick disk or, on
the other hand, coincides with the flattened,
rotation-supported halo subsystem (Prochaska
et al., 2000), or both.
Further investigation will provide
additional support, or contradict,
the results of the current attempt.

\section{Conclusion} \label{conc}

The empirical differential metallicity
distribution (EDMD) in the thick disk
has been deduced from two different samples
involving (WG95) local thick disk stars derived
from Gliese and scaled in situ samples
within the range, $-1.20\le$[Fe/H]$\le-0.20$
(Wyse and Gilmore, 1995), and (CB00) $N=46$
likely metal-weak thick disk stars within the
range, $-2.20\le$[Fe/H]$\le-1.00$ (Chiba and
Beers, 2000).   The EDMD in the thin disk has
been deduced from (RM96) $N=287$ chemically
selected G dwarfs within 25 kpc from the sun,
with the corrections performed in order to
take into account the stellar scale height
(Rocha-Pinto and Maciel, 1996).   To this
aim, two alternative [O/H]-[Fe/H] dependences
have been used, Eqs.\,(\ref{eq:gra}) and
(\ref{eq:isa}), according to earlier attempts
where the EDMD was deduced in halo field
stars (C01), the globular cluster subsystem
(C07), and the bulge (C07).

The metal-poor and metal-rich EDMD related
to the thick disk, have shown analogies
with their halo and bulge counterparts,
respectively.   For this reason, the thick
disk has been conceived as made of two
distinct regions: the halo-like and
bulge-like thick disk, and the related
EDMD has been deduced.   The data have
been fitted, to an acceptable extent, by
both homogeneous and inhomogeneous simple
models of chemical evolution.   For the
thin disk, only inhomogeneous models have
suceeded in reproducing the data, as already
shown in earlier attempts (Malinie at al.,
1993; C00).

Under the assumption of an universal initial
mass function (IMF) and a value of the true
yield equal to the linear fit related to the thin
disk solar neighbourhood, inhibition of
thick disk star formation (implying gas
outflow) has been required to reproduce
the EDMD.   A power-law IMF has been considered,
$\phi(\widetilde{m})\propto\widetilde{m}^
{-p}$, within the range, $2.35\le p\le2.9$,
and special effort has been given to the
limiting cases, (i) $p=2.9$,
which is acceptably close to Scalo (1986) IMF for
$m\appgeq {\rm m}_\odot$, and (ii)
$p=2.35$, which is the Salpeter (1955) IMF.

The EDMD related to the disk has been determined
by weighting the mass, where each distribution
has been assumed to be typical for the corresponding
subsystem.   Following a similar procedure, the
EDMD related to the Galaxy has been computed
using the results of an earlier attempt (C07)
with regard to the bulge and the halo.   In any
case, it has been inferred that a more refined
model involving an initially increasing
star formation efficiency (but not necessarily
implying gas infall) while assembling Galactic
subsystems, could provide a better agreement
with the data, including the right amount of
mass budget needed for the formation of the
bulge-like thick disk and the thin disk.
The theoretical differential
metallicity distribution (TDMD) related to
both homogeneous and inhomogeneous simple
models has provided an acceptable fit to the
EDMD related to the Galaxy and, in particular,
a non-monotonic trend has been reproduced.

An indicative comparison has been performed
between the EDMD deduced for the thick + thin
disk both in presence and in absence of
[O/Fe] plateau, and its counterpart computed
for (RA07) $N=523$ nearby ($d<$150 pc) stars
with metallicity in the range, $-1.5<$[Fe/H]
$<0.5$, for which the oxygen abundance has
been determined both in presence and in absence
of the LTE approximation (Ramirez et al., 2007).
It has been found that both distributions exhibit
a similar trend, though systematic differences
exist, which has made the fit acceptable to a
first extent.

The empirical age-metallicity distribution
(EAMR), taken from a subsample $(N=223)$
of the RA07 sample for
which age determination is available (Ramirez
et al., 2007), has been compared with the
theoretical age-metallicity relation (TAMR),
predicted by models related to the halo,
the halo-like thick disk, the bulge-like thick disk, and
the thin disk.   The fit has shown to be
unsatisfactory for the following reason.
The models of chemical evolution used in
the current paper have been selected to
match the data from earlier samples for
both the thick disk (WG95 and CB00) and
the thin disk (RM96), where a small amount
of oxygen-rich (supersolar) objects were
found using 
Eqs.\,(\ref{eq:gra}) and (\ref{eq:isa}).
Conversely, a consistent number of
oxygen-rich objects has
been found in the RA07 sample which, on
the other hand, is biased towards
oxygen-poor objects (Ramirez et al., 2007).

%\section*{Acknowledgements}
%Thanks are due to F. De Angeli for making available
%values of absolute ages, errors, and additional
%data, related to the quoted reference De Angeli
%et al. (2005), and for stimulating e-mail
%correspondence.  In addition, J. Fulbright
%and M. Rich are also thanked for making available a preprint
%of the quoted reference Fulbright, McWilliam
%and Rich (2005) and, together with D. Terndrup,
%for fruitful e-mail correspondence.  Finally,
%thanks are due to J. Silk for useful explanations
%on the quoted reference Wang and Silk (1993).

%\appendix
%\section*{Appendix}

%\section{Mass and number element abundance}
%\label{a:manua}

%Let A be a generic element heavier than He.

%\section{Correspondence between homogeneous and
%inhomogeneous simple models}\label{a:corhi}

%For a generic system, let $\mu_o$,

%\section{The inhomogeneous simple model}
%\label{a:insmo}

%For an interested reader, it could be useful

%\section{Caption of symbols}\label{a:simbo}

%To help the reader, the following list includes


\begin{thebibliography}{}
%\bibitem{} autore-i, n. anno, riv. num., pag.
%\bibitem[Abia \& Rebolo]{abr89} Abia, C., Rebolo, R.:
% 1989, \apj 347, 186
\bibitem{adl96} Adams, F.C., Fatuzzo, M., 1996. ApJ 464, 256.
%\bibitem[Adams \& Laughlin]{adl96} Adams, F.C., Laughlin, G.:
% 1996, \apj~468, 586
\bibitem{} Allende-Prieto, C., Lambert, D.L., Asplund M., 2001. ApJ 556, L63.
\bibitem{} Anders, E., Grevesse, N.,  1989. Geochim. Cosmochim. Acta 53, 197.
\bibitem{} Asplund, M., Grevesse, N., Sauval, A.J., et al., 2004. A\&A 417,
           751.
\bibitem{} Asplund, M., Grevesse, N., Sauval, A.J., 2005. In Cosmic
           Abundances as Records of Stellar Evolution and Nucleosynthesis
           ASP, ed. F.N. Bash and T.G. Barnes, 25.
%\bibitem[Armandroff]{arm89} Armandroff, T.E.: 1989,
% \aj~97, 375
\bibitem{} Ballero, S.K., Kroupa, P., Matteucci, F., 2007a. A\&A 467, 117.
\bibitem{} Ballero, S.K., Matteucci, F., Origlia, L., Rich, R.M., 2007b.
           A\&A 467, 123.
%\bibitem{} Barbuy, B., 1988.  A\&A 191, 121.
\bibitem{} Barbuy, B., Nissen, P.E.,
Peterson, R., Spite, F. (eds.), Proceedings of Oxygen
abundances in stars and implications to nucleosynthesis and cosmology
(IAU Joint Discussion 8), 2001. New Astron. Rev. 45, 509.
%\bibitem[Beers \& Sommer-Larsen]{bes95} Beers, T.C.,
% Sommer-Larsen, J.: 1995, \apjs~96, 175
\bibitem{} Beers, T.C., Drilling, J.S., Rossi, S., et al., 2002. AJ 124, 931.
\bibitem{} Bensby, T., Feltzing, S., Lundstr\"om, I., 2004. A\&A 415, 155.
%\bibitem[Benvenuto \& Althaus]{bea99} Benvenuto, O.G.,
% Althaus, L.G.: 1999, \mnras 303, 30
%\bibitem[Bertelli et al.,]{bea94} Bertelli, P., Bressan, A.,
% Chiosi, C., Fagotto, F., Nasi, E.: 1994, \a&aSupp. Ser. 106, 275
\bibitem{} Binney, J., 1999. MNRAS 307, L27.
%\bibitem{} Binney, J., Gerhard, O., Silk, J., 2001. MNRAS 321, 471.
%\bibitem[Binney \& Tremaine]{bit87} Binney, J.,
% Tremaine, S.: 1987, Galactic Dynamics (Princeton
% Univ. Press)
%\bibitem[Boesgaard et al.,]{boa99} Boesgaard, A.M., King, J.R.,
%Deliyannis, C.P., Vogt, S.S.: 1999, \aj 117, 492
%\bibitem[Bond]{bon81} Bond, H.E.: 1981, \apj~248, 606
%\bibitem{} Baugh, C.M., Lacey, C.G., Frenk, C.S., et al., 2005.
%           MNRAS 356, 1191.
\bibitem{} Bromm, V., Loeb, A., 2003. Nat 425, 812.
\bibitem{} Bromm, V., 2004. PASP 116, 103.
\bibitem{} Bromm, V., Larson, R.B., 2004. ARA\&A 42, 79.
\bibitem{} Burrows, A., Hubbard, W.B., Saumon, D., Lunine, J.I., 1993.
           ApJ 406, 158.
%\bibitem{} Caimmi, R., 1978a. ApSS 54, 453.
\bibitem{} Caimmi, R., 1978a. ApSS 59, 109.
\bibitem{} Caimmi, R., 1978b. ApSS 59, 413.
\bibitem{} Caimmi, R., 1981. ApSS 79, 87.
\bibitem{} Caimmi, R., 1982. ApSS 84, 373.
%\bibitem[Caimmi]{cai95} Caimmi, R.: 1995, \a&a 295, 335
%\bibitem[Caimmi]{cai96} Caimmi, R.: 1996, \a&a 312, 797
%\bibitem[Caimmi]{cai97} Caimmi, R.: 1997, Astron.~Nachr.~318, 339
%\bibitem[Caimmi]{cai98} Caimmi, R.: 1998, Astron.~Nachr.~319, 285
\bibitem{} Caimmi, R., 2000. AN 321, 323 (C00). 
\bibitem{} Caimmi, R., 2001a. AN 322, 65 (C00, erratum).
\bibitem{} Caimmi, R., 2001b. AN 322, 241 (C01).
\bibitem{} Caimmi, R., 2007. NewA 12, 289 (C07).
%\bibitem[Carigi]{car96} Carigi, L.: 1996,
% Rev.~Mex.~Astron.~Astrophys.~32, 179
%\bibitem[Carlberg et al.,]{caa85} Carlberg, R.G.,
%\bibitem{} Carney, B.W., Latham, D.W., Laird, J.B., 1990. AJ 99, 572.
%\bibitem[Carraro et al.,]{caa98} Carraro, G., Ng, Y.K.,
% Portinari, L.: 1998, \mnras 296, 1045
%\bibitem[Carraro et al.,]{caa99} Carraro, G., Girardi, L.,
% Chiosi, C.: 1999, \mnras 309, 430
%\bibitem{} Carretta, E., Gratton, R., 1997. A\&AS 121, 95. (CG)
\bibitem{} Carretta, E., Gratton, R.,  Sneden, C., 2000. A\&A 356, 238.
%\bibitem{} Carretta, E., Cohen, J.G., Gratton, R.,
%Behr, B.B., 2001. AJ 122, 1469.
%\bibitem[Cassisi et al.,]{cas99} Cassisi, S., Castellani, V.,
% Degl'\,Innocenti, S., Salaris, M., Weiss, A: 1999,
% \a&aSupp. Ser. 134, 103
%\bibitem[Cavallo et al.,]{caa97} Cavallo, R.M., Pilachowski, C.A.,
% Rebolo, R.: 1997, \pasp 109, 226
%\bibitem[Chamcham \& Hendry]{chh96} Chamcham, K.,
% Hendry, M.A.: 1996, \mnras 279, 1083
%\bibitem[Chiappini et al.,]{cha01} Chiappini, C., Matteucci, F.,
%Romano, D.: 2001, \apj 554, 1044
\bibitem{} Chiba, M., Beers, T.C., 2000. AJ 119, 2843.
%\bibitem[Clayton]{cla88} Clayton, D.D.: 1988, \mnras 234, 1
%\bibitem[Chaboyer]{cha95} Chaboyer, B.: 1995, \apj 444, L9
\bibitem{} Christlieb, N., Bessell, M.S., Beers, T.C., et al.,
2002. Nat 419, 904.
%\bibitem[Da Costa \& Armandroff]{dca95} Da Costa, G.S.,
% Armandroff, T.E.: 1995, \aj 109, 2533
%\bibitem[Davies et al.,]{daa98a} Davies, R.I., Sugai,
% H., Ward, M.J.: 1998a, \mnras 295, 43
%\bibitem[Davies et al.,]{daa98b} Davies, R.I., Sugai,
% H., Ward, M.J.: 1998b, \mnras 300, 388
% Dawson, P.C., Hsu, T., van den Bergh, D.A.:
% 1985, \apj 294, 674
%\bibitem{} De Angeli, F., Piotto, G., Cassisi, S., et al.,
%2005. AJ 130, 116.
%\bibitem{} De Angeli, F., 2005. Private communication.
%\bibitem{} Doane, J.S., Mathews, W.G., 1993. ApJ 419, 573.
%\bibitem{} Edvardsson, B., Andersen,
%J., Gustaffson, B.,
%Lambert, D.L., Nissen, P.E., Tomkin, J.
%et al.,, 1993. A\&A 275, 101.
%\bibitem[Eggen et al.,]{ega62} Eggen, O.J., Lynden-Bell,
% D., Sandage, A.R.: 1962, \apj 136, 748
\bibitem{} Favata, F., Micela, G., Sciortino, S., 1997. A\&A 323, 809.
%\bibitem{} Feltzing, S., Holmberg, J, Hurley, J.R., 2001. A\&A 377, 911.
\bibitem{} Ferreras, I., Wyse, R.F.G., Silk, J., 2003. MNRAS 345, 1381. 
%\bibitem[Fields et al.,]{fia97} Fields, B.D., Mathews, G.J.,
% Schramm, D.N.: 1997, \apj 483, 625
\bibitem{} Frebel, A., Christlieb, N., Norris, J.E., et al.,
2006. ApJ 638, L17.
\bibitem{} Fulbright J.P., Rich R.M., McWilliam A., 2005.
NPA 758, 197. Available from $<$astro-ph/0411041$>$.
%\bibitem{} Fulbright J.P., McWilliam A., Rich R.M., 2006.
%ApJ 636, 821.
\bibitem{} Garcia Perez, A.E., Asplund, M., Primas, F., et al., 2006. A\&A
           451, 621.
%\bibitem[Gibson \& Mould]{gim97} Gibson, B.K., Mould, J.R.:
% 1997, \apj 482, 98
%\bibitem[Gilmore \& Wyse]{giw86} Gilmore, G., Wyse, R.F.G.:
% 1986, Nat.~322, 806
%\bibitem[Gilmore et al.,]{gia89} Gilmore, G., Wyse, R.F.G.,
% Kuijken, K: 1989, \araa 27, 555
\bibitem{} Gilmore, G., Wyse, R.F.G., Jones, J.B., 1995. AJ 109, 1095.
%\bibitem[Gratton et al.,]{gra00} Gratton, R., Carretta, E.,
% Matteucci, F, Sneden, C: 2000, \a&a 358, 671
\bibitem{} Gray, D.F., 2005. The Observation and Analysis of Stellar
           Photospheres, Cambridge University Press, UK.
\bibitem{} Grevesse, N., Sauval, A.J., Dishoeck, E.F., 1984. A\&A 141, 10.
%\bibitem[Hansen]{han99} Hansen, B.M.S.: 1999, \apj 520, 680
%\bibitem{} Harris, W.E., 1996. AJ 112, 1487.
\bibitem{} Hartwick, F.D.A., 1976. ApJ 209, 418.
\bibitem{} Haywood, M., 2001. MNRAS 325, 1365.
\bibitem{} Haywood, M., 2006. MNRAS 371, 1760.
\bibitem{} Henry, R.B.C., Worthey, G., 1999. PASP 111, 919.
%\bibitem[Holweger]{hol01} Holweger, H., 2001, in ``Solar and
%Galactic Composition'', ed. R.F. Wimmer - Schweingruber
%(Berlin: Springer), in press
%\bibitem{} Huchra, J.P., Brodie, J.P., Kent, S.M., 1991.
%ApJ 370, 495.
\bibitem{} Ibata, R.A., Gilmore, G.F., 1995. MNRAS 275, 605.
%\bibitem{} Ibukiyama, A., Arimoto, N., 2002. A\&A 394, 927.
%\bibitem[Israelian et al.,]{isa98} Israelian, G., Garcia-Lopez,
% R.J., Rebolo, R.: 1998, \apj 507, 805
\bibitem{} Israelian, G., Rebolo, R., Garcia-Lopez, R.J., et al.,
%Bonifacio, P., Molaro, P., Basri, G., Shchukina, N.
2001a. ApJ 551, 833.
\bibitem{} Israelian, G., Rebolo, R., Garcia-Lopez, R.J., et al.,
2001b. ApJ 560, 535.
\bibitem{} Iwamoto, N., Umeda, H., Tominaga, N., et al., 2005. Sci 309, 451.
\bibitem{} Jonsell, K., Edvardsson, B., Gustafsson, B., et al., 2005. A\&A
           440, 321.
\bibitem{} J\o\phantom{}rgensen, B.R., 2000. A\&A 363, 947.
%\bibitem{} Karatas, Y., Bilir, S., Schuster, W.J., 2005. MNRAS 360, 1345.
\bibitem{} Karlsson, T., 2005. A\&A 439, 93.
%\bibitem{} Kent, S.M., 1992. ApJ 387, 181.
%\bibitem[King \& Boesgaard]{kib95} King, J.R., Boesgaard, A.M.:
%1995, \aj 109, 383
%\bibitem[Kinman et al.,]{kia94} Kinman, T.D., Suntzeff, N.B.,
\bibitem{} Kotoneva, E., Flynn, C., Chiappini, C., Matteucci, F., 2002.
           MNRAS 336, 879.
%\bibitem[Kraft]{kra79} Kraft, R.P.: 1979, \araa 17, 309
% Kraft, R.P.: 1994, \aj 108, 1722
%\bibitem[Kraft et al.,]{kra92} Kraft, R.P., Sneden, C., Langer, G.E.,
%Prosser, C.F.: 1992, \aj 104, 645
%\bibitem[Knox et al.,]{kna99} Knox, R.A., Hawkins, M.R.S.,
% Hambly, N.C: 19979, \mnras 306, 736
%\bibitem[J{\o}rgensen]{jor00} J{\o}rgensen B.R.%
%\footnote{Bjarne Rosenkilde}%
%: 2000, \a&a 363, 947
\bibitem{} Landi, E., Feldman, U., Doschek, G.A., 2007. ApJ 659, 743.
\bibitem{} Larson, R.B., 1974. MNRAS 166, 585.
%\bibitem{} Larson, R.B., 1998. MNRAS 301, 569.
\bibitem{} Larson, R.B., 2005. MNRAS 359, 211.
%\bibitem[Layden]{lay95} Layden, A.C.: 1995, \aj 110, 2288
%\bibitem[Legget et al.,]{lea98} Legget, S.K., Ruiz, M.T.,
% Bergeron, P.: 1998, \apj 497, 294
%\bibitem{} Lee, Y.-W., 1990. ApJ 363, 159.
%\bibitem{} Lee, Y.-W., Demarque, P., Zinn, R., 1994. ApJ 423, 248.
%\bibitem{} Li, Y., Burstein, D., 2003. ApJ 598, L103.
%\bibitem{} Liu, W.M., Chaboyer, B., 2000. ApJ 544, 818.
%\bibitem[Low \& Lynden-Bell]{lol76} Low, C., Lynden-Bell,
% D.: 1976, \mnras 176, 367
%\bibitem{} Lu, Y., Zhao, G., Deng, L.C.,
%Cen, M.R., Lieng, Y.C., 2001. A\&A 367, 277.
\bibitem{} Lynden-Bell, D., 1975. Vistas Astron. 19, 299.
%\bibitem{} Mackey, A.D., Gilmore, G., 2004. MNRAS 355, 504. 
\bibitem{} Mackey, A.D., van den Bergh, S., 2005. MNRAS 360, 631. 
\bibitem{} Malinie, G., Hartmann, D.H.,
           Clayton, D.D., Mathews, G.J., 1993. ApJ 413, 633.
%\bibitem{} Marigo, P., Girardi, L., Chiosi, C., Wood, P.R.,
%2001. A\&A 371, 152.
\bibitem{} Martin, J.C., Morrison, H.L., 1998. AJ 116, 1724.
\bibitem{} Melendez, J., 2004. ApJ 615, 1042.
\bibitem{} Melendez, J., Shchukina, N.G., Vasiljeva, I.E., Ramirez, I., 2006.
           ApJ 642, 1082.
%\bibitem[Mera et al.,]{mea96} Mera, D., Chabrier, G.,
% Baraffe, I.: 1996, \apj 459, L87
%\bibitem[Mera et al.,]{mea98a} Mera, D., Chabrier, G.,
% Schaeffer, R.: 1998a, \a&a 330, 937
%\bibitem[Mera et al.,]{mea98b} Mera, D., Chabrier, G.,
% Schaeffer, R.: 1998b, \a&a 330, 953
\bibitem{} Meusinger, H., Stecklum, B., Reimann, H.-G.,
1991. A\&A 245, 57.
%\bibitem[Meusinger \& Stecklum]{mes92} Meusinger, H.,
% Stecklum, B.: 1992, \a&a 256, 415
%\bibitem[Meusinger]{meu94} Meusinger, H.: 1994, Astron.~Nachr.~315, 279
\bibitem{} Miller, G.E., Scalo, J.M., 1979. ApJS 41, 513.
%\bibitem[Mishenina et al.,]{mia00} Mishenina, T., Korotin, S.,
%Klochkova, V., Panchuk, V.: 2000, \a&a 353, 978
%\bibitem[Montgomery et al.,]{moa99} Montgomery, M.H.,
% Klumpe, E.W., Winget, D.E., Wood, M.A.:
% 1999, \apj 525, 482
%\bibitem[Morrison \& Harding]{moh93} Morrison, H.L.,
%Harding, P.: 1993, \pasp 105, 977
%\bibitem{} Nagashima, M., Lacey, C.G., Baugh, C.M., et al., 2005.
%           MNRAS 358, 1247.
%\bibitem[Nakamura \& Umemura]{nau99} Nakamura, F.,
% Umemura, M.: 1999, \apj 515, 239
%\bibitem[Ng \& Bertelli]{ngb98} Ng, Y.K., Bertelli, G.:
% 1998, \a&a 329, 943
%\bibitem[Nissen \& Zomkin]{niz93} Nissen, P.E., Zomkin, I.:
% 1993, \a&a 275, 101
\bibitem{} Nordstr\"om, B., Mayor, M., Andersen, J., et al.,
2004. A\&A 418, 989.
%\bibitem[Norris]{nor94} Norris, J.E.: 1994, \apj 431, 645
\bibitem{} Oey, M.S., 2003. MNRAS 339, 849.
%\bibitem[Oloffson]{olo95} Oloffson, K.: 1995, \a&a 293, 652
%\bibitem[Ostriker \& Thuan]{ost75} Ostriker, J.B.,
%Thuan, T.X.: 1975, \apj 202, 353
%\bibitem{} Pagel, B.E.J., 1987. In: Gilmore, G.,
%Carswell, R.F. (eds.) The Galaxy, Reidel, p.\,341.
\bibitem{} Pagel, B.E.J., 1989. The G-dwarf Problem and Radio-active
           Cosmochronology. In: Beckman J.E., Pagel B.E.J. (eds.)
           Evolutionary Phenomena in Galaxies, Cambridge Univ. Press, p.\,201.
%\bibitem[Pagel]{pag97} Pagel, B.E.J.: 1997, Nucleosynthesis
%and Chemical Evolution of Galaxies, Cambridge Univ.
%Press, Sect.\,8.3.2
\bibitem{} Pagel, B.E.J., Patchett, B.E., 1975. MNRAS 172, 13.
%\bibitem{} Perrett, K.M., Bridges, T.J., Hanes, D.A., et al.,,
%2002. AJ 123 2490.
%\bibitem{} Pilyugin, L.S., 1993. A\&A 277, 42.
%\bibitem{} Pilyugin, L.S., 1996. A\&A 313, 803.
\bibitem{} Pilyugin, L.S., Edmunds, M.G., 1996. A\&A 313, 792.
%\bibitem{} Piotto, G., et al., 2002. A\&A 391, 945.
%\bibitem[Portinari et al.,]{poa98} Portinari, L., Chiosi, C.,
%Bressan, A.: 1998, \a&a 334, 505
\bibitem{} Prantzos, N., 1994. A\&A 284, 477.
\bibitem{} Prantzos, N., 2003. A\&A 404, 211.
\bibitem{} Prantzos, N., 2007. Arxiv: astro-ph/0611476.
%\bibitem[Prantzos \& Aubert]{pra95} Prantzos, N., Aubert, O.:
% 1995, \a&a 302, 69
\bibitem{} Prantzos, N., Silk, J., 1998. ApJ 507, 229.
\bibitem{} Prochaska, J.X., Naumov, S.O., Carney, B.W., et al.,
%McWilliam, A., Wolfe, A.M.
2000. AJ 120, 2513.
%\bibitem{} Pritzl, B.J., Venn, K.A., Irwin, M., 2005. AJ 130, 2140.
\bibitem{} Ramirez, I., Allende Prieto, C., Lambert, D.L., 2007. A\&A 465,
           271.
\bibitem{} Reddy, D.E., Tomkin, J., Lambert, D.L., Allende Prieto, C., 2003.
           MNRAS 340, 304.
\bibitem{} Reddy, D.E., Lambert, D.L., Allende Prieto, C., 2006. MNRAS 367,
           1329.
%\bibitem[Rieke et al.,]{ria93} Rieke, G.H., Loken, K., Rieke, M.J.,
% Tamblyn, P.: 1993, \apj 412, 99
\bibitem{} Rocha-Pinto, H.J., Maciel, W.J., 1996. MNRAS 279, 447.
%\bibitem{} Rocha-Pinto, H.J., Maciel, W.J., 1997. A\&A 325, 523.
%\bibitem[Rocha-Pinto \& Maciel]{rpm97b} Rocha-Pinto, H.J.,
% Maciel, W.J.: 1997b, \mnras 289, 882
\bibitem{} Rocha-Pinto, H.J., Scalo, J.M., Maciel, W.J., Flynn, C., 2000.
           ApJ 531, L115.
%\bibitem{} Rocha-Pinto, H.J., Maciel,
%W.J., Scalo, J.M., Flynn, C., 2000. A\&A 358, 850.
%\bibitem[Rocha-Pinto et al.,]{rpa00c} Rocha-Pinto, H.J., Scalo, J.M.,
% Maciel, W.J., Flynn, C.: 2000c, \a&a in press
%\bibitem[Rosemberg]{ros00} Rosemberg, A.: 2000, \pasp 112, 575
%\bibitem{} Rosemberg, A., Piotto, G., Saviane, I., Aparicio, A.,
%2000a. A\&A 144, 5.
%\bibitem{} Rosemberg, A., Aparicio, A., Saviane, I., Piotto, G.,
%2000b. A\&A 145, 451.
\bibitem{} Ryan, S.G., Norris, J.E., 1991. AJ 101, 1865.
\bibitem{} Sadler, E.M., Rich, R.M., Terndrup, D.M., 1996. AJ 112, 171.
%\bibitem{} Salaris, M., Weiss, A., 2002. A\&A 388, 492.
%\bibitem[Salaris et al.,]{saa97} Salaris, M.,
% Degl'\,Innocenti, S., Weiss, A.: 1997, \apj 479, 665
\bibitem{} Salpeter, E.E., 1955. ApJ 121, 161.
%\bibitem{} Sarajedini, A., Chabojer, B., Demarque, P., 1997.
%PASP 109, 1321.
\bibitem{} Sauval, A.J., Grevesse, N., Brault, N., et al., 1984.
           ApJ 282, 330.
\bibitem{} Scalo, J.M., 1986. FCPh 11, 1.
%\bibitem[Schmidt]{sch59} Schmidt, M.: 1959, \apj 129, 243
\bibitem{} Schmidt, M., 1963. ApJ 137, 758.
\bibitem{} Searle, L., 1972. Star Formation and the Chemical History of
           Galaxies.   In: Cayrel de Strobel, G., Delplace, A.M. (eds.)
           L'Ages des Etoiles, Observatoire de Paris-Meudon, p.\,52.
\bibitem{} Searle, L., Sargent, W.L.W., 1972. ApJ 173, 25.
%\bibitem[Searle \& Zinn]{sez78} Searle, L.,
% Zinn, R.: 1978, \apj 225, 357
%\bibitem[Silk]{sil77} Silk, J.: 1977, \apj 214, 152
\bibitem{} Shchukina, N., Trusillo Bueno, J., Asplund, M., 2005. ApJ 618, 939.
%\bibitem{} Smith, D.A., Herter, T., Haynes,
% M.P., Beichman, C.A., Gautier, T.N. III, 1995. ApJ 439, 623.
%\bibitem{} Smith, D.A., Herter, T., Haynes, M.P., 1998. ApJ 494, 150.
\bibitem{} Socas-Navarro, H., Norton, A.A., 2007. ApJ 660, L153.
\bibitem{} Sofia, U.J., Meyer, P.M., 2001. ApJ 554, L221.
%\bibitem[Sommer-Larsen]{som91} Sommer-Larsen, J.: 1991,
% \mnras 249, 368
%\bibitem{} Stetson, P.B., Vandenberg, D.A., Bolte, M., 1996.
%PASP 108, 560.
%\bibitem[Suntzeff et al.,]{sua91} Suntzeff, N.B.,
% Kinman, T.D., Kraft, R.P.: 1991, \apj 367, 528
%\bibitem[Takada-Hidai et al.,]{taa01}  Takada-Hidai, M., Takeda, Y.,
%Sato, S., Sargent, W.L., Lu, L., Barlow, T., Jugeku, J.: 2001, 
%New Astron. Rev. 45, 549
%\bibitem[Talbot \& Arnett]{taa75} Talbot, R.J.,
% Arnett, W.D.: 1975, \apj 197, 551
\bibitem{} Thacker, R.J., Scannapieco, E., Davis, M., 2002. ApJ 581, 836.
\bibitem{} Tinney, C.G., 1993. ApJ 414, 279.
%\bibitem[Tomkin et al.,]{toa92} Tomkin, J., Lemke, M.,
%Lambert, D.L., Sneden, C.: 1992, \aj 104, 1568
%\bibitem[Truran \& Cameron]{trc71} Truran, J.W.,
% Cameron, A.G.W.: 1971, \apss 14, 179
%\bibitem[Tsujimoto et al.,]{tsa97} Tsujimoto, T., Yoshii, Y.,
% Nomoto, K., Matteucci, F., Thielemann, F., Hashimoto, M.:
% 1997, \apj 483, 228
%\bibitem[Twarog]{twa80a} Twarog, B.A.: 1980a, \apj 242, 242
%\bibitem[Twarog]{twa80b} Twarog, B.A.: 1980b, \apjs 44, 1
%\bibitem[Uehara et al.,]{uea96} Uehara, H., Susa, H., Nishi, R.,
% Yamada, M., Nakamura, T.: 1996, \apj 473, L95
\bibitem{} van den Bergh, S., 1962. ApJ 67, 486.
%\bibitem[van den Bergh]{vdb93} van den Bergh, S.: 1993, \aj 105, 971
%\bibitem[van den Bergh et al.,]{vaa96} van den Bergh, D.A.,
% Bolte, M., Stetson, P.B.: 1996, \araa 34, 461
%\bibitem[van den Hoek]{vah97} van den Hoek, L.B.:
% 1997, On the chemical and spectro-photometric
% evolution of nearby galaxies, thesis, Univ. of
% Amsterdam, The Netherlands.
%\bibitem{} van den Hoek, L.B., de Jong, T., 1997. A\&A 318, 231.
\bibitem{} Wang, B., Silk, J., 1993. ApJ 406, 580.
\bibitem{} Weidner, C., Kroupa, P., 2005. ApJ 625, 754.
\bibitem{} Wielen, R., 1977. A\&A 60, 263.
%\bibitem{} Wilmes, M., K\"oppen, J., 1995. A\&A 294, 47.
\bibitem{} Worthey, G., Dorman, B., Jones, L.A., 1996. AJ 112, 948.
\bibitem{} Wyse, R.F.G., Gilmore, G., 1992. AJ 104, 144.
\bibitem{} Wyse, R.F.G., Gilmore, G., 1995. AJ 110, 2771.
%\bibitem[Yoshii]{yos84} Yoshii, Y.: 1984, \aj 89, 1190
%\bibitem{} Yoshii, Y., Tsujimoto, T.,
% Kowara, K., 1998. ApJ 507, L113.
%\bibitem[Young et al.,]{yoa96} Young, J.S., Allen, L.,
% Kenney, J.D.P., Lesser, A., Rownd, B.: 1996,
% \aj 112, 1903
%\bibitem[Zaroubi et al.,]{zaa96} Zaroubi, S., Naim, A.,
% Hoffman, Y.: 1996, \apj 457, 50
%\bibitem{} Zinn, R., West, M.J., 1984. ApJS 55, 45. (ZW) 
%\bibitem{} Zinn, R., 1993. In: Smith G.H.,
% Brodie J.P. (eds.) ASP Conference Series, Vol.\,48,
% The Globular Cluster-Galaxy Connection, p.\,38.
%\bibitem{} Zoccali, M., Renzini, A., Ortolani, S., et al.,,
%2003. A\&A 399, 931.
\end{thebibliography}
\end{document}